\documentclass[reprint,nofootinbib,amsmath,amssymb,aps,eqsecnum,showpacs,twocolumn,superscriptaddress,prb,amsart]{revtex4-2}
\usepackage{graphicx}
\usepackage[nointegrals]{wasysym} 
\usepackage[export]{adjustbox}
\usepackage{amsmath,amsfonts,amssymb,latexsym}
\usepackage{hhline}
\usepackage{bm}
\usepackage{verbatim}
\usepackage{enumitem}
\usepackage{mathrsfs}
\usepackage{slashed}
\usepackage{empheq}
\usepackage{mwe}

\def\({\left(}
\def\){\right)}
\def\[{\left[}
\def\]{\right]}
\newcommand{\ie}{\begin{equation}\begin{aligned}}
\newcommand{\fe}{\end{aligned}\end{equation}}

\usepackage{subcaption}
\usepackage[usenames,dvipsnames]{xcolor}
\definecolor{shadecolor}{gray}{0.9}
\usepackage{framed}

\usepackage{hyperref} 
\hypersetup{final}
\hypersetup{colorlinks, citecolor=red, linkcolor=red, urlcolor=red}

\usepackage{braket}

\usepackage{dcolumn}
\usepackage{pifont}
\usepackage{ulem}

\begin{document}
\hfill MIT-CTP/5617
\title{Classification of Dipolar Symmetry-Protected Topological Phases: \\
Matrix Product States, Stabilizer Hamiltonians and Finite Tensor Gauge Theories}

\author{Ho Tat Lam}
\email[Electronic address:$~~$]{htlam@mit.edu}
\affiliation{Department of Physics, Massachusetts Institute of Technology, Cambridge, Massachusetts 02139, USA}

\date{\today}

\begin{abstract}
We classify one-dimensional symmetry-protected topological (SPT) phases protected by dipole symmetries. A dipole symmetry comprises two sets of symmetry generators:\ charge and dipole operators, which together form a non-trivial algebra with translations. Using matrix product states (MPS), we show that for a $G$ dipole symmetry with $G$ a finite abelian group, the one-dimensional dipolar SPTs are classified by the group $H^2[G\times G,U(1)]/H^2[G,U(1)]^2$. Because of the symmetry algebra, the MPS tensors exhibit an unusual property, prohibiting the fractionalization of charge operators at the edges. For each phase in the classification, we explicitly construct a stabilizer Hamiltonian to realize the SPT phase and derive the response field theories by coupling the dipole symmetry to background tensor gauge fields. These field theories generalize the Dijkgraaf-Witten theories to twisted finite tensor gauge theories.
\end{abstract}

\maketitle

\tableofcontents

\section{Introduction}

Classification of phases of matter has been a long term program in condensed matter physics. It has been recognized that in the presence of global symmetry, even the simplest short-ranged entangled, symmetric gapped phases admit multiple distinct phases, known as symmetry-protected topological (SPT) phases \cite{Chen2011-et,Turner2011-zi,fidkowski2011topological,Schuch2011-jx,Chen2011-kz,Pollmann2012-lv,Chen2012-oa,senthil15}. In recent years, significant progress has been made in the classification of these SPT phases (see e.g.\ \cite{Levin2012-dv,Vishwanath2013-pb,Yao2013-vc,else2014classifying,Gu2014-lj,chen14,Kapustin:2014tfa,Kapustin:2014dxa,Gaiotto2016-ba,Freed:2016rqq,cheng-PRB18,PhysRevX.8.011055,PhysRevX.10.031055} and the reference therein).

Recently, a new kind of symmetries, known as modulated symmetries, has featured prominently in many different contexts ranging from quantum dynamics \cite{PhysRevX.9.021003,feldmeier20,sala2020ergodicity,gromov20,nandkishore21,Nandkishore-multipole,sala2022dynamics,glorioso22,glorioso2023goldstone,han2023scaling,Jain:2023nbf,Gliozzi:2023zth} to fracton topological orders \cite{Chamon2005-fc,Haah2011-ny,PhysRevB.94.235157,Prem_2019} and fractonic phases of matter \cite{Paramekanti_2002,Prem-Pretko,He:2019ktp,Yuan_2020,Gorantla:2022eem,lake1,Jensen:2022iww,lake2,feldmeier,lake2023non,chen2023manybody} (see \cite{Nandkishore_2019,Pretko_2020} for a review on these subjects). They are also realized experimentally by ultracold atoms placed in strongly tilted optical lattices \cite{sachdev02,pielawa11,bakr20,aidelsburger21,kohlert2021experimental,weitenberg22}. Unlike conventional symmetries which act uniformly in space, modulated symmetries are symmetries that act in a non-uniform, spatially modulated manner. Examples of modulated symmetries encompass multipole symmetries acting polynomially in space \cite{pretko18,gromov2019towards,Seiberg:2019vrp,Dubinkin_2021,Stahl_2022,Gorantla:2022eem,bulmash2023multipole,Burnell:2023fsr}, exponential symmetries acting with exponential modulations \cite{watanabe23,watanabe2022ground,delfino20232d}, subsystem symmetries acting on rigid sub-manifolds \cite{PhysRevB.72.045137, Nussinov_2009,you2018subsystem,paper1,paper2,paper3,Gorantla:2020xap,Distler:2021qzc,Burnell:2021reh,Rayhaun:2021ocs}, fractal symmetries acting on fractals \cite{PhysRevE.60.5068,Castelnovo_2012,PhysRevB.88.125122,PhysRevB.94.155128,Devakul_2019}, $L$-cycle symmetries whose symmetry operators vary periodically over $L$ sites \cite{Lcyclesym,Stephen_2019,Stephen:2022zyt} and many more. 

It is natural to ask whether there are distinct topological phases protected by modulated symmetries, and if so, what their classifications are. This is not an isolated problem, as gauging a modulated SPT often leads to fracton topological orders \cite{Dominic-gauging,PhysRevB.94.235157,pretko18,Shirley_2019}. Therefore, classifying modulated SPTs can help shed light on the classification of fracton orders. In addition, as modulated symmetries naturally couple to tensor gauge fields, classifying modulated SPTs is equivalent to classifying twisted tensor gauge theories, which are analogs of Dijkgraaf-Witten theories \cite{Dijkgraaf:1989pz} for tensor gauge fields.

Much of our understanding of modulated SPTs has been centered around subsystem SPTs \cite{you2018subsystem,You:2018zhj,Shirley_2020_twisted,devakul2018classification,Stephen_2019_SSPT,Devakul_2020,Hsin:2021mjn} and fractal SPTs \cite{Devakul_2019,devakul2019classifying}. Recently, a broad class of modulated SPTs protected by symmetries such as dipole, quadrupole and exponential symmetries has been systematically constructed by generalizing the idea of decorated domain walls to modulated symmetries \cite{Han:2023fas}. 
In this work, we focus on the classification of the simplest modulated SPTs -- one-dimensional dipolar SPTs. 
We approach the problem through a variety of different methods, including constructions of stabilizer Hamiltonians, matrix product states (MPS), and response field theories. These methods can also be adapted to classifying SPTs protected by other types of modulated symmetries. 

A general lesson we learn is that not every anomaly of the symmetry operators can be realized on the boundary of a modulated SPT. Consider a dipolar SPT with a finite abelian $G$ dipole symmetry, which comprises two sets of symmetry generators:\ charge and dipole operators. From the boundary perspective, these symmetry operators act as an ordinary $G\times G$ symmetry; however, not all possible $G\times G$ anomalies, or equivalently $G\times G$ projective representations, are permitted. Notably, the charge operators always act linearly on the boundaries, owing to the non-trivial algebraic relation between the charge and dipole operators. In the end, utilizing dipole-symmetric MPS, $G$ dipolar SPTs with $G$ a finite abelian group are classified by the group
\ie\label{eq:classification-group-intro} 
\mathcal{C}[G]=\frac{H^2[G\times G,U(1)]}{H^2[G,U(1)]^2}~,
\fe
which is a quotient of $H^2[G\times G,U(1)]$, the classification group of an ordinary $G\times G$ SPT. This classification is further substantiated by the other constructive approaches.

The rest of the paper is organized as follows. In Sec.\ \ref{sec:Stabilizer}, we follow a constructive approach and establish families of stabilizer Hamiltonians that possess non-trivial dipolar SPT orders. In Sec.\ \ref{sec:MPS}, we employ MPS to show that distinct $G$ dipolar SPTs with $G$ a finite abelian group are in one-to-one correspondence with the elements of the group in \eqref{eq:classification-group-intro} and the constructions in Sec.\ \ref{sec:Stabilizer} exhaust all possible dipolar SPTs. 
In Sec.\ \ref{sec:field-theory}, we couple the dipole symmetry to background tensor gauge fields, construct quantized gauge invariant response field theories and match them with the classification proposed in Sec.\ \ref{sec:MPS}. 

Some technical details and background materials are provided in the appendices. In App.\ \ref{app:U(1)}, we give an overview on $U(1)$ dipole symmetry. App.\ \ref{app:quadrupole} provides more details on the quadrupolar SPTs discussed in Sec.\ \ref{sec:Stabilizer}. App.\ \ref{app:sym-MPS} proves an unusual property of the dipole-symmetric MPS used in Sec.\ \ref{sec:MPS}. App.~\ref{app:projective} reviews projective representations of finite abelian groups and their classifications based on the second group cohomology. App.~\ref{app:DW_theory} reviews the Dijkgraaf-Witten theories for finite abelian gauge groups in 1+1 dimensions. App.~\ref{app:compactness} provides a rigorous treatment of the compactness of tensor gauge fields coupled to dipole symmetries using the modified Villain formalism developed in \cite{Sulejmanpasic:2019ytl,Gorantla:2021svj,Fazza:2022fss}. Finally, App.\ \ref{app:derivation-response-theory} derives the response field theories in Sec.\ \ref{sec:field-theory} directly from the stabilizer Hamiltonians constructed in Sec.\ \ref{sec:Stabilizer}. 

\section{Stabilizer Hamiltonians}
\label{sec:Stabilizer}
One-dimensional SPTs protected by a $\mathbb{Z}_N$ dipole symmetry were constructed in \cite{Han:2023fas} using the method of decorated domain walls. In this section, we review the construction and then generalize it to SPTs protected by a general finite abelian dipole symmetry, which introduces new features.

\subsection{$\mathbb{Z}_N$ dipolar models}
Consider a $\mathbb{Z}_N$ spin chain. At each site, there are $N$ states $|h\rangle$, $h\in\mathbb{Z}_N$ and the $X,Z$ operators that act as 
 \ie
 Z|h\rangle=e^{\frac{2\pi ih}{N}}|h\rangle~,\quad X|g\rangle=|h+1\rangle~.
 \fe
A $\mathbb{Z}_N$ dipole symmetry is abstractly characterized by its $\mathbb{Z}_N$ charge and dipole operators, $Q$ and $D$, and the algebra they form with the lattice translation operator $\mathcal{T}$,
\ie
\mathcal{T}^{-1} Q \mathcal{T} = Q~,\quad \mathcal{T}^{-1} D \mathcal{T} = Q D~.
\fe
We can pick a specific realization of the charge and dipole symmetry operators as
\ie\label{eq:dipole-charge}
Q= \prod_j X_j , \quad D = \prod_j ( X_j )^j~.
\fe
Here, $X_j$ can be thought of as the single-site charge operator at site $j$;  then, $Q$ and $D$ are the overall charge and the overall dipole moment of the charge distribution, respectively.\footnote{The dipole operator is well-defined on an infinite chain, but it is no longer compatible with the periodic boundary condition $X_j=X_{j+L}$ on a closed chain unless the length $L$ is an integer multiple of $N$ \cite{Gorantla:2022eem}. One can nevertheless still make sense of the dipole symmetry when $L$ is not divisible by $N$ by resorting to the notion of bundle symmetries. See \cite{Han:2023fas} for more discussions. To avoid this subtlety, we will mostly work on either an infinite chain or an open chain, unless specified.}~These $\mathbb{Z}_N$ symmetry operators can be motivated by drawing analogy with the $U(1)$ dipole symmetry, as reviewed in App.\ \ref{app:U(1)}.

The $\mathbb{Z}_N$ dipolar model introduced in \cite{Han:2023fas} is an example of $\mathbb{Z}_N$ dipolar SPTs protected by the dipole symmetry in \eqref{eq:dipole-charge}. Its Hamiltonian, 
\ie\label{eq:ZN_Hamiltonian}
&H =-\sum_{j} \left( a_j+\text{h.c.}\right)~,
\\
&a_j=(Z_{j-1} Z^\dag_j)^\eta\cdot X_j\cdot (Z^\dag_j Z_{j+1})^\eta~,
\fe
is a stabilizer Hamiltonian, where every term in the Hamiltonian commutes with each other. The stabilizers $a_j$ are the single-site charge operator $X_j$ decorated by the dipole domain wall operator $(Z_{j-1} Z^\dag_{j})^\eta\cdot(Z^\dag_j Z_{j+1})^\eta$. This resembles the decorated domain wall construction for the ordinary SPTs \cite{chen14}. The parameter $\eta$ is valued in
\ie
\eta\in\mathbb{Z}_N~,
\fe 
suggesting a $\mathbb{Z}_N$ classification for $\mathbb{Z}_N$ dipolar SPTs. This proposal will be confirmed by the MPS classification in Sec.\ \ref{sec:MPS} and the response field theory analysis in Sec.\ \ref{sec:field-theory}.

On a close periodic chain, the Hamiltonian has a unique short-ranged entangled ground state preserving the dipole symmetry in \eqref{eq:dipole-charge}. Since the stabilizers all commute, the ground state $|\Psi\rangle$ is an eigenstate of all of them,  $a_j|\Psi\rangle=|\Psi\rangle$. The ground state has non-zero expectation values for two string order parameters 
\ie
S_Q &=\prod_{n=1}^{m}a_{j+n} = {V_{j}^\eta}^\dagger\cdot \prod_{n=1}^{m} X_{j+n} \cdot V_{j+m}^{\eta}~, \nonumber \\ 
S_D &= \prod_{n=1}^{m}a_{j+n}^n
={W_{j}^{\eta}}^\dagger\cdot \prod_{n=1}^{m} X_{j+n}^{n} \cdot W_{j+m}^{\eta} V_{j+m}^{m\eta} ~,
\fe
indicating a non-trivial dipolar SPT order. The string order parameters are strings of charge and dipole operators dressed with dipoles $V_j=Z^\dagger_{j} Z_{j+1}$ and charges $W_j=Z_j^\dagger$ at the end points, respectively. The dipolar SPT order is also evident in the degeneracy observed in the ground state entanglement spectrum of a finite region \cite{Han:2023fas}.

When placed on an open chain $1\leq j\leq L$ of length $L$, the symmetry operators fractionalize into products of edge operators, $Q\sim{\cal L}_Q \times {\cal R}_Q$ and $D\sim{\cal L}_Q \times {\cal R}_Q$. These edge operators are
\begin{alignat}{2}
{\cal L}_Q &= X_1 V_1^{\eta}~,\quad &&{\cal R}_Q = {V_{L-1}^{\eta\dagger}} X_L~,\nonumber  \\
{\cal L}_D &
= {\cal L}_Q \cdot W_1^{\eta}~, \quad 
&&{\cal R}_D 
=
{\cal R}_Q ^L\cdot W_L^{\eta\dagger}~.
\end{alignat}
It is straightforward to check that operators related by $\sim$ differ by only products of the stabilizers, thus acting on the ground states in the same way. This also implies that the edge operators that are far separated should independently commute with the Hamiltonian.
These edge operators obey the algebra,
\ie\label{eq:edge-algebra-ZN}
{\cal L}_Q {\cal L}_D = e^{\frac{2\pi i\eta}{N}} {\cal L}_D {\cal L}_Q ,\quad {\cal R}_Q {\cal R}_D = e^{-\frac{2\pi i\eta}{N} } {\cal R}_D {\cal R}_Q~,
\fe
that the charge and dipole operators do not commute at the edges.
This enforces a minimal $N/\gcd(N,\eta)$ number of protected edge modes per edge, that transform as a projective representation under the charge and dipole operators.

\subsection{Charge-dipole dipolar SPTs}
We now proceed to generalize the $\mathbb{Z}_N$ dipolar model. As a warm-up, consider the case when $G=\mathbb{Z}_N\times\mathbb{Z}_M$. The models are defined on a spin chain where every site has a $\mathbb{Z}_N$ and a $\mathbb{Z}_M$ spin with their own $X,Z$ operators denoted by $X,Z$ and $\tilde X,\tilde Z$ respectively. We choose a realization of the $\mathbb{Z}_N\times\mathbb{Z}_M$ dipole symmetry as
\ie\label{eq:dipole-charge-2} 
&Q=\prod_j X_j~,\quad D=\prod_j X_j^j~,
\\
&\tilde Q=\prod_j \tilde X_j~,\quad \tilde D=\prod_j \tilde X_j^j~,
\fe 
where $Q,D$ belongs to the $\mathbb{Z}_N$ dipole symmetry, while $\tilde Q,\tilde D$ belongs to the $\mathbb{Z}_M$ dipole symmetry.

A direct generalization of the $\mathbb{Z}_N$ model in \eqref{eq:ZN_Hamiltonian} leads to the following dipole-symmetric stabilizer Hamiltonian
\ie\label{eq:ZN_ZM_Hamiltonian}
&H=-\sum_j\left(a_j+\tilde a_j+\text{h.c.}\right)~,
\\
&a_j=V_{j-1}^{\eta\dagger}\cdot(\tilde V^{\dagger}_{j-1})^{\frac{M\zeta}{K}}\cdot X_j\cdot \tilde V_{j}^{\frac{M\zeta}{K}}\cdot V_{j}^{\eta}~,
\\
&\tilde a_j=\tilde V_{j-1}^{\tilde\eta\dagger}\cdot  (V_{j-1}^\dagger)^{\frac{N\zeta}{K}}\cdot \tilde X_j\cdot V_{j}^{\frac{N\zeta}{K}}\cdot \tilde V_{j}^{\tilde\eta}~.
\fe
Here, $V_j=Z^\dagger_{j} Z_{j+1}$, $\tilde V_j=\tilde Z^\dagger_{j} \tilde Z_{j+1}$, $K=\gcd(N,M)$, and the parameters are valued in
\ie
\{\eta,\tilde\eta,\zeta\}\in \mathbb{Z}_N\times \mathbb{Z}_M\times\mathbb{Z}_{K}~.
\fe
The stabilizers are single-site charge operators $X_j$, $\tilde X_j$ decorated by the dipole domain wall operators:\ $\mathcal{W}_j=V^\dagger_{j-1}\cdot V_j$ and $\tilde{\mathcal{W}}_j=\tilde V^\dagger_{j-1}\cdot \tilde V_j$. Notably, the decoration of $X_j$ by $\tilde{\mathcal{W}}_j$ and  $\tilde{X}_j$ by $\mathcal{W}_{j}$ are correlated such that the stabilizers commute with each other.  This crucially ensures the gappedness of the Hamiltonian and the uniqueness of the ground state (assuming there is no redundancy among the stabilizers) on a close periodic chain.

On an open chain of length $L$, the symmetry operators in \eqref{eq:dipole-charge-2} fractionalize into products of edge operators. The left edge operators obey the algebra,
\begin{alignat}{2}\label{eq:charge-dipole-frac-ZNZM}
&{\cal L}_Q {\cal L}_D = e^{\frac{2\pi i\eta}{N}} {\cal L}_D {\cal L}_Q ~,\quad &&\tilde {\cal L}_{Q} \tilde{\cal L}_{ D} = e^{\frac{2\pi i\tilde \eta}{M}} \tilde{\cal L}_{ D} \tilde{\cal L}_{ Q} ~,\nonumber
\\
&{\cal L}_Q \tilde{\cal L}_{D} = e^{2\pi i \zeta\over K} \tilde{\cal L}_{D} {\cal L}_Q ~,
&&\tilde{\cal L}_{ Q} {\cal L}_D = e^{2\pi i\zeta\over K} {\cal L}_D \tilde{\cal L}_{ Q} ~,
\end{alignat}
where trivial commutation relations between commuting operators are omitted. The right edge operators obey a similar algebra but with the opposite projectivity. Note that the commutations between $\mathcal{L}_Q$ and $\tilde{\cal L}_D$, and between $\tilde{\cal L}_Q$ and ${\cal L}_D$, are correlated and both determined by the parameter $\zeta$. In \eqref{eq:charge-dipole-frac-ZNZM}, all the fractionalization at the edges are due to the noncommutativity between the charge and dipole operators. We will refer to such kind of dipolar SPTs as \textit{charge-dipole dipolar SPTs}. It is natural to wonder whether the charge or dipole operators themselves can be noncommutative at the edges. We will soon answer this question in the next subsection.


For now, let's extend the above construction to more general $G$ charge-dipole dipolar SPTs, with $G=\prod \mathbb{Z}_{N_I}$ a finite abelian group. The Hamiltonians are a sum of commuting stabilizers, which are single-site charge operators decorated by dipole domain wall operators. We can decorate the single-site charge operator of the $I$-th cyclic factor in $G$ by the dipole domain wall operator of the $J$-th cyclic factor. This decoration needs to be correlated with the decoration of the $J$-th single-site charge operator by the $I$-th dipole domain wall operator so that the stabilizers commute. The parameter for such decoration is valued in $\mathbb{Z}_{N_{IJ}}$, with $N_{IJ}\equiv \text{gcd}(N_I,N_J)$. All together, the charge-dipole dipolar SPTs constitute a $\prod_{I\leq J} \mathbb{Z}_{{N}_{IJ}}$ class of $G$ dipolar SPTs.

\subsection{Dipole-dipole dipolar SPTs}
We now return to the question whether the dipole operators themselves can be noncommutative at the edges of a dipolar SPT. We give an affirmative answer to this question by constructing the following stabilizer Hamiltonian on the $\mathbb{Z}_N\times\mathbb{Z}_M$ spin chain,
\ie\label{eq:quandrupole_Hamiltonian}
    &H =-\sum_{j} (a_{j}+\tilde a_{j}+\text{h.c.})~, 
    \\
    &a_{j}= (\tilde{Z}_{j-1} \tilde{Z}_{j}^{3\dagger})^{\frac{M\xi}{K}} \cdot X_{j+1}^\dagger \cdot (\tilde{Z}_{j+1}^3 \tilde{Z}_{j+2}^{\dagger})^{\frac{M\xi}{K}}~,
    \\
    &\tilde a_{j}= (Z_{j-1} Z_{j}^{3\dagger})^{\frac{N\xi}{K}}\cdot \tilde X_{j}\cdot (Z_{j+1}^3 Z_{j+2}^{\dagger})^{\frac{N\xi}{K}} ~,
\fe
where the parameter $\xi$ is valued in 
\ie
\xi\in\mathbb{Z}_K~.
\fe
The Hamiltonian is symmetric with respect to the $\mathbb{Z}_N\times\mathbb{Z}_M$ dipole symmetry in \eqref{eq:dipole-charge-2}, which in fact is part of a larger $\mathbb{Z}_N\times\mathbb{Z}_M$ quadrupole symmetry with quadrupole operators
\ie\label{eq:quadrupole-charge}
P = \prod_j ( X_j )^{j^2}~,\quad \tilde{P} = \prod_j ( \tilde{X}_{j} )^{j^2} ~.
\fe 
Applying a lattice translation $\mathcal{T}$ on the quadrupole operators generate the charge and dipole operators in \eqref{eq:dipole-charge-2}
\ie
\mathcal{T}^{-1}P\mathcal{T}=QD^2P~.
\fe
The Hamiltonian in \eqref{eq:quandrupole_Hamiltonian} defines an SPT protected by the $\mathbb{Z}_N\times\mathbb{Z}_M$ quadrupole symmetry, generalizing the quadrupolar SPT constructed in \cite{Han:2023fas}.  The stabilizers are single-site charge operators decorated by quadrupole domain wall operators, e.g.\ $Z_{j-1}Z_j^{3\dagger} Z^3_{j+1} Z_{j+2}^\dagger$. The decorations in the stabilizers, $a_j$ and $\tilde a_j$, are correlated so that they commute with each other.

Despite the enhanced quadrupole symmetry, we will neglect the quadrupole operators $P$, $\tilde P$, and view the Hamiltonian in \eqref{eq:quandrupole_Hamiltonian} as a $\mathbb{Z}_N\times\mathbb{Z}_M$ dipolar SPT. In particular, it means that continuous deformations preserving the dipole symmetry but violating the quadrupole symmetry are allowed. For completeness, in App.\ \ref{app:quadrupole}, we include more details about the Hamiltonian \eqref{eq:quandrupole_Hamiltonian} when interpreted as a quadrupolar SPT and discuss the relation between the dipolar SPT and the quadrupolar SPT perspectives of the Hamiltonian. 

When placed on an open chain $1\leq j\leq L$, the $\mathbb{Z}_N\times\mathbb{Z}_M$ dipole symmetry operators fractionalize at the edges. The symmetry operators localized on the left edge are
\begin{alignat}{2}
&\mathcal{L}_{Q}=X_1X_2(\tilde{Z}_1\tilde{Z}_2^{2\dagger}\tilde Z_3)^{\frac{M\xi}{K}}~,\quad &&\tilde{\mathcal{L}}_{ Q}=\tilde X_1({Z}_1^\dagger{Z}_2^{2} Z_3^\dagger)^{\frac{N\xi}{K}}~,\nonumber
\\
&\mathcal{L}_{D}=X_1X_2^2(\tilde{Z}_1^3\tilde{Z}_2^{5\dagger}\tilde Z_3^2)^{\frac{M\xi}{K}}~,\quad &&\tilde{\mathcal{L}}_{ D}=\tilde X_1({Z}_1^{2\dagger}{Z}_2^{3} Z_3^\dagger)^{\frac{N\xi}{K}}~.\nonumber
\end{alignat}
They obey the algebra,
\ie\label{eq:algebra-dipole-dipole}
\mathcal{L}_{D}\tilde{\mathcal{L}}_{{D}}=e^{-\frac{2\pi i\xi}{K}}\tilde{\mathcal{L}}_{{D}}\mathcal{L}_D~,
\fe
that only the dipole operators do not commute at the edges. We will refer to such kind of dipolar SPTs as \textit{dipole-dipole dipolar SPTs}. 

Let's now extend the construction to more general $G$ dipole-dipole dipolar SPTs, with $G=\prod\mathbb{Z}_{N_{I}}$ a finite abelian group. The stabilizers in the Hamiltonians are single-site charge operators decorated by quadrupole domain walls. Notably, not all possible decorations lead to commuting stabilizers. First of all, decorating the single-site charge operators by the quadrupole domain walls from the same cyclic factor in $G$ leads to noncommuting stabilizers, e.g. $a_j=(Z_{j-1} Z_{j}^{3\dagger})^{\xi}\cdot X_{j} \cdot(Z_{j+1}^3 Z_{j+2}^{\dagger})^{\xi}$ with $[a_j,a_{j+1}]\neq 0$. Next, the decoration of the $I$-th single-site charge operator by the $J$-th quadrupole domain wall needs to be correlated with the opposite decoration so that the stabilizers commute. Such decorations are parameterized by  $\mathbb{Z}_{N_{IJ}}$, with $N_{IJ}\equiv\text{gcd}(N_I,N_J)$. All together, the dipole-dipole dipolar SPTs form a $\prod_{I<J} \mathbb{Z}_{{N}_{IJ}}$ class of $G$ dipolar SPTs.

In summary, there are two types of dipolar SPTs:\ the charge-dipole dipolar SPTs with the charge and dipole operators not commuting at the edges, and the dipole-dipole dipolar SPTs with the dipole operators themselves not commuting at the edges. They generate a
\ie\label{eq:C[G]-abelian}
\mathcal{C}[G]=\prod_{I\leq J}\mathbb{Z}_{N_{IJ}}\times \prod_{I< J}\mathbb{Z}_{N_{IJ}}
\fe
class of $G=\prod\mathbb{Z}_{N_I}$ dipolar SPTs. The first factor classifies charge-dipole dipolar SPTs, while the second factor classifies dipole-dipole dipolar SPTs. What about charge-charge dipolar SPTs with the charge operators not commuting at the edges? We will argue in next section that they are forbidden, and the dipolar SPTs we constructed in this section exhaust all the possibilities.

\section{MPS classification}
\label{sec:MPS}
Matrix product state (MPS) is an efficient way to encode the ground state of a one-dimensional translation invariant gapped Hamiltonians \cite{Fannes1992}. It also provides a classification scheme for ordinary one-dimensional SPTs \cite{Chen2011-et,Pollmann2012-lv}. See \cite{bridgeman17,Cirac21} for a review on MPS. Here, we employ the machinery of MPS to classify one-dimensional dipolar SPTs.

An MPS on a periodic closed chain
\ie
|\psi\rangle=\sum_{\mathbf{h}}\text{Tr}[A^{h_1}A^{h_2}\cdots A^{h_L}]|\mathbf{h}\rangle~,
\fe
is constructed by contracting the MPS tensor $A^h_{ij}$ where $h$ is the physical label and $i,j$ are the internal indices. 

\begin{figure*}
\centering
\captionsetup[subfigure]{justification=centering}
\begin{subfigure}[c]{0.45\textwidth}
\centering
\includegraphics[scale=0.6]{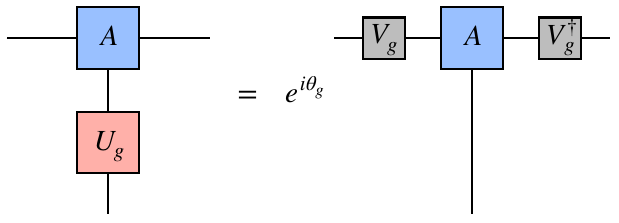}
\subcaption{Push $U_g$ through the MPS tensor via \eqref{eq:MPS-push}.}\label{fig:MPS-push}
\end{subfigure}
\qquad
\begin{subfigure}[c]{0.45\textwidth}
\centering
\includegraphics[scale=0.6]{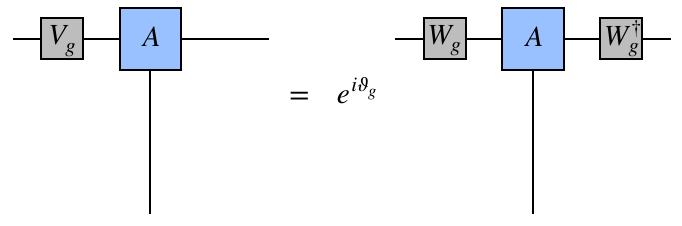}
\subcaption{Push $V_g$ through the MPS tensor via \eqref{eq:MPS-push-virtual}.}
\label{fig:MPS-V-push}
\end{subfigure}
\begin{subfigure}[c]{1\textwidth}
\centering
\includegraphics[scale=0.6]{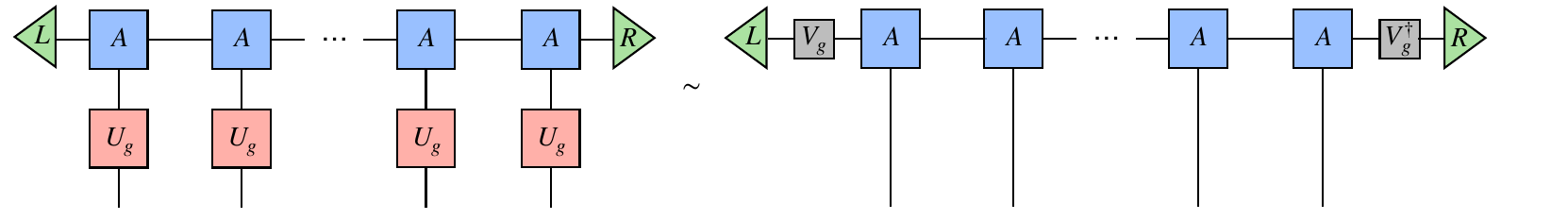}
\includegraphics[scale=0.6]{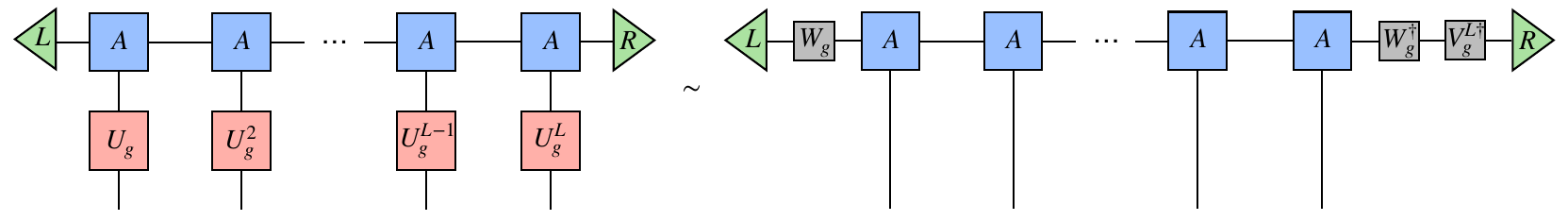}
\subcaption{Push the charge operator $Q_g=\prod_xU_g(x)$ (top figure) and the dipole operator $D_g=\prod_x[U_g(x)]^x$ (bottom figure) through the MPS on an open chain as in \eqref{eq:pushing-Q-D-through}.}
\label{fig:MPS-push-open-chain}
\end{subfigure}
\caption{Diagrammatic representations of \eqref{eq:MPS-push}, \eqref{eq:MPS-push-virtual} and \eqref{eq:pushing-Q-D-through} describing how to push various operators through the MPS tensor or the MPS on an open chain. }\label{fig:1}
\end{figure*}

Suppose the MPS is symmetric with respect to a $G$ dipole symmetry, with $G$ a finite abelian group and the charge and dipole operators defined as
\ie
Q_g=\prod_x U_g(x)~,\quad D_g=\prod_x [U_g(x)]^x~,
\fe
where $U_g$ is a unitary operator labeled by $g\in G$.
The invariance under the charge operators implies that the unitary operator $U_g$ can be pushed through the MPS tensor as follows:
\ie\label{eq:MPS-push}
U_g\cdot A_{ij}^{h}=e^{i\theta_g}(V_gA^{h}V_g^\dagger)_{ij}~,
\fe
as illustrated in Fig.~\ref{fig:MPS-push} \cite{P_rez_Garc_a_2008}. We review the proof in App.\ \ref{app:sym-MPS}. Using this property, we can push the dipole operator $D_g$ through the MPS tensor and get
\begin{align}
D_g|\psi\rangle&\,\dot{=}\sum_{\mathbf{h}}\text{Tr}[V_gA^{h_1}V_g^\dagger\cdot V_g^2A^{h_2}V_g^{2\dagger}\cdots V_g^LA^{h_L}V_g^{L\dagger}]|\mathbf{h}\rangle\nonumber
\\
&\,\dot{=}\sum_{\mathbf{h}}\text{Tr}[V_gA^{h_1}V_gA^{h_2}\cdots V_g A^{h_L}V_g^{L\dagger}]|\mathbf{h}\rangle~,
\end{align}
where $\dot{=}$ means that the two states are the same up to an overall phase factor. Recall that the dipole operator is generically incompatible with the periodic boundary condition \cite{Gorantla:2022eem}. Hence, on a periodic chain, we only require the dipole-symmetric MPS to be symmetric with respect to the compatible dipole operators $D_g$, which has $U_g^L=V_g^L=1$; the dipole-symmetric MPS obeys
\ie\label{eq:MPS-constraint}
|\psi\rangle\,\dot{=}\, D_g|\psi\rangle
\,\dot{=}\,{\sum_{\mathbf{h}}\text{Tr}[V_gA^{h_1}V_gA^{h_2}\cdots V_gA^{h_L}]|\mathbf{h}\rangle~,}
\fe
where $V_g^L$ is dropped from the trace.
This condition holds for any length $L$ if and only if the virtual operator $V_g$ can be pushed through the MPS tensor as follows:
\ie\label{eq:MPS-push-virtual}
(V_gA^h)_{ij}=e^{i\vartheta_g}(W_gA^hW_g^\dagger)_{ij}~,
\fe
as illustrated in Fig.\ \ref{fig:MPS-V-push}.
It is straightforward to check that \eqref{eq:MPS-push-virtual} implies \eqref{eq:MPS-constraint}. In App.\ \ref{app:sym-MPS}, we prove the other direction of the statement.

We are now ready to discuss how the symmetry fractionalizes on an open chain. On an open chain, the MPS 
\ie
|B\rangle=\sum_{\mathbf{h}}\text{Tr}[BA^{h_1}A^{h_2}\cdots A^{h_L}]|\mathbf{h}\rangle
\fe
are labeled by the operator $B=|R\rangle\langle L|$ formed by the virtual boundary state $|L\rangle/|R\rangle$ for the left/right boundary. Using the two properties of the MPS tensor in \eqref{eq:MPS-push}, \eqref{eq:MPS-push-virtual}, we can push the charge and dipole operators through the MPS and get
\ie\label{eq:pushing-Q-D-through}
&Q_g\left|B\right\rangle\dot{=} \left|V_g^\dagger BV_g\right\rangle
\\
&D_g\left|B\right\rangle\dot{=} \left|W_g^\dagger V_g^{L\dagger} BW_g\right\rangle~,
\fe
as illustrated in Fig.\ \ref{fig:MPS-push-open-chain}.

Although the MPS $|B\rangle$ are in a linear representation of the dipole symmetry, the virtual operators $V_g, W_g$ can act projectively on the virtual boundary states $|L\rangle$, $|R\rangle$; their projectivity characterizes different dipolar SPTs. As the charge and dipole operators form a $G\times G$ symmetry group, the most general projective representations of the virtual operators are labeled by the elements of the second group cohomology
$H^2(G\times G,U(1))$. (See App.\ \ref{app:projective} for a review on projective representations and its relation to group cohomology.) 
However, as we will explain below, the linearity of the charge and dipole operators on the physical states precludes some projective representations of the virtual operators. 

Being a finite abelian group, the projective representations of $G\times G$ can be uniquely characterized by the commutation relations of the virtual operators,
\ie\label{eq:projective}
V_gV_h&=e^{i\phi_{v}(g,h)}V_h V_{g}~,
\\
W_gW_h&=e^{i\phi_{w}(g,h)}W_{h} W_g~,
\\
V_gW_h&=e^{i\varphi(g,h)}W_{h} V_g~.
\fe
As linear representations, the $G\times G$ symmetry operators should commute when acting on the MPS $|B\rangle$. The commutativity of the charge operators follows automatically from \eqref{eq:projective},
\ie
Q_{g}Q_{h}|B\rangle&=|V_g^\dagger V_h^\dagger BV_hV_g\rangle
\\
&=|V_h^\dagger V_g^\dagger BV_gV_h\rangle=Q_{h}Q_{g}|B\rangle~.
\fe
On the other hand, the commutativity of the charge operators with the dipole operators, 
\ie
Q_{g}D_{h}|B\rangle&=|V_g^\dagger W_h^\dagger V_h^{L\dagger} BW_hV_g\rangle
\\
&=e^{i\phi_{v}(g,h)L}|W_h^\dagger V_h^{L\dagger} V_g^\dagger B V_g W_h\rangle
\\
&=e^{i\phi_{v}(g,h)L}D_{h} Q_{g}|B\rangle~,
\fe
and the commutativity of the dipole operators,
\ie
D_{g}D_{h}|B\rangle=e^{i[\phi_{v}(g,h)L+\varphi(g,h)-\varphi(h,g)]L}D_{h}D_{g}|B\rangle~,
\fe
for any $L$, imposes the following conditions
\ie
e^{i\phi_v(g,h)}=1~,\quad e^{i\varphi(g,h)}=e^{i\varphi(h,g)}~.
\fe
The first condition implies that the virtual boundary states are in a linear representation of $V_g$. In other words, there is no charge-charge dipolar SPTs with the charge operators acting projectively at the edges. Therefore, we should quotient out all projective representations of $V_g$ that form the group $H^2(G,U(1))$. The second condition implies that commutations between $V_g$ and $W_h$, and between $V_h$ and $W_g$, are correlated. As a result, the projectivity of the diagonal subgroup $(g,g)\in G$ of the $G\times G$ symmetry group is fixed by the projectivity of the charge and dipole operators as
\ie
V_gW_g\cdot V_hW_h
=e^{i\phi_{v}(g,h)+i\phi_w(g,h)} V_hW_h\cdot V_gW_g~.
\fe
Thus, we should quotient out another $H^2(G,U(1))$ that classifies the projective representations of the diagonal subgroup $G$.

In summary, not every projective representation can exist at the boundaries of dipolar SPTs; the allowed ones are in one-to-one correspondence to the quotient group
\ie\label{eq:classification}
\mathcal{C}[G]=\frac{H^2(G\times G,U(1))}{[H(G,U(1))]^2}~.
\fe
For a general finite abelian group $G=\prod\mathbb{Z}_{N_I}$,
\ie\label{eq:classification-explicit}
\mathcal{C}[G]=\prod_{I\leq J}\mathbb{Z}_{N_{IJ}}\times\prod_{I<J}\mathbb{Z}_{N_{IJ}}~,
\fe
where $N_{IJ}\equiv\text{gcd}(N_I,N_J)$ and in particular $N_{II}=N_I$. (See App.\ \ref{app:projective} for the detailed computation.) It is identical to \eqref{eq:C[G]-abelian}, confirming that the explicit constructions in Sec.\ \ref{sec:Stabilizer} cover all possible dipolar SPTs. 

\section{Response field theories}
\label{sec:field-theory}
In this section, we develop the response field theories for the dipolar SPTs. Inherited from the spin chain setup, we will work with a discrete spatial lattice and a continuous Lorentzian time.

To construct the response field theories, we couple the $G=\prod\mathbb{Z}_{N_I}$ dipole symmetry to background gauge fields. A $\mathbb{Z}_{N_I}$ dipole symmetry naturally couples to a $\mathbb{Z}_{N_I}$ tensor gauge field defined on sites\footnote{We refer to this gauge field as a tensor gauge field, since its higher-dimensional generalization $A_{ij}$ is a symmetric rank-2 tensor of the rotation group. See \cite{xu2006novel,Gu:2006vw,Xu_2006,Xu_2008,Gu:2009jh,rasmussen2016stable,pretko17,Pretko_2017_generalizeEM,Slagle:2017wrc,Pretko2017-ej,Ma:2018nhd,Bulmash:2018lid,Bulmash:2018knk,Slagle:2018kqf,pretko18,You:2019bvu,fractonCS,Wang:2019aiq,Wang:2019cbj,paper1,paper2,paper3,Nguyen:2020yve,Gorantla:2020xap,Dubinkin:2020kxo,Qi_2021,Gorantla:2020jpy,Rudelius:2020kta,Gorantla:2021svj,Burnell:2021reh,Du:2021pbc,Oh_2022_rank2toric,Gorantla:2022eem,Oh_2022_dipolarfieldtheory,Pace:2022wgl,Ebisu:2023ayu,Gorantla:2022mrp,Gorantla:2022ssr,Radzihovsky_lifshitz_2022,Gorantla:2022pii,Ohmori:2022rzz,Oh:2023bnk,Du:2023zdq,Nguyen:2023quf} for a partial list of works on tensor gauge theories.}
 \ie\label{eq:tensor_gauge_field}
A^I_t \rightarrow  A^I_{t}+\partial_t \alpha^I~,
\quad A^I_{xx}\rightarrow  A^I_{xx}+\Delta_x^2\alpha^I~,
\fe 
where the lattice difference operators are defined as
\ie
&\Delta_x \alpha(x) = \alpha(x+1)-\alpha(x)~,
\\
&\Delta_x^2 \alpha(x) = \alpha(x+1)-2\alpha(x)-\alpha(x-1)~.
\fe
These $\mathbb{Z}_N$ tensor gauge fields can be motivated by drawing analogy with the $U(1)$ dipole symmetry, as discussed in App.\ \ref{app:U(1)}.
Following \cite{Gorantla:2022eem}, we embed these $\mathbb{Z}_{N_I}$ tensor gauge fields into their $U(1)$ counterparts. In doing so, we introduce compact scalars $\phi^I\sim \phi^I+2\pi$ as the Lagrange multipliers that restrict the $U(1)$ gauge fields to $\mathbb{Z}_N$ gauge fields. This leads to the following Lagrangian for the response field theory
\ie\label{eq:full-action}
\mathcal{L}[A^I]+\sum_I \frac{N_I}{2\pi} \phi^I(x) \big[\partial_t A^I_{xx}(x)-\Delta_x^2A^I_t(x)\big]~,
\fe
where $\mathcal{L}[A^I]$ is a functional of $A^I$ that depends on specific dipolar SPTs.
Given that the tensor gauge field $A^I$ is compact, the Lagrangian should be accompanied by some integer gauge fields. We omit them here to avoid cluttering the equations. See App.\ \ref{app:compactness} for a precise treatment of the compactness of these gauge fields using the modified Villain formalism developed in \cite{Sulejmanpasic:2019ytl,Gorantla:2021svj,Fazza:2022fss}.

Our goal is to find the response functional $\mathcal{L}[A^I]$ for each dipolar SPT. The most direct approach is to derive the response field theories directly from the stabilizer Hamiltonians constructed in Sec.\ \ref{sec:Stabilizer}. As the derivations can be quite technical, we defer them to App.\ \ref{app:derivation-response-theory}. In the main text, we pursue an alternative path by enumerating as many valid response functionals as possible and matching them with the classification in \eqref{eq:classification-explicit}. There are two guiding principles:\ first, the response action should be gauge invariant on a closed chain; second, the coefficient of the response functional $\mathcal{L}[A^I]$ should be quantized. These quantized gauge invariant actions can be thought of as the generalizations of Dijkgraaf-Witten actions \cite{Dijkgraaf:1989pz} for the tensor gauge fields in \eqref{eq:tensor_gauge_field}. For comparison, in App.\ \ref{app:DW_theory}, we review the Dijkgraaf-Witten theories in 1+1 dimensions for finite abelian gauge groups.

\subsection{Charge-dipole dipolar SPTs}

The first type of response functional is
\ie\label{eq:charge-dipole-term}
\mathcal{L}[A^I]=\sum_{I,J}\frac{K_{IJ} N_I N_J}{2\pi N_{IJ}} A^I_t(x) A^J_{xx}(x)~,
\fe
where $N_{IJ}\equiv\text{gcd}(N_I,N_J)$ and $K_{IJ}$ is a symmetric matrix. Under the gauge transformation \eqref{eq:tensor_gauge_field}, it is shifted by
\begin{align}
&\sum_{I,J}\frac{K_{IJ} N_I N_J}{2\pi N_{IJ}}\ \times
\\
&\left[\partial_t \alpha^I(x) A^J_{xx}(x)+A^I(x) \Delta_x^2\alpha^J(x)+\partial_t \alpha^I(x)\Delta_x^2\alpha^J(x)\right]\,.\nonumber
\end{align}
The last term is a total derivative so it vanishes upon summation on a closed chain. To cancel the remaining gauge transformation, the Lagrange multiplier $\phi^I$ should also transform under the gauge symmetry as
\ie
\phi^I\rightarrow  \phi^I+\sum_J\frac{K_{IJ}N_J}{N_{IJ}}\alpha^J~.
\fe
Since both $\phi^I\sim \phi^I+2\pi$ and $\alpha^J\sim\alpha^J+2\pi$ are compact scalars, the coefficients ${K_{IJ}N_J}/{N_{IJ}}$ in the gauge transformation have to be integers to preserve the periodicities. This quantizes $K_{IJ}$ to be integers. In addition, there is an identification $K_{IJ}\sim K_{IJ}+N_{IJ}$ because $A^I$ are $\mathbb{Z}_{N_I}$ tensor gauge fields rendering $\frac{1}{2\pi}\sum_x N_I N_JA^I_t(x) A^J_{xx}(x)$ integer multiples of $2\pi$. The quantization and the identification are more manifest in the modified Villain formalism, as explained in App.\ \ref{app:compactness}.

The response functional in \eqref{eq:charge-dipole-term} describes a charge-dipole dipolar SPT. As a check, the integer symmetric matrix $K_{IJ}$ indeed parameterizes $\prod_{I\leq J}\mathbb{Z}_{N_{IJ}}$, the classification group of charge-dipole dipolar SPTs. To see the charge-dipole fractionalization at the edges, we place the response field theory on an infinite half chain $1\leq x$ with the sum in the action running from $x=2$ to $x=\infty$. In the presence of the boundary, the theory is no longer gauge invariant but transforms with a boundary term generating an anomaly inflow,
\ie\label{eq:inflow-1}
\sum_{I,J} \frac{K_{IJ} N_I N_J}{2\pi N_{IJ}}\Big[\alpha^I(1)A^J_t(2)-\alpha^J(2)\big[A^I_t(1)+\partial_t\alpha^I(1)\big]\Big]~.
\fe
Denote the unitary single-site charge operator that $A_t^I(x)$ couples to by $U_I(x)$. Inserting $U_I(x_0)$ at time $t_0$ sets 
\ie
A_t^I(x)=\frac{2\pi}{N_I}\delta_{x,x_0}\delta(t-t_0)~.
\fe 
Now imagine a configuration with $U_I(1)$ inserted at time $t_0$ and ${U}_J(2)$ inserted at time $t_1<t_0$. We can move $U_J(2)$ to a later time $t_2>t_0$ by doing a gauge transformation with gauge parameter 
\ie
\alpha^J(2)=\frac{2\pi}{N_J}\left[\Theta(t-t_2)-\Theta(t-t_1)\right]~.
\fe
Because of the inflow action in \eqref{eq:inflow-1}, this process generates a phase $e^{2\pi i K_{IJ}/N_{IJ}}$ so the single-site charge operators are noncommutative and form the algebra
\ie\label{eq:U_I-algebra-1}
U_J(2)U_I(1)=e^{\frac{2\pi i K_{IJ}}{N_{IJ}}}U_I(1)U_J(2)~,
\fe
with trivial commutation relations omitted.
The charge and dipole operator, $Q_I$ and $D_I$, acts on the boundary as $\mathcal{L}_{Q_I}=U_I(1)U_I(2)$ and $\mathcal{L}_{D_I}=U_I(1)U_I(2)^2$, respectively. So the algebra in \eqref{eq:U_I-algebra-1} leads to the following algebra for the edge operators,
\ie
\mathcal{L}_{Q_I}\mathcal{L}_{D_J}=e^{-\frac{2\pi i K_{IJ}}{N_{IJ}}}\mathcal{L}_{D_J}\mathcal{L}_{Q_I}~,
\fe
with trivial commutation relations omitted. This exactly reproduces the charge-dipole fractionalization at the edges.

On a closed chain, we can activate the background gauge field $A_{xx}^J$ by twisting the periodic boundary condition to $\mathcal{O}(x+L)=(U_J)^x\mathcal{O}(x)(U_J^{\dagger})^x$. The response functional in \eqref{eq:charge-dipole-term} then implies that the ground states in the twisted Hilbert space transforms under the charge operator $Q_I$ by a phase $e^{2\pi i K_{IJ}/N_{IJ}}$. 

This phenomenon can be explicitly verified in the $\mathbb{Z}_N$ dipolar model \eqref{eq:ZN_Hamiltonian}, which is shown in App.\ \ref{app:derivation-response-theory} to be described by the response functional in \eqref{eq:charge-dipole-term}, specialized to the case that $G=\mathbb{Z}_N$ and $K_{IJ}=-\eta$. In this model, the twisted boundary condition is $Z_{j+L}=e^{2\pi i j/N}Z_{j}$. It does not affect the stabilizers $a_j$ for $j=1,..., L-1$ but introduces a phase to
\ie
a_L&=(Z_{L-1} Z^\dag_L)^\eta\cdot X_L\cdot (Z^\dag_L Z_{L+1})^\eta
\\
&=e^{2\pi i \eta/N}(Z_{L-1} Z^\dag_L)^\eta\cdot X_L\cdot (Z^\dag_L Z_{1})^\eta~.
\fe
The ground state $|\Psi\rangle=a_j|\Psi\rangle$ in the twisted Hilbert space then transforms under the charge operator $Q=e^{-2\pi i\eta/N}\prod_j a_j$ by a phase $e^{-2\pi i\eta/N}$.

\subsection{Dipole-dipole dipolar SPTs}

The second type of response functional is
\ie\label{eq:dipole-dipole-term}
\mathcal{L}[A^I]=\sum_{I,J}\frac{G_{IJ} N_I N_J}{2\pi N_{IJ}} \Delta_x A^I_t(x-\theta_{I,J}) A^J_{xx}(x)~,
\fe
where $G_{IJ}$ is an antisymmetric matrix and $\theta_{I,J}$ is a step function that equals to 1 if $I>J$ and vanishes otherwise.
In order to maintain the gauge invariance on a closed chain, the Lagrange multiplier $\phi^I$ should transform under the gauge symmetry as
\ie
\phi^I\rightarrow  \phi^I-\sum_{J}\frac{G_{IJ}N_J}{N_{IJ}}\Delta_x\alpha^J(x-\theta_{J,I})~.
\fe
The compactness of $\phi^I$ and $\alpha^J$ then quantizes $G_{IJ}$ to be integers. Similar to $K_{IJ}$,  there is also an identification on $G_{IJ}\sim G_{IJ}+N_{IJ}$. The quantization and the identification are more manifest in the modified Villain formalism, as explained in App.\ \ref{app:compactness}.

The response functional in \eqref{eq:dipole-dipole-term} describes dipole-dipole dipolar SPTs. As a check, the antisymmetric matrix $G_{IJ}$ indeed parameterizes $\prod_{I< J}\mathbb{Z}_{N_{IJ}}$, the classification group of dipole-dipole dipolar SPTs. On an infinite half chain $1\leq x$, the response field theory is no longer gauge invariant but transforms with a boundary term that generates an anomaly inflow,
\begin{align}
&\sum_{I<J}\frac{G_{IJ}N_J}{N_{IJ}}\ \times
\\
&\Big[\Delta_x\alpha^I(1)\Delta_x A_t^J(1)-\Delta_x\alpha^J(1)\big[\Delta_x A_t^I(1)+\Delta_x\partial_t \alpha^I(1)\big]\Big]\,,\nonumber
\end{align}
The inflow action implies that 
\ie
U_J(1)U_I(1)&=e^{\frac{2\pi i G_{IJ}}{N_{IJ}}}U_I(1)U_J(1)~,
\\
U_J(2)U_I(1)&=e^{-\frac{2\pi i G_{IJ}}{N_{IJ}}}U_I(1)U_J(2)~,
\\
U_J(2)U_I(2)&=e^{\frac{2\pi i G_{IJ}}{N_{IJ}}}U_I(2)U_J(2)~.
\fe
This further leads to the following algebra of the fractionalized symmetry operators, $\mathcal{L}_{Q_I}=U_I(1)U_I(2)$ and $\mathcal{L}_{D_J}=U_J(1)U_J(2)^2$,  at the boundary:
\ie
\mathcal{L}_{D_I}\mathcal{L}_{D_J}=e^{-\frac{2\pi i G_{IJ}}{N_{IJ}}}\mathcal{L}_{D_J}\mathcal{L}_{D_I}~,
\fe
where trivial commutation relations are omitted. This exactly reproduces the dipole-dipole fractionalization at the boundary.

On a closed chain, the response functional in \eqref{eq:dipole-dipole-term} implies that the ground states in the twisted Hilbert space with boundary condition $\mathcal{O}(x+L)=(U_J)^x\mathcal{O}(x)(U_J^{\dagger})^x$ transforms under the dipole operator $D_I$ by a phase $e^{2\pi i G_{IJ}/N_{IJ}}$. Here, for the dipole operator to be well-defined, we take $L$ to be divisible by $N_I$. We can explicitly verify this phenomenon in the stabilizer model in \eqref{eq:quandrupole_Hamiltonian}, which is shown in App.\ \ref{app:derivation-response-theory} to be described by the response functional in \eqref{eq:dipole-dipole-term} for $G=\mathbb{Z}_N\times\mathbb{Z}_M$ with
\ie
G_{IJ}=\left[\begin{array}{cc}
     0& \xi \\
    -\xi & 0
\end{array}\right]~.
\fe
With the twisted boundary condition $Z_{j+L}=e^{2\pi i j/N}Z_{j}$, the stabilizers, $a_j$ for all $j$ and $\tilde a_j$ for $j=1,..., L-2$, are unaffected, but $\tilde a_{L-1}$ and $\tilde a_{L}$ acquire a phase and become $\tilde a_{L-1} e^{-2\pi i\xi/K}$  and $\tilde a_{L}e^{2\pi i\xi/K}$, respectively. As a result, the dipole operator $\tilde D\,\dot{ = }\, \prod_j (a_j)^j$ also acquires a phase $e^{-2\pi i\xi/K}$ when acting on the ground state $|\Psi\rangle=a_j|\Psi\rangle=\tilde a_j|\Psi\rangle$ of the twisted Hilbert space.

\section{Conclusion and outlook}
In this work, we propose a classification of one-dimensional dipolar SPTs based on matrix product states and finite tensor gauge theories. Furthermore, we provide a concrete realization for every dipolar SPT in the proposed classification using stabilizer Hamiltonians. There are several natural directions for future exploration. One direction is to extend the classification to higher-dimensional dipolar SPTs. To this end, it would be useful to develop a framework for dipolar SPTs akin to the one established by Else and Nayak for ordinary SPTs \cite{Else:2014vma}. Another interesting direction is the classifications of SPTs protected by other modulated symmetries, e.g.\ multiple symmetries, exponential symmetries and more \cite{Han:2023fas}. It would also be of interest to investigate the phase transitions between these dipolar SPTs, especially those between charge-dipole and dipole-dipole dipolar SPTs. Some preliminary DMRG numerics were performed for the phase transitions between the $\mathbb{Z}_N$ dipolar model and the trivial phase in \cite{Han:2023fas}.

\acknowledgments I am grateful to Jung Hoon Han, Ethan Lake, Ruben Verresen and Yizhi You for collaborations on a related work \cite{Han:2023fas}, and Jie Wang for discussions. I thank Jung Hoon Han for comments on a draft.
I also thank the Croucher Summer Course "Quantum Entanglement and Topological Order" and the ``Paths to Quantum Field Theory 2023" workshop for hospitality during the course of this project. I am supported in part by a Croucher fellowship from the Croucher Foundation, the Packard Foundation and the Center for Theoretical Physics at MIT.

\onecolumngrid
\appendix
\section{$U(1)$ dipole symmetry}
\label{app:U(1)}
In this appendix, we give an overview of $U(1)$ dipole symmetry following \cite{Gorantla:2022eem}. The purpose of this appendix is to draw the analogy between $U(1)$ and $\mathbb{Z}_N$ dipole symmetry, and motivate the symmetry operators \eqref{eq:dipole-charge} and the background gauge fields \eqref{eq:tensor_gauge_field} for $\mathbb{Z}_N$ dipole symmetry. 

A $U(1)$ dipole symmetry on a one-dimensional chain is generated by the current $J_t,J_{xx}$ that obeys the current conservation equation
\ie\label{eq:conservation_equation}
\partial_t J_t-\Delta_x^2 J_{xx}=0~.
\fe
The charge and the dipole charge of the $U(1)$ dipole symmetry is given by
\ie
Q_C=\sum_x J_t(x)~,\quad
Q_D=\sum_x x J_t(x)~.
\fe
They form a non-trivial algebra with lattice translation $\mathcal{T}$
\ie
\mathcal{T}^{-1}Q_C\mathcal{T}=Q_C~,\quad \mathcal{T}^{-1}Q_D\mathcal{T}=Q_C+Q_D~.
\fe
Exponentiating them gives the corresponding symmetry operators, 
\ie
Q(\theta)=e^{i\theta Q_C}=\prod_x e^{i\theta J_t(x)}~,\quad D(\theta)=e^{i\theta Q_D}=\prod_x \big[e^{i \theta J_t(x)}\big]^x~,
\fe
which are the $U(1)$ counterpart of the $\mathbb{Z}_N$ charge and dipole operators in \eqref{eq:dipole-charge}. 
These charges are conserved because of the non-standard current conservation equation \eqref{eq:conservation_equation}:
\ie
&\partial_t Q_C=\sum_x \partial_t J_t(x)=\sum_x \Delta_x^2 J_{xx} =0~,
\\
&\partial_t Q_D=\sum_x x\partial_t J_t(x)=\sum_x x\Delta_x^2 J_{xx}=\sum_x \Delta_x (x\Delta_x J_{xx}-J_{xx})=0~.
\fe
Here, for simplicity, we work on an infinite chain. The dipole symmetry is more subtle on periodic chains (see \cite{Gorantla:2022eem,Han:2023fas} for more details). 

The $U(1)$ dipole symmetry current naturally couples to a $U(1)$ tensor gauge field
\ie\label{eq:U(1)-tensor-field}
A_t \rightarrow  A_{t}+\partial_t \alpha~,
\quad A_{xx}\rightarrow  A_{xx}+\Delta_x^2\alpha~,
\fe
through the coupling 
\ie
\sum_x \left[J_t(x) A_t(x)+J_{xx}(x)A_{xx}(x)\right]~,
\fe
where $\alpha$ is the gauge parameter. The gauge invariance of the coupling is ensured by the current conservation equation \eqref{eq:conservation_equation}. Drawing an analogy, a $\mathbb{Z}_N$ dipole symmetry should couple to the $\mathbb{Z}_N$ version of the tensor gauge field \eqref{eq:U(1)-tensor-field}.

\section{Additional details on quadrupolar SPT}
\label{app:quadrupole}
In this appendix, we provide more details on the quadrupolar model in \eqref{eq:quandrupole_Hamiltonian}:
\ie
    &H =-\sum_{j} (a_{j}+\tilde a_{j}+\text{h.c.})~, 
    \\
    &a_{j}= (\tilde{Z}_{j-1} \tilde{Z}_{j}^{3\dagger})^{\frac{M\zeta}{K}} X_{j+1}^\dagger (\tilde{Z}_{j+1}^3 \tilde{Z}_{j+2}^{\dagger})^{\frac{M\zeta}{K}}~,
    \\
    &\tilde a_{j}= (Z_{j-1} Z_{j}^{3\dagger})^{\frac{N\zeta}{K}} \tilde X_{j} (Z_{j+1}^3 Z_{j+2}^{\dagger})^{\frac{N\zeta}{K}} ~. 
\fe
We will give a complete account of the fractionalization of the $\mathbb{Z}_N\times\mathbb{Z}_M$ quadrupole symmetry at the edges.

On an open chain $1\leq j\leq L$, the Hamiltonian includes only stabilizers $a_j,\tilde a_j$ for $2\leq j\leq L-1$:
\ie\label{eq:open-chain-quadrupole}
H =-\sum_{j=2}^{L-1} (a_{j}+\tilde a_{j}+\text{h.c.})~.
\fe
The six symmetry operators for the $\mathbb{Z}_N\times\mathbb{Z}_M$ quadrupole symmetry fractionalize on the edges. The edge operators localized on the left edge are
\begin{alignat}{2}
&\mathcal{L}_{Q}=X_1X_2(\tilde{Z}_1\tilde{Z}_2^{2\dagger}\tilde Z_3)^{\frac{M\xi}{K}}~,\quad &&\tilde{\mathcal{L}}_{ Q}=\tilde X_1({Z}_1^\dagger{Z}_2^{2} Z_3^\dagger)^{\frac{N\xi}{K}}~,\nonumber
\\
&\mathcal{L}_{D}=X_1X_2^2(\tilde{Z}_1^3\tilde{Z}_2^{5\dagger}\tilde Z_3^2)^{\frac{M\xi}{K}}~,\quad &&\tilde{\mathcal{L}}_{ D}=\tilde X_1({Z}_1^{2\dagger}{Z}_2^{3} Z_3^\dagger)^{\frac{N\xi}{K}}~,\nonumber
\\
&\mathcal{L}_{P}=X_1X_2^4(\tilde{Z}_1^9\tilde{Z}_2^{11\dagger}\tilde Z_3^4)^{\frac{M\xi}{K}}~,\quad &&\tilde{\mathcal{L}}_{ {P}}=\tilde X_1({Z}_1^{4\dagger}{Z}_2^{3} Z_3^\dagger)^{\frac{N\xi}{K}}~.\nonumber
\end{alignat}
They obey the following algebra
\begin{alignat}{2}
&\mathcal{L}_{Q}\tilde{\mathcal{L}}_{{P}}=e^{\frac{4\pi i}{K}}\tilde{\mathcal{L}}_{{P}}\mathcal{L}_Q~,\quad &&\mathcal{L}_{P}\tilde{\mathcal{L}}_{{Q}}=e^{\frac{4\pi i}{K}}\tilde{\mathcal{L}}_{{{Q}}}\mathcal{L}_{P}~,\nonumber
\\
&\mathcal{L}_{D}\tilde{\mathcal{L}}_{{P}}=e^{\frac{2\pi i}{K}}\tilde{\mathcal{L}}_{{P}}\mathcal{L}_D~,\quad && \mathcal{L}_{P}\tilde{\mathcal{L}}_{{D}}=e^{-\frac{2\pi i}{K}}\tilde{\mathcal{L}}_{{{D}}}\mathcal{L}_{P}~,\nonumber
\\
&\mathcal{L}_{D}\tilde{\mathcal{L}}_{{D}}=e^{-\frac{2\pi i}{K}}\tilde{\mathcal{L}}_{{D}}\mathcal{L}_D~,\quad &&\mathcal{L}_{P}\tilde{\mathcal{L}}_{{P}}=e^{\frac{2\pi i}{K}}\tilde{\mathcal{L}}_{{P}}\mathcal{L}_{P}~,\nonumber
\end{alignat}
with trivial commutation relations omitted. The right edge operators obey a similar algebra but with an opposite projectivity. The edge operators can be paired up to form three independent Heisenberg algebras as following
\ie
    \mathcal{L}_{D}\tilde{\mathcal{L}}_{D}&=e^{-\frac{2\pi i}{K}} \tilde{\mathcal{L}}_{D}\mathcal{L}_{D}~,
    \\
    \mathcal{L}_{Q}\cdot\tilde{\mathcal{L}}_{ P}\tilde{\mathcal{L}}_{D}&=e^{\frac{4\pi i}{K}}\tilde{\mathcal{L}}_{P}\tilde{\mathcal{L}}_{D}\cdot\mathcal{L}_{Q}~,
    \\
    \mathcal{L}_{P}\mathcal{L}_{D}^{\dagger}\cdot\tilde{\mathcal{L}}_{Q}&=e^{\frac{4\pi i}{K}}\tilde{\mathcal{L}}_{Q}\cdot\mathcal{L}_{P}\mathcal{L}_{D}^{\dagger}~. 
\fe
These algebras enforce a minimal ground state degeneracy of $K^3/\text{gcd}(K,2)^2$ per edge.

The Hamiltonian \eqref{eq:open-chain-quadrupole} however has $N^2M$ modes localize on the left edge due to the following three independent Heisenberg algebra:
\ie
\mathcal{L}_{Q}\cdot Z_1^2Z_2^\dagger&=e^{-\frac{2\pi i}{N}}Z_1^2Z_2^\dagger\cdot\mathcal{L}_{Q}~,
\\
\tilde{\mathcal{L}}_{Q}\cdot\tilde Z_1&=e^{-\frac{2\pi i}{M}}\tilde Z_1\cdot\tilde{\mathcal{L}}_{Q}~,
\\
\mathcal{L}_{D}\cdot Z_1Z_2^\dagger&=e^{\frac{2\pi i}{N}}Z_1Z_2^\dagger\cdot\mathcal{L}_{D}~,
\fe
where all the operators in the algebra commute with the Hamiltonian \eqref{eq:open-chain-quadrupole}. We can lift the ground state degeneracy to the minimal protected value $K^3/\text{gcd}(K,2)^2$ by adding the following edge perturbation to the existing Hamiltonian
\ie\label{eq:edge-perturbation}
H_{\text{edge}}=-\left[(\mathcal{L}_{Q})^{\frac{K}{\gcd(K,2)}}+(\tilde{\mathcal{L}}_{Q})^{\frac{K}{\gcd(K,2)}}+(\mathcal{L}_{D})^K+\text{h.c.}\right]~.
\fe
These edge perturbations preserve the quadrupole symmetry and commute with the existing Hamiltonian. The remaining $K^3/\text{gcd}(K,2)^2$ ground states $|\Psi\rangle$ are stabilized by these three terms in the perturbation, i.e. $|\Psi\rangle=(\mathcal{L}_{Q})^{\frac{K}{\gcd(K,2)}}|\Psi\rangle=(\tilde{\mathcal{L}}_{Q})^{\frac{K}{\gcd(K,2)}}|\Psi\rangle=(\mathcal{L}_{D})^K|\Psi\rangle$.

If we further break the quadrupole symmetry down to the dipole symmetry, the algebra \eqref{eq:algebra-dipole-dipole} only protects a $K$ ground state degeneracy per edge. Indeed, we can further lift the ground state degeneracy to the protected value $K$ by adding the following edge perturbation on top of the existing one \eqref{eq:edge-perturbation}
\ie
H'_{\text{edge}}=-(\mathcal{L}_{Q}+\tilde{\mathcal{L}}_{Q}+\text{h.c.})~.
\fe
The perturbation commutes with the Hamiltonian \eqref{eq:open-chain-quadrupole} as well as the edge perturbation \eqref{eq:edge-perturbation}. It preserves the dipole symmetry but violates the quadrupole symmetry.

\section{Symmetric MPS}
\label{app:sym-MPS}

In this appendix, we first revisit the proof of the push-through property \eqref{eq:MPS-push} for MPS that are symmetric with respect to an ordinary symmetry, following \cite{bridgeman17} (see \cite{P_rez_Garc_a_2008} for the original proof). Then, we generalize the result to prove the new push-through property \eqref{eq:MPS-constraint} for dipole-symmetric MPS. To make the proofs more accessible, we will use diagrammatic representations of MPS. We will work with injective MPS in the left canonical form, i.e.\ the largest eigenvalue of the transfer matrix $T(x)\equiv \sum_h A^h X A^{h\dagger}$ is 1 with identity the unique left eigenvector and the full-rank density matrix $\rho$ the unique right eigenvector. In diagrammatic representation, it means
\ie\label{eq:canonical-form}
\vcenter{\hbox{\includegraphics[scale=0.6]{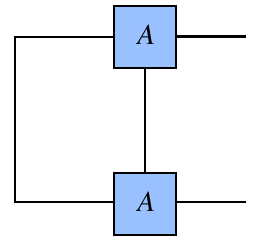}}}\ =\ \vcenter{\hbox{\includegraphics[scale=0.6]{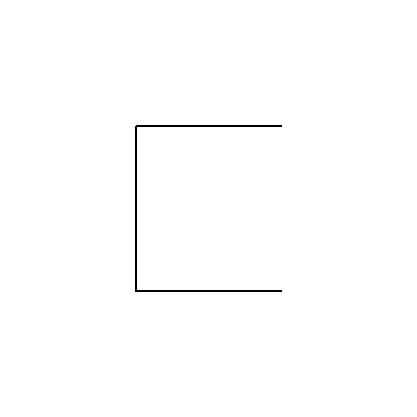}}}~,\qquad
\vcenter{\hbox{\includegraphics[scale=0.6]{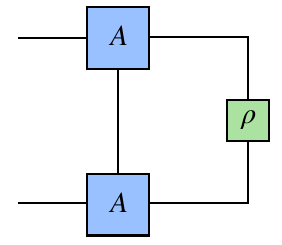}}}\ =\ \vcenter{\hbox{\includegraphics[scale=0.6]{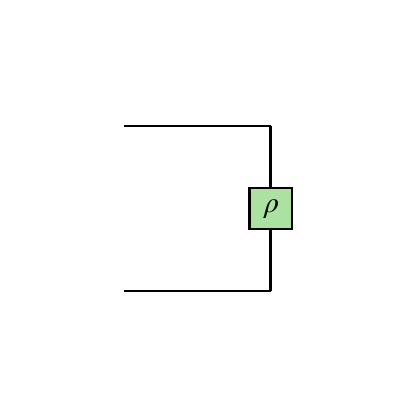}}}~.
\fe
We will refer to $\rho$ as the canonical right eigenvector of the transfer matrix $T$.

\subsection{Ordinary symmetry}\label{app:sym-constraint}

\textbf{Theorem 1.} An MPS $|\Psi\rangle$ is symmetric with respect to an ordinary symmetry $Q_g=\prod_x U_g(x)$, i.e.\ $Q_g|\Psi\rangle=e^{i\phi_g}|\Psi\rangle$, if and only if the unitary operator $U_g$ can be pushed through the MPS tensor as 
\ie\label{eq:pushing-through}
\vcenter{\hbox{\includegraphics[scale=0.6]{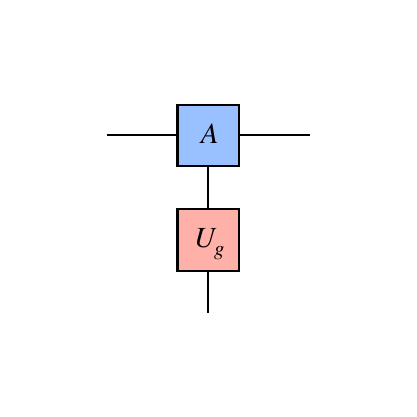}}}\ =\ e^{i\theta_g}\ \vcenter{\hbox{\includegraphics[scale=0.6]{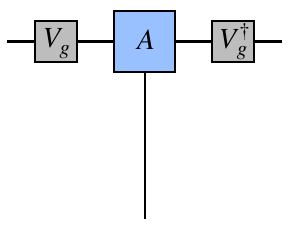}}}~,
\fe
with $V_g$ a unitary operator.
\\

\textbf{Proof.} The if direction is straightforward to verify by pushing $Q_g$ through the tensor using \eqref{eq:pushing-through}
\ie
&\vcenter{\hbox{\includegraphics[scale=0.6]{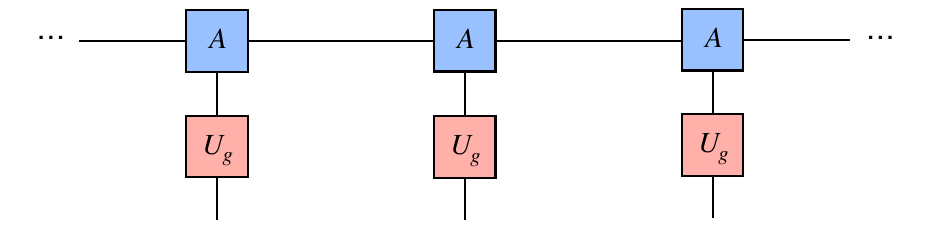}}}
\\ 
= \ e^{i\phi_g}\ &\vcenter{\hbox{\includegraphics[scale=0.6]{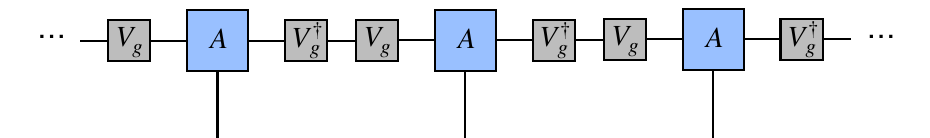}}}
\\
= \ e^{i\phi_g}\ &\, \vcenter{\hbox{\includegraphics[scale=0.6]{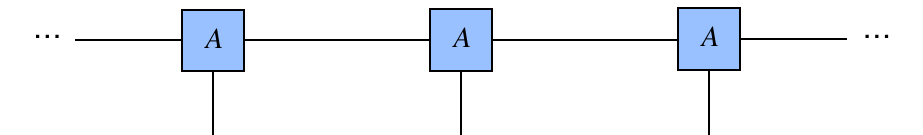}}}
\fe
where $\phi_g$ is proportional to $\theta_g$.

Next, we prove the only if direction. As the MPS $|\Psi\rangle$ is symmetric with respect to $Q_g$, the inner product $\langle\Psi|Q_g|\Psi\rangle$ is always a phase that does not decay with the size of the chain
\ie
\langle\Psi|Q_g|\Psi\rangle\ =\ \vcenter{\hbox{\includegraphics[scale=0.6]{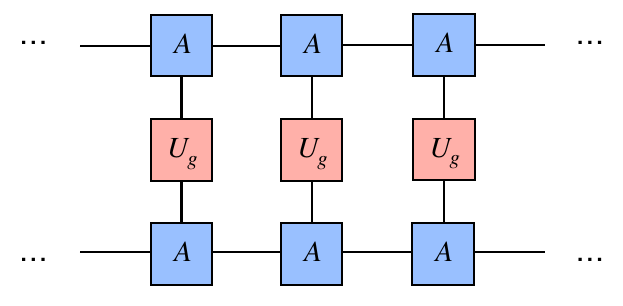}}}\ =\ e^{i\phi_g}~.
\fe
It is possible only if the largest eigenvalue of the transfer matrix $T_g(X)\equiv\sum_h (U_g)_{hh'} A^h X A^{h'\dagger} $ has unit modulus. Denote the corresponding eigenvector by $V_g^\dagger$ (here we do not assume $V_g$ to be unitary). Then, we have
\ie
\vcenter{\hbox{\includegraphics[scale=0.6]{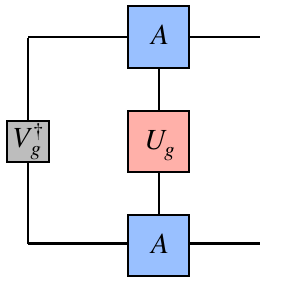}}}\ = e^{i\theta_g}\ \vcenter{\hbox{\includegraphics[scale=0.6]{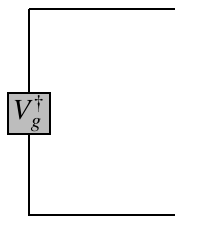}}}~.
\fe
Let's act both sides of the equation on $V_g$ and $\rho$, the canonical right eigenvector of the transfer matrix $T$:
\ie\label{eq:inner-product-CS}
e^{i\theta_g}\ \vcenter{\hbox{\includegraphics[scale=0.6]{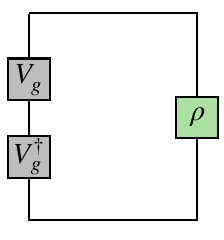}}}\ = \
\vcenter{\hbox{\includegraphics[scale=0.6]{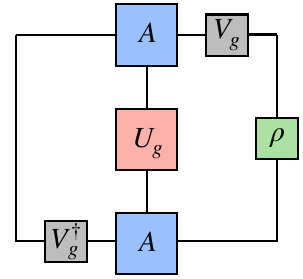}}}\ = \
\vcenter{\hbox{\includegraphics[scale=0.6]{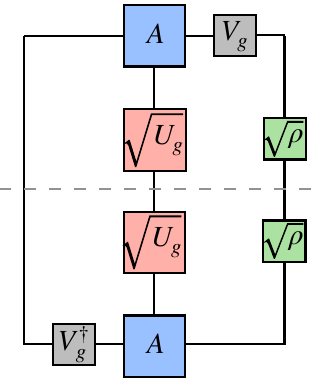}}}~.
\fe
In the second equality, we decompose the expression in the middle into the inner product between two vectors. These two vectors share the same norm, whose square is identical to the magnitude of their inner product as computed below:
\ie\label{eq:norm-CS}
\vcenter{\hbox{\includegraphics[scale=0.6]{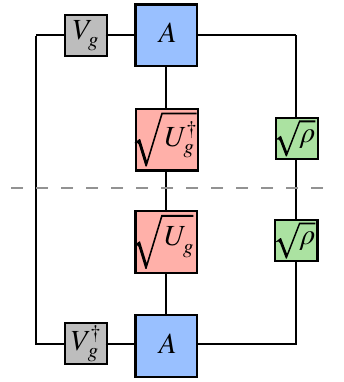}}} \ = \ \vcenter{\hbox{\includegraphics[scale=0.6]{correlation-function-1.pdf}}} \ = \
\vcenter{\hbox{\includegraphics[scale=0.6]{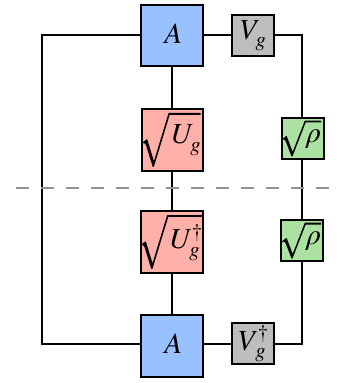}}}
~.
\fe
These two equalities follow from the left canonical form \eqref{eq:canonical-form} of the MPS tensors.
In the first equality, we use the fact that $\rho$ is the eigenvalue 1 right eigenvector of the transfer matrix $T$, while in the second equality, we use the fact that the identity is the eigenvalue 1 left eigenvector of the transfer matrix $T$. To proceed, let's recall the Cauchy-Schwarz inequality. It states that the magnitude of the inner product of two vectors is upper bounded by the product of the two vectors'~norms; moreover, the equality holds only if the two vectors are proportional to each other. Comparing \eqref{eq:inner-product-CS} and \eqref{eq:norm-CS}, we learn that the two vectors should be proportional to each other. Furthermore, as they share the same norm, the proportionality constant has to be a phase. This gives
\ie
\vcenter{\hbox{\includegraphics[scale=0.6]{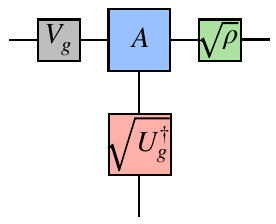}}} \ =\ e^{i\alpha_g} \  \vcenter{\hbox{\includegraphics[scale=0.6]{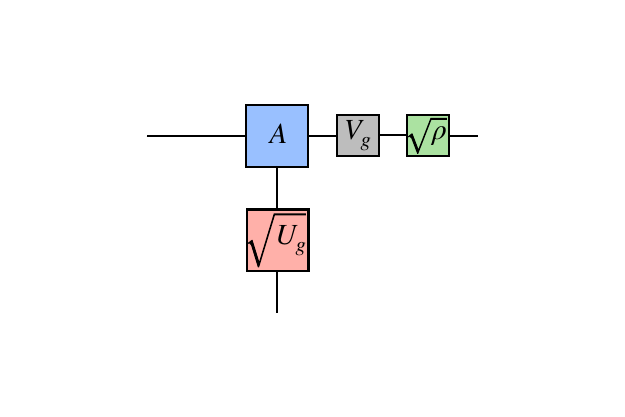}}}~.
\fe
Since $\rho$ is a full rank matrix, we can remove it from the equality and simplify the equality into
\ie\label{eq:simplify-equality}
\vcenter{\hbox{\includegraphics[scale=0.6]{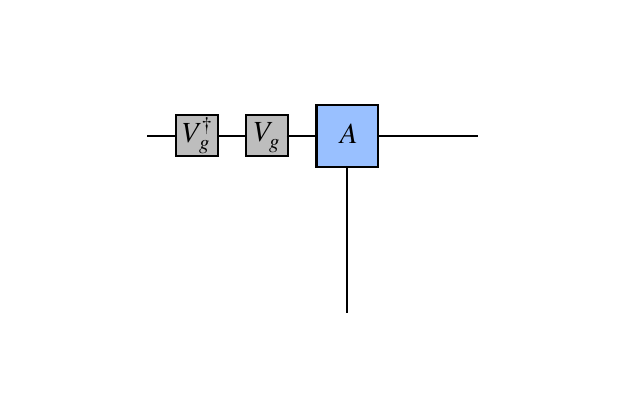}}}  \ =\ e^{i\alpha_g} \ \vcenter{\hbox{\includegraphics[scale=0.6]{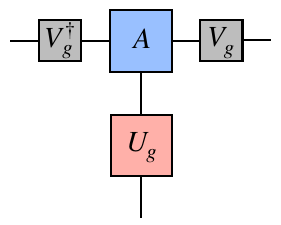}}}~.
\fe
It implies that $V_gV_g^\dagger$ is a left eigenvector of the transfer matrix $T$ with eigenvalue $e^{i(\theta_g-\alpha_g)}$,
\ie
\vcenter{\hbox{\includegraphics[scale=0.6]{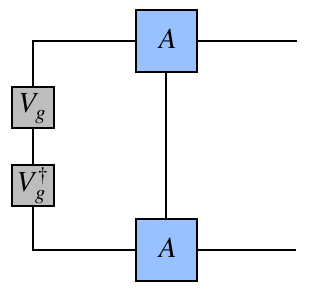}}}
\ = \ e^{i\alpha_g}\
\vcenter{\hbox{\includegraphics[scale=0.6]{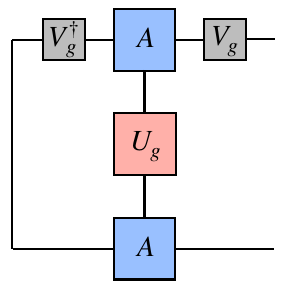}}} \ =\ e^{i(\theta_g+\alpha_g)} \ \vcenter{\hbox{\includegraphics[scale=0.6]{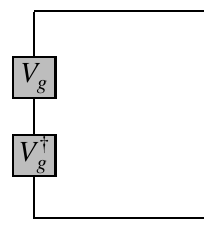}}}~.
\fe
Recall that the MPS is injective and we are working in the left canonical form \eqref{eq:canonical-form} so the only left eigenvector of the transfer matrix $T$ with unit modulus eigenvalue is the identity matrix and the eigenvalue is 1. This then implies that $V_g$ is a unitary operator $V_g V_g^\dagger=1$ and $e^{i\alpha_g}=e^{-i\theta_g}$. We can then multiply \eqref{eq:simplify-equality} by $V_g$ and $V_g^\dagger$ from the two sides and substitute $e^{i\alpha_g}=e^{-i\theta_g}$. This reproduces the push-through property \eqref{eq:pushing-through} and concludes the proof.

\subsection{Dipole symmetry}\label{app:dipole-sym-constraint}

\textbf{Theorem 2.} An MPS $|\Psi\rangle$ is symmetric with respect to a dipole symmetry with symmetry generator $Q_g=\prod_x U_g(x)$ and $D_g=\prod_x [U_g(x)]^x$ if and only if the unitary operator $U_g$ can be pushed through the MPS tensor as 
\ie\label{eq:pushing-through-vertical}
\vcenter{\hbox{\includegraphics[scale=0.6]{ug-tensor.pdf}}}\ =\ e^{i\theta_g}\ \vcenter{\hbox{\includegraphics[scale=0.6]{push-through.pdf}}}~,
\fe
with $V_g$ a unitary operator, and $V_g$ can be pushed through the MPS tensor as 
\ie\label{eq:pushing-through-horizontal}
\vcenter{\hbox{\includegraphics[scale=0.6]{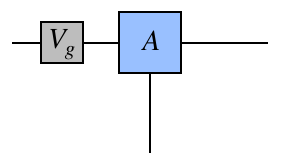}}}\ =\ e^{i\vartheta_g}\ \vcenter{\hbox{\includegraphics[scale=0.6]{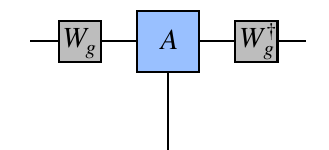}}}~,
\fe
with $W_g$ a unitary operator.
\\

\textbf{Proof.} The if direction is straightforward to verify. As already been checked in App.~\ref{app:sym-constraint}, the push-through property \eqref{eq:pushing-through-vertical} implies $Q_g|\Psi\rangle=e^{i\phi_g}|\Psi\rangle$ so we only need to check that \eqref{eq:pushing-through-vertical} and \eqref{eq:pushing-through-horizontal} implies $D_g|\Psi\rangle=e^{i\varphi_g}|\Psi\rangle$. The details of the computation are shown below:
\ie
&\vcenter{\hbox{\includegraphics[scale=0.6]{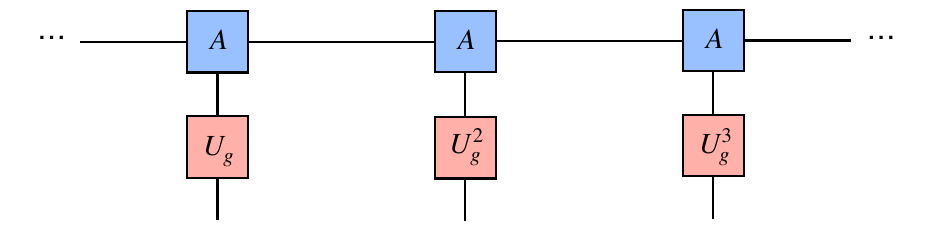}}}
\\ 
= \ e^{i\phi_g'}\ &\vcenter{\hbox{\includegraphics[scale=0.6]{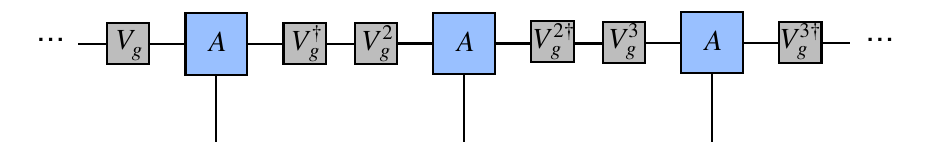}}}
\\
= \ e^{i\phi_g'}\ &\, \vcenter{\hbox{\includegraphics[scale=0.6]{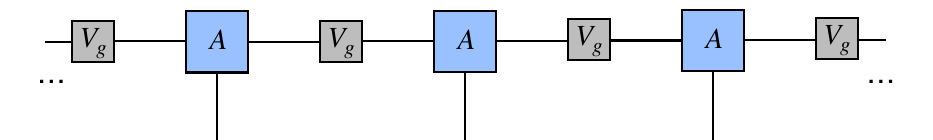}}}
\\
= \ e^{i\varphi_g}\ &\, \vcenter{\hbox{\includegraphics[scale=0.6]{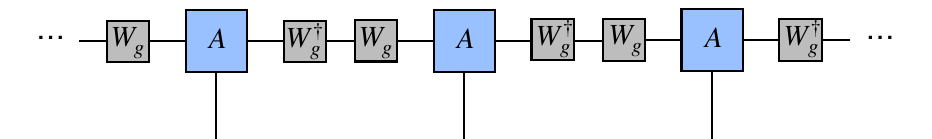}}}
\\
= \ e^{i\varphi_g}\ &\, \vcenter{\hbox{\includegraphics[scale=0.6]{symmetry-3a.pdf}}}
\fe
where $\phi_g'$ is proportional to $\theta_g$ and $\varphi_g-\phi'_g$ is proportional to $\vartheta_g$. Here for simplicity we work on an infinite chain. If we are working on a periodic chain of length $L$, we only need to check that the MPS is symmetric with respect to those dipole operators $D_g$ that are compatible with the periodic boundary condition, i.e.\ those $D_g$ built from $U_g$ satisfying $U_g^L=1$. In such cases, $V_g^L=1$, so $V_g^L$ can be dropped in the second line and the rest of the calculation goes through.

We now prove the only if direction. As already been proven in App.\  \ref{app:sym-constraint}, $Q_g|\Psi\rangle=e^{i\phi_g}|\Psi\rangle$ implies the push-through property \eqref{eq:pushing-through-vertical} with $V_g$ a unitary operator, so we only need to prove that $D_g|\Psi\rangle=e^{i\varphi_g}|\Psi\rangle$ together with \eqref{eq:pushing-through-vertical} implies the second push-through property \eqref{eq:pushing-through-horizontal} with $W_g$ a unitary operator. The idea is similar to the proof in App.\ \ref{app:sym-constraint} so we will be brief. Using \eqref{eq:pushing-through-vertical}, the inner product $\langle\Psi|D_g|\Psi\rangle$ can be decomposed into products of the transfer matrix $T'_g$ defined as
\ie
T'_g(X)=\sum_h A^h V_g X A^{h\dagger}=\vcenter{\hbox{\includegraphics[scale=0.6]{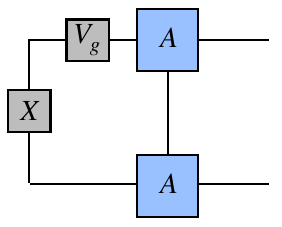}}}~.
\fe
As the inner product is always a phase that does not decay with the size of the chain, the largest eigenvalue of the transfer matrix $T'_{g}$ is unit modulus. Denote the left eigenvector by $W_g^\dagger$ (here we don't assume $W_g$ is unitary). Then, we have
\ie\label{eq:W-eigenvector}
\vcenter{\hbox{\includegraphics[scale=0.6]{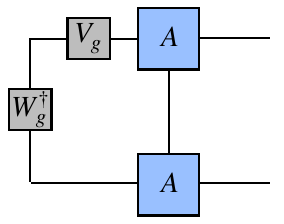}}}\ = \ e^{i\vartheta_g}\ \vcenter{\hbox{\includegraphics[scale=0.6]{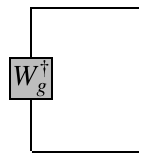}}}~.
\fe
Let's act both sides of the equation on $W_g$ and $\rho$, the right canonical eigenvector of the transfer matrix $T$, and decompose the expression into the inner product of two vectors:
\ie
e^{i\vartheta_g} \ \vcenter{\hbox{\includegraphics[scale=0.6]{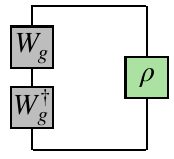}}}\ = \ \ \vcenter{\hbox{\includegraphics[scale=0.6]{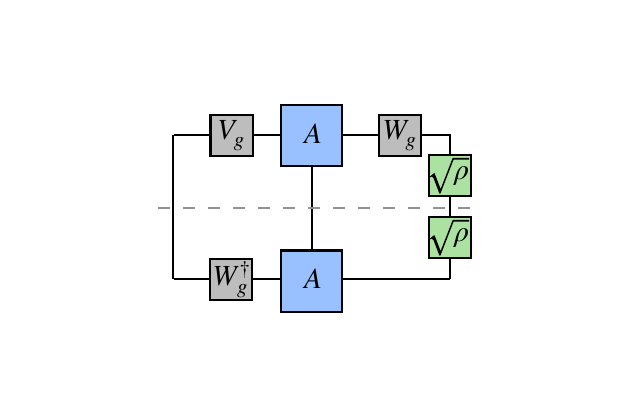}}}~.
\fe
The square of these two vectors' norm is identical to the magnitude of their inner product, as computed below:
\ie
\vcenter{\hbox{\includegraphics[scale=0.6]{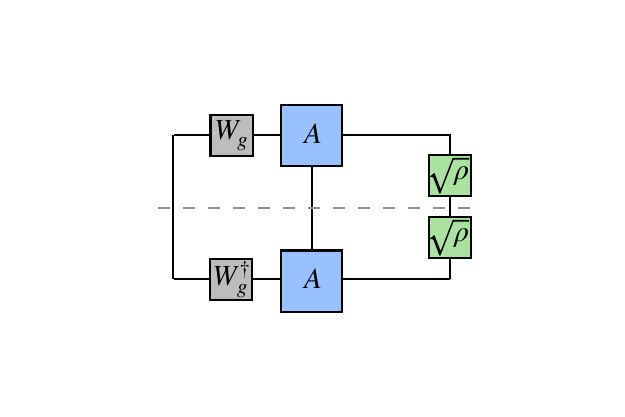}}}\ = \ \vcenter{\hbox{\includegraphics[scale=0.6]{inner-product-W-1.pdf}}}\ = \ 
\vcenter{\hbox{\includegraphics[scale=0.6]{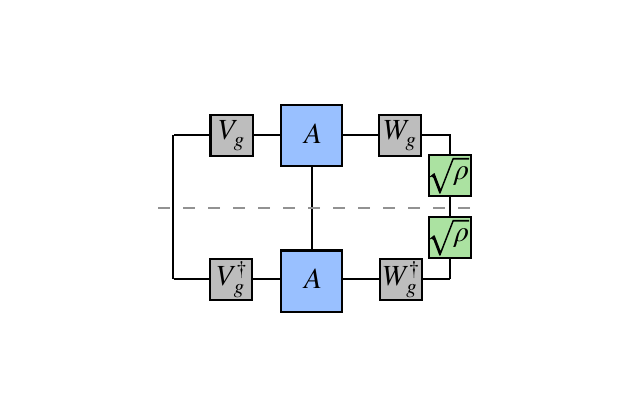}}}~.
\fe
In the first equality, we use the fact that $\rho$ is the eigenvalue 1 right eigenvector of the transfer matrix $T$. In the second equality, we use the fact that $V_g$ is unitary to remove $V_gV_g^\dagger$  and then use the fact that the identity is the eigenvalue 1 left eigenvector of the transfer matrix $T$. Next, we use the Cauchy-Schwarz inequality. It implies that the two vectors are proportional to each other. Because they share the same norm, the proportionality constant is a phase $e^{i\beta_g}$,
\ie
\vcenter{\hbox{\includegraphics[scale=0.6]{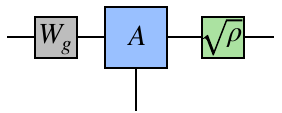}}}\ = \ e^{i\beta_g}\ \vcenter{\hbox{\includegraphics[scale=0.6]{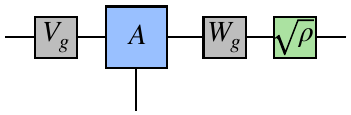}}}~.
\fe 
We can remove the full rank density matrix $\rho$ from the equality and multiply $W_g^\dagger$ from the left. This gives
\ie\label{eq:equality-W}
\vcenter{\hbox{\includegraphics[scale=0.6]{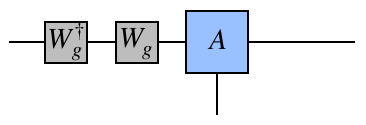}}}\ = \ e^{i\beta_g}\ \vcenter{\hbox{\includegraphics[scale=0.6]{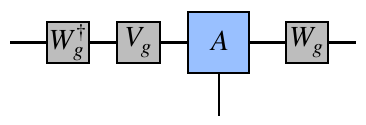}}}~.
\fe
The equality implies that $W_g W_g^\dagger$ is a left eigenvector of the transfer matrix $T$ with a unit modulus eigenvalue. Below is the diagrammatic proof:
\ie
\vcenter{\hbox{\includegraphics[scale=0.6]{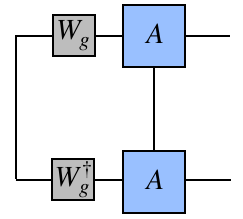}}}\ = \ e^{i\beta_g} \
\vcenter{\hbox{\includegraphics[scale=0.6]{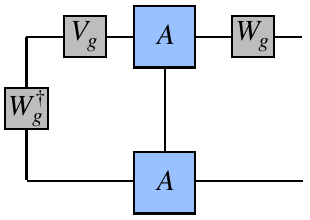}}}\ = \ e^{i(\vartheta_g+\beta_g)} \ \vcenter{\hbox{\includegraphics[scale=0.6]{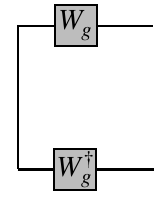}}}~,
\fe
where in the second equality, we use the fact that $W_g^\dagger$ is a left eigenvector of $T'_g$, as illustrated in \eqref{eq:W-eigenvector}. Since the transfer matrix $T$ has the identity matrix as its unique left eigenvector with unit modulus eigenvalue and the eigenvalue is 1, we learn that $W_g$ is a unitary operator and $e^{i\beta_g}=e^{-i\vartheta_g}$. Substituting these back to \eqref{eq:equality-W} and multiplying $W_g$ and $W_g^\dagger$ from the two sides, we recover the second push-through property \eqref{eq:pushing-through-horizontal}. This concludes the proof.

\section{Projective representations and group cohomology}
\label{app:projective}
In this appendix, we review the relation between projective representations and group cohomology and then compute the group cohomology for a general finite abelian group.

In a projective representation of a group $G$, the operators act as
\ie
V_gV_h=e^{i\omega(g,h)}V_{gh}~.
\fe
The associativity of the group action 
\ie
V_gV_hV_k&=e^{i\omega(h,k)}V_gV_{hk}=e^{i\omega(h,k)+i\omega(g,hk)}V_{ghk}
\\
&=e^{i\omega(g,h)}V_{gh}V_{k}=e^{i\omega(g,h)+i\omega(gh,k)}V_{ghk}~,
\fe
imposes a consistency condition that
\ie
e^{i\omega(h,k)+i\omega(g,hk)-i\omega(g,h)-i\omega(gh,k)}=1~,
\fe
which is precisely the cocycle condition for a 2-cocycle. Since we can always redefine
\ie
V_g\rightarrow e^{i\beta(g)}V_g~,
\fe
the 2-cocycle has an ambiguity
\ie\label{eq:co-boundary}
e^{i\omega(g,h)}\rightarrow e^{i\omega(g,h)}e^{i\beta(gh)-i\beta(g)-i\beta(h)} ~,
\fe
that amounts to shifting the 2-cocycle by a 2-coboundary. Hence, inequivalent projective representations are in one-to-one correspondence with the second group cohomology $H^2(G,U(1))$.

For a finite abelian group 
\ie\label{eq:finite-abelian-group}
G=\prod_I\mathbb{Z}_{N_I}~,
\fe
its second group cohomology is
\ie\label{eq:second-group-cohom}
H^2(G,U(1))=\prod_{I<J}\mathbb{Z}_{N_{IJ}}~,
\fe
where $N_{IJ}\equiv\text{gcd}(N_I,N_J)$. We now explicitly demonstrate how this group computed from the second group cohomology is in one-to-one correspondence with the projective representations of the finite abelian group $G$. As the group $G$ is abelian, the group elements commute $gh=hg$. It is then useful to define
\ie
e^{i\phi(g,h)}=e^{i\omega(g,h)-i\omega(h,g)}~,
\fe
that is free of the coboundary ambiguity in \eqref{eq:co-boundary},
\ie
e^{i\phi(g,h)}\rightarrow e^{i\phi(g,h)}e^{i\beta(gh)-i\beta(hg)}=e^{i\phi(g,h)}~.
\fe
Thus, it is more convenient to use this phase $e^{i\phi(g,h)}$ to characterize the projective representations for the finite abelian group $G$.
This phase enters into the commutation relations of the operators
\ie
V_gV_h=e^{i\phi(g,h)}V_hV_g~.
\fe
Denote the generator of the $\mathbb{Z}_{N_I}$ subgroup of $G$ by $V_I$. The most general commutation relations are
\ie
V_IV_J=e^{{2\pi i K_{IJ}}/{N_{IJ}}}V_JV_I~,
\fe
where $K_{IJ}$ is an integer antisymmetric matrix with period $K_{IJ}\sim K_{IJ}+N_{IJ}$ that exactly parameterizes the second group cohomology \eqref{eq:second-group-cohom}. The commutation relations are constrained by the requirement that $V_I^{N_I}=1$.

Using \eqref{eq:second-group-cohom}, we can compute the classification group $\mathcal{C}[G]$ of the $G$ dipolar SPTs, given by \eqref{eq:classification}, for the finite abelian group $G=\prod \mathbb{Z}_{N_I}$:
\ie
\mathcal{C}[G]&=\frac{H^2(G\times G,U(1))}{[H(G,U(1))]^2}
=\frac{\prod_{I<J}\mathbb{Z}_{N_{IJ}}\times \prod_{I,J}\mathbb{Z}_{N_{IJ}} \times\prod_{I<J}\mathbb{Z}_{N_{IJ}}}{\prod_{I<J}\mathbb{Z}_{N_{IJ}}\times \prod_{I<J}\mathbb{Z}_{N_{IJ}}}
=\prod_{I\leq J}\mathbb{Z}_{N_{IJ}}\times \prod_{I<J}\mathbb{Z}_{N_{IJ}}~.
\fe
This gives the formula in \eqref{eq:classification-explicit}.

\section{Dijkgraaf-Witten theories in 1+1d}
\label{app:DW_theory}

In this appendix, we review the Dijkgraaf-Witten theories \cite{Dijkgraaf:1989pz} for finite abelian gauge groups in 1+1 dimensions. We will present the theories in their continuum formulations following \cite{Kapustin:2014gua}. 

Dijkgraaf-Witten theories are topological gauge theories for finite gauge groups $G$. When the gauge fields are classical, the Dijkgraaf-Witten actions characterize the response of SPTs protected by $G$ global symmetry. In 1+1 dimensions, they are classified by the second group cohomology $H^2(G,U(1))$. For a finite abelian group $G=\prod\mathbb{Z}_{N_I}$, $H^2(G,U(1))=\prod_{I<J}\mathbb{Z}_{N_{IJ}}$. In the continuum formulation, the abelian Dijkgraaf-Witten theories are described the following Lagrangian
\ie\label{eq:DW-action}
\mathcal{L}=\sum_I \frac{N_I}{2\pi}\phi_I d A_I+ \sum_{I<J}\frac{G_{IJ}N_I N_J}{2\pi N_{IJ}} A_I\wedge A_J~.
\fe
Here, the Lagrangian is written using differential forms; $\phi_I\sim\phi_I+2\pi$ are compact scalars and $A_I$ are $U(1)$ gauge fields; $G_{IJ}$ is an antisymmetric matrix and $N_{IJ}\equiv \gcd(N_I,N_J)$. The scalars $\phi_I$ act as Lagrange multipliers that restrict the $U(1)$ gauge fields $A_I$ to $\mathbb{Z}_{N_I}$ gauge fields. There is a gauge symmetry
\ie
A_I&\rightarrow A_I+d\alpha_I~,
\\
\phi_I&\rightarrow \phi_I -\sum_J \frac{G_{IJ} N_J}{N_{IJ}} \alpha_J~.
\fe
Since both $\phi_I$ and $\alpha_I$ are $2\pi$-periodic scalars, the coefficients ${G_{IJ} N_J}/{N_{IJ}}$ appear in the gauge transformation have to be integers. This quantizes $G_{IJ}$ to be integer-valued. Furthermore, since $A_I$ are effectively $\mathbb{Z}_{N_I}$ gauge fields, the integral $\frac{1}{2\pi}\int dx^2\, N_I N_J A_I\wedge A_J$ is an integer multiple of $2\pi$. Therefore, $G_{IJ}$ has the identification $G_{IJ}\sim G_{IJ}+N_{IJ}$. In summary, $G_{IJ}$ is an integer antisymmetric matrix with identification $G_{IJ}\sim G_{IJ}+N_{IJ}$, which exactly parametrizes $H^2(G,U(1))=\prod_{I<J}\mathbb{Z}_{N_{IJ}}$, the classification group of the 1+1d abelian Dijgraaf-Witten theories. 

The finite tensor gauge theories constructed in Sec.~\ref{sec:field-theory} can be thought of as the generalizations of these Dijkgraaf-Witten theories. Similar Lagrangian in 2+1 dimensions,
\ie
\mathcal{L}=\frac{N}{2\pi}\phi(\partial_t A_{xy}-\partial_x\partial_y A_t)+\frac{N}{2\pi}A_tA_{xy}~,
\fe
for a hollow symmetric tensor gauge field,
\ie
A_t\sim A_t+\partial_t\alpha~,\quad A_{xy}\sim A_{xy}+\partial_x\partial_y \alpha~,
\fe
has also appeared in \cite{Burnell:2021reh}, which were used to cancel the anomalies of 1+1d theories with subsystem symmetries via anomaly inflow.

\section{Compactness of tensor gauge fields}
\label{app:compactness}

In this appendix, we give a detailed treatment on the compactness of the tensor gauge fields in \eqref{eq:tensor_gauge_field}. This will explain why the parameters $K_{IJ}$ and $G_{IJ}$ in the response functional \eqref{eq:charge-dipole-term}, \eqref{eq:dipole-dipole-term} are quantized and have the $N_{IJ}$ identifications. Specifically, we will formulate the tensor gauge theory \eqref{eq:full-action} with the response functionals \eqref{eq:charge-dipole-term}, \eqref{eq:dipole-dipole-term} in the modified Villain formulation developed in \cite{Sulejmanpasic:2019ytl,Gorantla:2021svj,Fazza:2022fss}. The modified Villain formalism is a modification of the traditional Villain formalism \cite{Villain:1974ir} that suppresses the fluxes of the integer Villain gauge fields. It has the advantage of preserving the global symmetries and dualities on the lattice \cite{Gorantla:2021svj}. 

Different from the main text, we will first work on a square spacetime lattice with a discrete space and a discrete Lorentzian time. Later, we will take a continuum limit in the time direction to make contact with the setup in the main text. The sites on the spacetime lattice are labeled by the coordinates $(x,t)$. We label the links and plaquettes by their vertices with the smallest coordinates $(x,t)$. Various real-valued fields are defined at
\ie
\text{sites:}\ \ \phi^I, \alpha^I, A_{xx}^I~,\quad \text{$t$-links:}\ \ A_{t}^I~.
\fe
In order to make these real-valued gauge fields compact or circle-valued, we introduce additional integer-valued Villain gauge fields at
\ie
\text{$t$-links:}\ \ n_t^I, n_{txx}^I~,\quad \text{$x$-links:}\ \ n_x^I~.
\fe
Following the modified Villain formalism, we further introduce real-valued field to suppress the fluxes of the integer-valued Villain gauge field
\ie
\text{sites:}\ \ \tilde\phi^I~.
\fe

After introducing these fields, we are now ready to write down the full Lagrangian of the response field theory. For simplicity, we will consider the case when the diagonal entries of the symmetric matrix $K_{IJ}$ are all even. We can then define a new integer symmetric matrix 
\ie
\hat K_{IJ}=\begin{cases}
    \frac{1}{2}K_{IJ}~,\quad &I=J
    \\
    K_{IJ}~,\quad &I\neq J
\end{cases}~.
\fe
The full Lagrangian is
\ie\label{eq:villain-discrete}
\mathcal{L}=&\,\sum_I\frac{N_I}{2\pi}\phi^I(x,t)\left[\Delta_t A_{xx}^I(x,t)-\Delta_x^2 A_t^I(x,t)\right]
\\
&+\sum_{I\leq J}\frac{\hat K_{IJ} N_I N_J}{2\pi N_{IJ}} \left[A^I_t(x,t) A^J_{xx}(x,t)+A^J_t(x,t-1) A^I_{xx}(x,t)\right]
\\
&+\sum_{I<J}\frac{G_{IJ} N_I N_J}{2\pi N_{IJ}} \left[\Delta_x A^I_t(x,t) A^J_{xx}(x,t)-\Delta_x A^J_t(x-1,t-1) A^I_{xx}(x,t)\right]
\\
&+\sum_I N_I\left[n^I_t(x,t-1) A^I_{xx}(x,t)+n^I_{xx}(x,t) A^I_t(x,t)-\phi^I(x,t) n^I_{txx}(x,t)\right]
\\
&+\sum_I\tilde \phi^I(x,t+1)\left[\Delta_t n^I_{xx}(x,t)-\Delta_{xx}n^I_t(x,t)\right]
\\
&+\sum_{I\leq J} \hat K_{IJ}\left[\tilde\phi^I(x,t+1) n^J_{txx}(x,t)+\phi^J(x,t+1) n^I_{txx}(x,t+1)\right]
\\
&-\sum_{I< J} G_{IJ}\left[\tilde\phi^I(x,t+1) \Delta_x n^J_{txx}(x-1,t)-\phi^J(x,t+1) n^I_{txx}(x,t+1)\right]~.
\fe
The first three lines are the discrete time version of the Lagrangian in \eqref{eq:full-action} with $\mathcal{L}[A^I]$ given by \eqref{eq:charge-dipole-term} and \eqref{eq:dipole-dipole-term}. The fourth line is the coupling between the real-valued fields and the integer Villain gauge fields that effectively makes the real-valued fields compact, while the last three lines are the Lagrange multiplier terms that suppress the fluxes of the integer Villain gauge fields. Let's define
\ie
F^I[\alpha](x,t)
=\,\sum_{J\geq I}\frac{N_J}{ N_{IJ}}\left[\hat K_{IJ}\alpha^J(x,t)-G_{IJ}\Delta_x\alpha^J(x-1,t)\right]
+\sum_{J\leq I}\frac{N_J}{ N_{IJ}}\left[\hat K_{IJ}\alpha^J(x,t+1)-G_{IJ}\Delta_x\alpha^J(x,t+1)\right]~.
\fe
In additional to the standard gauge symmetries
\ie
A_t^I(x,t)&\rightarrow  A_t^I(x,t)+\Delta_t\alpha^I(x,t)~,
\\
A_{xx}^I(x,t)&\rightarrow  A_{xx}^I(x,t)+\Delta_x^2\alpha^I(x,t)~,
\\
\phi^I(x,t)&\rightarrow  \phi^I(x,t)+F^I[\alpha](x,t)~,
\\
\tilde\phi^I(x,t)&\rightarrow  \tilde\phi^I(x,t)+N^I\alpha^I(x,t)~,
\fe
there are also integer gauge symmetries that effectively make the real-valued fields compact
\ie\label{eq:integer-gauge-symmetry}
\phi^I(x,t)&\rightarrow  \phi^I(x,t)+2\pi M^I(x,t)~,
\\
\tilde\phi^I(x,t)&\rightarrow  \tilde\phi^I(x,t)+2\pi \tilde M^I(x,t)~,
\\
A_t^I(x,t)&\rightarrow  A_t^I(x,t)+2\pi M_t^I(x,t)~,
\\
A_{xx}^I(x,t)&\rightarrow  A_{xx}^I(x,t)+2\pi M_{xx}^I(x,t)~,
\\
n^I_t(x,t)&\rightarrow  n^I_t(x,t)+\Delta_t M^I(x,t)-F^I[M_t](x,t)~,
\\
n^I_{xx}(x,t)&\rightarrow  n^I_{xx}(x,t)+\Delta_x^2 M^I(x,t)-F^I[M_{xx}](x,t)~,
\\
n^I_{txx}(x,t)&\rightarrow  n^I_{txx}(x,t)+\Delta_t M_{xx}^I(x,t)-\Delta_x^2 M_t^I(x,t)~.
\fe
In order for the integer gauge symmetries \eqref{eq:integer-gauge-symmetry} to not spoil the integrality of $n^I_t$ and $n^I_{xx}$, the parameters $\hat K_{IJ}$ and $G_{IJ}$ have to be integers.

We can significantly simplify the Lagrangian \eqref{eq:villain-discrete} by gauge fixing $\tilde \phi^I=0$ using the gauge parameter $\alpha^I$ and integrating out all the Villain integer gauge fields $n_t^I$, $n^I_{txx}$, $n_x^I$ to constrain
\ie
A^I_t = \frac{2\pi}{N_I} m^I_t~,\quad A^I_{xx} = \frac{2\pi}{N_I} m^I_{xx}~, \quad A^I_t = \frac{2\pi}{N_I} m^I_t~,
\fe
to be ${2\pi}/{N_I}$ multiple of integer fields $m^I_t$, $m^I_{xx}$, $m^I_t$. The new Lagrangian is
\ie
\mathcal{L}=&\,\sum_I\frac{2\pi}{N_I}m^I(x,t)\left[\Delta_t m_{xx}^I(x,t)-\Delta_x^2 m_t^I(x,t)\right]
\\
&+\sum_{I\leq J}\frac{2\pi \hat K_{IJ} }{N_{IJ}} \left[m^I_t(x,t) m^J_{xx}(x,t)+m^J_t(x,t-1) m^I_{xx}(x,t)\right]
\\
&+\sum_{I<J}\frac{2\pi G_{IJ} }{N_{IJ}} \left[\Delta_x m^I_t(x,t) m^J_{xx}(x,t)-\Delta_x m^J_t(x-1,t-1) m^I_{xx}(x,t)\right]~.
\fe
Shifting $\hat K_{IJ}$ or $G_{IJ}$ by $N_{IJ}$ shifts the Lagrangian by an integer multiple of $2\pi$. Therefore, they should be viewed as an identification on $\hat K_{IJ}$ and $G_{IJ}$.

Starting from the Lagrangian \eqref{eq:villain-discrete}, we can further take the continuum limit in the time direction to derive a Lagrangian in the setup used in the main text. For simplicity, we will assume the time direction is non-compact. Before taking the continuum time limit, we use the integer gauge symmetry \eqref{eq:integer-gauge-symmetry} of $M^I$ and $M_{xx}^I$ to gauge fix $n^I_{t}$ and $n^I_{txx}$ to zero and then integrate out $\tilde \phi$ to constrain $n^I_{xx}$ to be time-independent. Let's introduce a lattice constant $a_t$ in the time direction. To take the continuum time limit, we replace
\ie
t\rightarrow a_t t~,\quad \Delta_t\rightarrow a_t\partial_t~,\quad A_t\rightarrow a_t A_t~.
\fe
and take the limit $a_t\rightarrow 0$. An overall factor of $a_t$ can be absorbed into the sum over $t$ turning it into an integral over $t$. In the end, the continuum Lagrangian is
\ie\label{eq:villain-continuum}
\mathcal{L}=&\,\sum_I\frac{N_I}{2\pi}\phi^I(x,t)\left[\partial_t A_{xx}^I(x,t)-\Delta_x^2 A_t^I(x,t)\right]+\sum_I  N_In^I_{xx}(x)A^I_t(x,t)
\\
&\,+\sum_{I,J}\frac{ K_{IJ}N_I N_J}{2\pi N_{IJ}} A^I_t(x,t) A^J_{xx}(x,t)+\sum_{I<J}\frac{G_{IJ} N_I N_J}{2\pi N_{IJ}} \left[\Delta_x A^I_t(x,t) A^J_{xx}(x,t)-\Delta_x A^J_t(x-1,t) A^I_{xx}(x,t)\right]~.
\fe
The residue gauge symmetries are
\ie\label{eq:gauge-symmetries-continuum}
\phi^I(x,t)\rightarrow &\, \phi^I(x,t)+\sum_{J}\frac{ N_J}{ N_{IJ}}\left[K_{IJ}\alpha^J(x,t)-G_{IJ}\Delta_x\alpha^J(x,t)\right]+2\pi M^I(x)~,
\\
A_t^I(x,t)\rightarrow &\, A_t^I(x,t)+\partial_t \alpha^I(x,t)~,
\\
A_{xx}^I(x,t)\rightarrow &\, A_{xx}^I(x,t)+\Delta_x^2\alpha^I(x,t)+2\pi M_{xx}^I(x)~,
\\
n^I_{xx}(x)\rightarrow &\, n^I_{xx}(x)+\Delta_x^2 M^I(x)-\sum_{J}\frac{ N_J}{ N_{IJ}}\left[K_{IJ}M_{xx}^J(x)-G_{IJ}\Delta_xM_{xx}^J(x)\right]~.
\fe
The parameters $K_{IJ}$ and $G_{IJ}$ are quantized to be integers for the gauge symmetries \eqref{eq:gauge-symmetries-continuum} to preserve the integrality of $n^I_{xx}$. Curiously, the continuum villain action \eqref{eq:villain-continuum} realizes all integer-valued $K_{IJ}$ while the lattice villain action \eqref{eq:villain-discrete} as it stands only realizes $K_{IJ}$ with even diagonal entries.

\section{Derivation of response field theories from stabilizer Hamiltonians}
\label{app:derivation-response-theory}

In this appendix, we derive the response functionals \eqref{eq:charge-dipole-term}, \eqref{eq:dipole-dipole-term} starting from the stabilizer Hamiltonians in \eqref{eq:ZN_Hamiltonian}, \eqref{eq:ZN_ZM_Hamiltonian} and \eqref{eq:quandrupole_Hamiltonian}. Our strategy is to first reformulate the stabilizer Hamiltonian in the Lagrangian formalism, then couple it to the background tensor gauge field in \eqref{eq:tensor_gauge_field}, and then derive the response field theory by integrating out the dynamical fields. 

\subsection{$\mathbb{Z}_N$ response field theory}

Zooming into the space of ground states, the stabilizer Hamiltonian in \eqref{eq:ZN_Hamiltonian} is described by the Lagrangian
\ie\label{eq:ZN-stabilizer-action}
\frac{N}{2\pi}a_{xx}(x,t) \partial_t b(x,t)+\frac{N}{2\pi}a_t(x,t) \left[a_{xx}(x,t)+\eta\Delta_x^2 b(x,t)-2\pi m_{xx}(x)\right]~.
\fe
Here $a_{xx}$ and $b$ are compact scalars, $a_t$ is a real-valued field and $m_{xx}$ is an integer-valued field. As $a_{xx}$ and $b$  are compact scalars, there are integer gauge symmetries
\ie
a_{xx}(x,t)&\rightarrow  a_{xx}(x,t)+2\pi k(x)~,
\\
b(x,t)&\rightarrow  b(x,t)+2\pi l(x)~,
\\
m_{xx}(x)&\rightarrow  m_{xx}(x)+k(x)+\eta\Delta_x l(x)~.
\fe
There are also real-valued gauge symmetries
\ie
a_t(x,t)&\rightarrow  a_t(x,t)+\partial_t \gamma(x,t)~,
\\
a_{xx}(x,t)&\rightarrow  a_{xx}(x,t)+\eta\Delta_x^2\gamma(x,t)~,
\\
b(x,t)&\rightarrow  b(x,t)-\gamma(x,t)~.
\fe
When $\eta=1$, this is in fact the same Lagrangian as \eqref{eq:villain-continuum} specialized to $G=\mathbb{Z}_N$ and $K=\eta$, except that all the fields here are dynamical.
The first line of the Lagrangian \eqref{eq:ZN-stabilizer-action} describes a $\mathbb{Z}_N$ quantum mechanics at every site where $a_{xx}(x)$ is the conjugate momentum of $b(x)$. Upon quantization, $a_{xx}(x)$ and $b(y)$ become noncommutative
\ie
[a_{xx}(x),b(y)]=\frac{2\pi i}{N}\delta_{x,y}~,
\fe
and we obtain a $N$-dimensional degenerate Hilbert space at every site $i=x$ with operators
\ie
X_i=e^{ia_{xx}(x)}~,\quad Z_i=e^{ib(x)}~,
\fe
obeying the Heisenberg algebra
\ie
Z_i X_j = e^{{2\pi i\delta_{ij}}/{N}} X_i Z_j~.
\fe
The second part of the Lagrangian \eqref{eq:ZN-stabilizer-action} is a Lagrange multiplier term. Integrating out $a_t$ imposes a constraint 
\ie
a_{xx}+\eta \Delta_x^2 b=0\text{ mod }2\pi~.
\fe
Exponentiating it gives 
\ie
a_i=(Z_{i-1} Z_i^\dagger)^\eta X_i (Z_i^\dagger Z_{i+1})^\eta=1~,
\fe
which exactly reproduces the stabilizer conditions $a_i|\Psi\rangle =|\Psi\rangle$ on the ground states $|\Psi\rangle$ of the stabilizer Hamiltonian in \eqref{eq:ZN_Hamiltonian}.

Now we couple the Lagrangian \eqref{eq:ZN-stabilizer-action} to the tensor gauge fields in \eqref{eq:tensor_gauge_field}. As the charge and dipole operators for the dipole symmetry are products of the $X=\exp(ia_{xx})$ operators, we should couple $A_t$ to $a_{xx}$. In order to preserve the gauge symmetry, $A_{xx}$ should couple to $\eta a_t$. In the end, the coupling adds to the Lagrangian \eqref{eq:ZN-stabilizer-action} the following terms
\ie
\frac{N}{2\pi} \left[A_t(x,t) a_{xx}(x,t)+\eta A_{xx}(x,t) a_t(x,t)\right]
+\frac{N}{2\pi} \phi(x,t) \big[\partial_t A_{xx}(x,t)-\Delta_x^2A_t(x,t)\big]+Nn_{xx}(x) A_t(x,t)~.
\fe
There are gauge symmetries
\ie
\phi(x,t)\rightarrow &\ \phi(x,t)+\eta\alpha(x,t)+\eta\gamma(x,t)+2\pi M(x)~,
\\
A_t(x,t)\rightarrow &\ A_t(x,t)+\partial_t \alpha(x,t)~,
\\
A_{xx}(x,t)\rightarrow &\ A_{xx}(x,t)+\Delta_x^2\alpha(x,t)+2\pi M_{xx}(x)~,
\\
n_{xx}(x)\rightarrow &\ n_{xx}(x)+\Delta_x^2 M(x)-k(x)~,
\\
b(x,t)\rightarrow &\ b(x,t)-\alpha(x,t)~,
\\
m_{xx}(x)\rightarrow  &\ m_{xx}(x)+\eta M_{xx}(x)~.
\fe
where $\alpha$ is real-valued and $M$, $M_{xx}$ are integer-valued.

To arrive at the response field theory, we integrate out the dynamical fields $a_{xx}$ and obtain
\ie
a_t=-\partial_t b-A_t ~.
\fe
Substituting it back to the Lagrangian, shifting $\phi\rightarrow \phi-\eta b$, $n_{xx}\rightarrow n_{xx}+m_{xx}$ and dropping the total derivative terms gives
\ie
\mathcal{L}=\frac{N}{2\pi} \phi(x,t) \big[\partial_t A_{xx}(x,t)-\Delta_x^2A_t(x,t)\big]-\frac{\eta N}{2\pi} A_t(x,t)A_{xx}(x,t)+Nn_{xx}(x) A_t(x,t) ~.
\fe
It is the same Lagrangian as \eqref{eq:villain-continuum} specialized to $G=\mathbb{Z}_N$ with $K_{IJ}=-\eta$.

\subsection{$\mathbb{Z}_N\times\mathbb{Z}_{M}$ response field theory}

Following similar steps as in the previous subsection, we can derive the response field theory for the stabilizer Hamiltonian in \eqref{eq:ZN_ZM_Hamiltonian}. Denote the tensor gauge fields that couple to the $\mathbb{Z}_N$ and $\mathbb{Z}_{M}$ dipole symmetry as $(A_t, A_{xx})$ and $(\tilde A_t,\tilde A_{xx})$. The Lagrangian for the response field theory is
\ie
\mathcal{L}=&\,\frac{N}{2\pi}\phi(x,t)\left[\partial_t A_{xx}(x,t)-\Delta_x^2 A_t(x,t)\right]
+Nn_{xx}(x)A_t(x,t)
\\
&\,+\frac{M}{2\pi}\tilde\phi(x,t)\left[\partial_t \tilde A_{xx}(x,t)-\Delta_x^2 \tilde A_t(x,t)\right]+M\tilde n_{xx}(x)\tilde A_t(x,t)
\\
&\,-\frac{\eta N}{2\pi} A_t(x,t)A_{xx}(x,t)-\frac{\tilde \eta M}{2\pi}\tilde A_t(x,t)\tilde A_{xx}(x,t)
-\frac{\zeta N M}{2\pi K} \left[A_t(x,t)\tilde A_{xx}(x,t)+\tilde A_t(x,t) A_{xx}(x,t)\right]
~.
\fe
It is the same Lagrangian as \eqref{eq:villain-continuum} specialized to $G=\mathbb{Z}_N\times\mathbb{Z}_M$ with $G_{IJ}=0$ and
\ie
K_{IJ}=-\left[\begin{array}{cc}
     \eta & \zeta \\
     \zeta & \tilde \eta
\end{array}\right]~.
\fe

Let's now turn to the stabilizer Hamiltonian in \eqref{eq:quandrupole_Hamiltonian} and derive its response field theory. The stabilizer Hamiltonian when acting on the space of ground states is described by the Lagrangian
\ie\label{eq:ZN-quadrupole-stabilizer-action}
&\frac{N}{2\pi}a(x,t) \partial_t b(x,t)+\frac{M}{2\pi}\tilde a(x,t) \partial_t \tilde b(x,t)
\\
&+\frac{M}{2\pi}a_t(x,t) \left[\tilde a(x,t)-\frac{\xi N }{K}\Delta_x^3  b(x,t)-2\pi m(x)\right]
+\frac{N }{2\pi}\tilde a_t(x,t) \left[a(x+1,t)+\frac{\xi M}{K}\Delta_x^3 \tilde b(x,t)-2\pi \tilde m(x)\right]~.
\fe
Here $a$, $\tilde a$, $b$ and $\tilde b$ are compact scalars, $a_t$, $\tilde a_t$ are real-valued fields, $m$, $\tilde m$ are integer-valued fields and the difference operator is defined as
\ie
\Delta_x^3 \alpha(x)=\alpha(x+2)-3\alpha(x+1)+3\alpha(x)-\alpha(x-1)~.
\fe

The Lagrangian as it stands has a $\mathbb{Z}_N\times\mathbb{Z}_M$ quadrupole symmetry. Let's couple the symmetry to its corresponding background tensor gauge fields
\ie\label{eq:quadrupole-background}
&B_t(x,t)\rightarrow  B_t(x,t)+\partial_t \beta(x,t)~,
\\
&B(x,t)\rightarrow  B(x,t)+\Delta_x^3 \beta(x,t)+2\pi L(x)~,
\\
&\tilde B_t(x,t)\rightarrow  \tilde B_t(x,t)+\partial_t \tilde \beta(x,t)~,
\\
&\tilde B(x,t)\rightarrow  \tilde B(x,t)+\Delta_x^3 \tilde \beta(x,t)+2\pi \tilde L(x)~.
\fe
The gauge field $(B_t,B)$ couples to the $\mathbb{Z}_N$ quadrupole symmetry while the gauge field $(\tilde B_t,\tilde B)$ couples to the $\mathbb{Z}_M$ subgroup. This leads to the coupling
\ie\label{eq:quadrupole-coupling}
&\frac{N}{2\pi} \left[B_t(x,t) a(x,t)-\frac{\xi M}{K} B(x,t) a_t(x,t)\right]+\frac{M}{2\pi} \left[\tilde B_t(x,t) \tilde a(x,t)+\frac{\xi N}{K}\tilde B(x,t) \tilde a_t(x,t)\right]
\\
&+\frac{N}{2\pi} \varphi(x,t) \big[\partial_t B(x,t)-\Delta_x^3B_t(x,t)\big]
+Nn(x) B_t(x,t)
+\frac{M}{2\pi} \tilde\varphi(x,t) \big[\partial_t \tilde B(x,t)-\Delta_x^3\tilde B_t(x,t)\big]+M\tilde n(x) \tilde B_t(x,t)~.
\fe
where $\varphi$ and $\tilde\varphi$ are compact scalars that constrain $(B_t,B)$ and $(\tilde B_t, \tilde B)$ to be $\mathbb{Z}_{N}$ and $\mathbb{Z}_M$ tensor gauge fields.
Integrating out the dynamical fields, we obtain the response field theories for the quadrupole symmetry
\ie
\mathcal{L}=
&\,\frac{\xi N M}{2\pi K} \left[B(x,t) \tilde B_t(x,t)-\tilde B(x,t) B_t(x+1,t)\right]
\\
&\,+\frac{N}{2\pi} \varphi(x,t) \big[\partial_t B(x,t)-\Delta_x^3B_t(x,t)\big]
+Nn(x) B_t(x,t)
\\
&\,+\frac{M}{2\pi} \tilde\varphi(x,t) \big[\partial_t \tilde B(x,t)-\Delta_x^3\tilde B_t(x,t)\big]+M\tilde n(x) \tilde B_t(x,t)~,
\fe

Instead of coupling the symmetry to the background tensor gauge fields \eqref{eq:quadrupole-background} for quadrupole symmetries, we can forget the quadrupole operators and view the symmetry only as a $\mathbb{Z}_N\times\mathbb{Z}_M$ dipole symmetry and then couple the dipole symmetry to background tensor gauge fields  $(A_t,A_{xx})$, $(\tilde A_t,\tilde A_{xx})$. This leads to the coupling
\ie
&\frac{N}{2\pi} \left[A_t(x,t) a(x,t)-\frac{\xi M}{K} \Delta_x A_{xx}(x,t) a_t(x,t)\right]
+\frac{M}{2\pi} \left[\tilde A_t(x,t) \tilde a(x,t)+\frac{\xi N}{K}\Delta_x\tilde A_{xx}(x,t) \tilde a_t(x,t)\right]
\\
&+\frac{N}{2\pi} \phi(x,t) \big[\partial_t A_{xx}(x,t)-\Delta_x^2A_t(x,t)\big]
+Nn_{xx}(x) A_t(x,t)
+\frac{M}{2\pi} \tilde\phi(x,t) \big[\partial_t \tilde A_{xx}(x,t)-\Delta_x^2\tilde A_t(x,t)\big]+M\tilde n_{xx}(x) \tilde A_t(x,t)\,.
\fe
The first line is related to the first line of \eqref{eq:quadrupole-coupling} by the substitution map
\ie
(B_t,B)\rightarrow(A_t,\Delta A_{xx})~,\quad (\tilde B_t,\tilde B)\rightarrow(\tilde A_t,\Delta \tilde A_{xx})~.
\fe
Integrating out the dynamical fields, we obtain the response field theory for the dipole symmetry
\ie
\mathcal{L}=
&\,\frac{\xi N M}{2\pi K} \left[\Delta_x A_{xx}(x,t) \tilde A_t(x,t)-\Delta_x\tilde A_{xx}(x,t) A_t(x+1,t)\right]
\\
&\,+\frac{N}{2\pi} \phi(x,t) \big[\partial_t A_{xx}(x,t)-\Delta_x^2A_t(x,t)\big]
+Nn_{xx}(x) A_t(x,t)
\\
&\,+\frac{M}{2\pi} \tilde\phi(x,t) \big[\partial_t \tilde A_{xx}(x,t)-\Delta_x^2\tilde A_t(x,t)\big]+M\tilde n_{xx}(x) \tilde A_t(x,t)~.
\fe
Upon summation by part, it matches with the general Lagrangian \eqref{eq:villain-continuum}, specialized to $G=\mathbb{Z}_N\times\mathbb{Z}_M$ with $K_{IJ}=0$ and 
\ie
G_{IJ}=\left[\begin{array}{cc}
     0& \xi \\
    -\xi & 0
\end{array}\right]~.
\fe

\bibliography{ref}

\begin{thebibliography}{141}%
\makeatletter
\providecommand \@ifxundefined [1]{%
 \@ifx{#1\undefined}
}%
\providecommand \@ifnum [1]{%
 \ifnum #1\expandafter \@firstoftwo
 \else \expandafter \@secondoftwo
 \fi
}%
\providecommand \@ifx [1]{%
 \ifx #1\expandafter \@firstoftwo
 \else \expandafter \@secondoftwo
 \fi
}%
\providecommand \natexlab [1]{#1}%
\providecommand \enquote  [1]{``#1''}%
\providecommand \bibnamefont  [1]{#1}%
\providecommand \bibfnamefont [1]{#1}%
\providecommand \citenamefont [1]{#1}%
\providecommand \href@noop [0]{\@secondoftwo}%
\providecommand \href [0]{\begingroup \@sanitize@url \@href}%
\providecommand \@href[1]{\@@startlink{#1}\@@href}%
\providecommand \@@href[1]{\endgroup#1\@@endlink}%
\providecommand \@sanitize@url [0]{\catcode `\\12\catcode `\$12\catcode
  `\&12\catcode `\#12\catcode `\^12\catcode `\_12\catcode `\%12\relax}%
\providecommand \@@startlink[1]{}%
\providecommand \@@endlink[0]{}%
\providecommand \url  [0]{\begingroup\@sanitize@url \@url }%
\providecommand \@url [1]{\endgroup\@href {#1}{\urlprefix }}%
\providecommand \urlprefix  [0]{URL }%
\providecommand \Eprint [0]{\href }%
\providecommand \doibase [0]{https://doi.org/}%
\providecommand \selectlanguage [0]{\@gobble}%
\providecommand \bibinfo  [0]{\@secondoftwo}%
\providecommand \bibfield  [0]{\@secondoftwo}%
\providecommand \translation [1]{[#1]}%
\providecommand \BibitemOpen [0]{}%
\providecommand \bibitemStop [0]{}%
\providecommand \bibitemNoStop [0]{.\EOS\space}%
\providecommand \EOS [0]{\spacefactor3000\relax}%
\providecommand \BibitemShut  [1]{\csname bibitem#1\endcsname}%
\let\auto@bib@innerbib\@empty
\bibitem [{\citenamefont {Chen}\ \emph
  {et~al.}(2011{\natexlab{a}})\citenamefont {Chen}, \citenamefont {Gu},\ and\
  \citenamefont {Wen}}]{Chen2011-et}%
  \BibitemOpen
  \bibfield  {author} {\bibinfo {author} {\bibfnamefont {X.}~\bibnamefont
  {Chen}}, \bibinfo {author} {\bibfnamefont {Z.-C.}\ \bibnamefont {Gu}},\ and\
  \bibinfo {author} {\bibfnamefont {X.-G.}\ \bibnamefont {Wen}},\ }\bibfield
  {title} {\bibinfo {title} {Classification of gapped symmetric phases in
  one-dimensional spin systems},\ }\href@noop {} {\bibfield  {journal}
  {\bibinfo  {journal} {Phys. Rev. B Condens. Matter}\ }\textbf {\bibinfo
  {volume} {83}},\ \bibinfo {pages} {035107} (\bibinfo {year}
  {2011}{\natexlab{a}})}\BibitemShut {NoStop}%
\bibitem [{\citenamefont {Turner}\ \emph {et~al.}(2011)\citenamefont {Turner},
  \citenamefont {Pollmann},\ and\ \citenamefont {Berg}}]{Turner2011-zi}%
  \BibitemOpen
  \bibfield  {author} {\bibinfo {author} {\bibfnamefont {A.~M.}\ \bibnamefont
  {Turner}}, \bibinfo {author} {\bibfnamefont {F.}~\bibnamefont {Pollmann}},\
  and\ \bibinfo {author} {\bibfnamefont {E.}~\bibnamefont {Berg}},\ }\bibfield
  {title} {\bibinfo {title} {Topological phases of one-dimensional fermions: An
  entanglement point of view},\ }\href
  {https://doi.org/10.1103/PhysRevB.83.075102} {\bibfield  {journal} {\bibinfo
  {journal} {Phys. Rev. B}\ }\textbf {\bibinfo {volume} {83}},\ \bibinfo
  {pages} {075102} (\bibinfo {year} {2011})}\BibitemShut {NoStop}%
\bibitem [{\citenamefont {Fidkowski}\ and\ \citenamefont
  {Kitaev}(2011)}]{fidkowski2011topological}%
  \BibitemOpen
  \bibfield  {author} {\bibinfo {author} {\bibfnamefont {L.}~\bibnamefont
  {Fidkowski}}\ and\ \bibinfo {author} {\bibfnamefont {A.}~\bibnamefont
  {Kitaev}},\ }\bibfield  {title} {\bibinfo {title} {Topological phases of
  fermions in one dimension},\ }\href
  {https://doi.org/10.1103/PhysRevB.83.075103} {\bibfield  {journal} {\bibinfo
  {journal} {Phys. Rev. B}\ }\textbf {\bibinfo {volume} {83}},\ \bibinfo
  {pages} {075103} (\bibinfo {year} {2011})}\BibitemShut {NoStop}%
\bibitem [{\citenamefont {Schuch}\ \emph {et~al.}(2011)\citenamefont {Schuch},
  \citenamefont {Perez-Garcia},\ and\ \citenamefont {Cirac}}]{Schuch2011-jx}%
  \BibitemOpen
  \bibfield  {author} {\bibinfo {author} {\bibfnamefont {N.}~\bibnamefont
  {Schuch}}, \bibinfo {author} {\bibfnamefont {D.}~\bibnamefont
  {Perez-Garcia}},\ and\ \bibinfo {author} {\bibfnamefont {I.}~\bibnamefont
  {Cirac}},\ }\bibfield  {title} {\bibinfo {title} {Classifying quantum phases
  using matrix product states and projected entangled pair states},\
  }\href@noop {} {\bibfield  {journal} {\bibinfo  {journal} {Phys. Rev. B
  Condens. Matter}\ }\textbf {\bibinfo {volume} {84}},\ \bibinfo {pages}
  {165139} (\bibinfo {year} {2011})}\BibitemShut {NoStop}%
\bibitem [{\citenamefont {Chen}\ \emph
  {et~al.}(2011{\natexlab{b}})\citenamefont {Chen}, \citenamefont {Liu},\ and\
  \citenamefont {Wen}}]{Chen2011-kz}%
  \BibitemOpen
  \bibfield  {author} {\bibinfo {author} {\bibfnamefont {X.}~\bibnamefont
  {Chen}}, \bibinfo {author} {\bibfnamefont {Z.-X.}\ \bibnamefont {Liu}},\ and\
  \bibinfo {author} {\bibfnamefont {X.-G.}\ \bibnamefont {Wen}},\ }\bibfield
  {title} {\bibinfo {title} {Two-dimensional symmetry-protected topological
  orders and their protected gapless edge excitations},\ }\href@noop {}
  {\bibfield  {journal} {\bibinfo  {journal} {Phys. Rev. B Condens. Matter}\
  }\textbf {\bibinfo {volume} {84}},\ \bibinfo {pages} {235141} (\bibinfo
  {year} {2011}{\natexlab{b}})}\BibitemShut {NoStop}%
\bibitem [{\citenamefont {Pollmann}\ \emph {et~al.}(2012)\citenamefont
  {Pollmann}, \citenamefont {Berg}, \citenamefont {Turner},\ and\ \citenamefont
  {Oshikawa}}]{Pollmann2012-lv}%
  \BibitemOpen
  \bibfield  {author} {\bibinfo {author} {\bibfnamefont {F.}~\bibnamefont
  {Pollmann}}, \bibinfo {author} {\bibfnamefont {E.}~\bibnamefont {Berg}},
  \bibinfo {author} {\bibfnamefont {A.~M.}\ \bibnamefont {Turner}},\ and\
  \bibinfo {author} {\bibfnamefont {M.}~\bibnamefont {Oshikawa}},\ }\bibfield
  {title} {\bibinfo {title} {Symmetry protection of topological phases in
  one-dimensional quantum spin systems},\ }\href
  {https://doi.org/10.1103/PhysRevB.85.075125} {\bibfield  {journal} {\bibinfo
  {journal} {Phys. Rev. B}\ }\textbf {\bibinfo {volume} {85}},\ \bibinfo
  {pages} {075125} (\bibinfo {year} {2012})}\BibitemShut {NoStop}%
\bibitem [{\citenamefont {Chen}\ \emph {et~al.}(2012)\citenamefont {Chen},
  \citenamefont {Gu}, \citenamefont {Liu},\ and\ \citenamefont
  {Wen}}]{Chen2012-oa}%
  \BibitemOpen
  \bibfield  {author} {\bibinfo {author} {\bibfnamefont {X.}~\bibnamefont
  {Chen}}, \bibinfo {author} {\bibfnamefont {Z.-C.}\ \bibnamefont {Gu}},
  \bibinfo {author} {\bibfnamefont {Z.-X.}\ \bibnamefont {Liu}},\ and\ \bibinfo
  {author} {\bibfnamefont {X.-G.}\ \bibnamefont {Wen}},\ }\bibfield  {title}
  {\bibinfo {title} {{Symmetry-Protected} topological orders in interacting
  bosonic systems},\ }\href@noop {} {\bibfield  {journal} {\bibinfo  {journal}
  {Science}\ }\textbf {\bibinfo {volume} {338}},\ \bibinfo {pages} {1604}
  (\bibinfo {year} {2012})}\BibitemShut {NoStop}%
\bibitem [{\citenamefont {Senthil}(2015)}]{senthil15}%
  \BibitemOpen
  \bibfield  {author} {\bibinfo {author} {\bibfnamefont {T.}~\bibnamefont
  {Senthil}},\ }\bibfield  {title} {\bibinfo {title} {Symmetry-protected
  topological phases of quantum matter},\ }\href
  {https://doi.org/10.1146/annurev-conmatphys-031214-014740} {\bibfield
  {journal} {\bibinfo  {journal} {Annual Review of Condensed Matter Physics}\
  }\textbf {\bibinfo {volume} {6}},\ \bibinfo {pages} {299} (\bibinfo {year}
  {2015})},\ \Eprint
  {https://arxiv.org/abs/https://doi.org/10.1146/annurev-conmatphys-031214-014740}
  {https://doi.org/10.1146/annurev-conmatphys-031214-014740} \BibitemShut
  {NoStop}%
\bibitem [{\citenamefont {Levin}\ and\ \citenamefont
  {Gu}(2012)}]{Levin2012-dv}%
  \BibitemOpen
  \bibfield  {author} {\bibinfo {author} {\bibfnamefont {M.}~\bibnamefont
  {Levin}}\ and\ \bibinfo {author} {\bibfnamefont {Z.-C.}\ \bibnamefont {Gu}},\
  }\bibfield  {title} {\bibinfo {title} {Braiding statistics approach to
  symmetry-protected topological phases},\ }\href
  {https://doi.org/10.1103/PhysRevB.86.115109} {\bibfield  {journal} {\bibinfo
  {journal} {Phys. Rev. B}\ }\textbf {\bibinfo {volume} {86}},\ \bibinfo
  {pages} {115109} (\bibinfo {year} {2012})}\BibitemShut {NoStop}%
\bibitem [{\citenamefont {Vishwanath}\ and\ \citenamefont
  {Senthil}(2013)}]{Vishwanath2013-pb}%
  \BibitemOpen
  \bibfield  {author} {\bibinfo {author} {\bibfnamefont {A.}~\bibnamefont
  {Vishwanath}}\ and\ \bibinfo {author} {\bibfnamefont {T.}~\bibnamefont
  {Senthil}},\ }\bibfield  {title} {\bibinfo {title} {Physics of
  three-dimensional bosonic topological insulators: Surface-deconfined
  criticality and quantized magnetoelectric effect},\ }\href
  {https://doi.org/10.1103/PhysRevX.3.011016} {\bibfield  {journal} {\bibinfo
  {journal} {Phys. Rev. X}\ }\textbf {\bibinfo {volume} {3}},\ \bibinfo {pages}
  {011016} (\bibinfo {year} {2013})}\BibitemShut {NoStop}%
\bibitem [{\citenamefont {Yao}\ and\ \citenamefont {Ryu}(2013)}]{Yao2013-vc}%
  \BibitemOpen
  \bibfield  {author} {\bibinfo {author} {\bibfnamefont {H.}~\bibnamefont
  {Yao}}\ and\ \bibinfo {author} {\bibfnamefont {S.}~\bibnamefont {Ryu}},\
  }\bibfield  {title} {\bibinfo {title} {Interaction effect on topological
  classification of superconductors in two dimensions},\ }\href
  {https://doi.org/10.1103/PhysRevB.88.064507} {\bibfield  {journal} {\bibinfo
  {journal} {Phys. Rev. B}\ }\textbf {\bibinfo {volume} {88}},\ \bibinfo
  {pages} {064507} (\bibinfo {year} {2013})}\BibitemShut {NoStop}%
\bibitem [{\citenamefont {Else}\ and\ \citenamefont
  {Nayak}(2014{\natexlab{a}})}]{else2014classifying}%
  \BibitemOpen
  \bibfield  {author} {\bibinfo {author} {\bibfnamefont {D.~V.}\ \bibnamefont
  {Else}}\ and\ \bibinfo {author} {\bibfnamefont {C.}~\bibnamefont {Nayak}},\
  }\bibfield  {title} {\bibinfo {title} {Classifying symmetry-protected
  topological phases through the anomalous action of the symmetry on the
  edge},\ }\href {https://doi.org/10.1103/PhysRevB.90.235137} {\bibfield
  {journal} {\bibinfo  {journal} {Phys. Rev. B}\ }\textbf {\bibinfo {volume}
  {90}},\ \bibinfo {pages} {235137} (\bibinfo {year}
  {2014}{\natexlab{a}})}\BibitemShut {NoStop}%
\bibitem [{\citenamefont {Gu}\ and\ \citenamefont {Wen}(2014)}]{Gu2014-lj}%
  \BibitemOpen
  \bibfield  {author} {\bibinfo {author} {\bibfnamefont {Z.-C.}\ \bibnamefont
  {Gu}}\ and\ \bibinfo {author} {\bibfnamefont {X.-G.}\ \bibnamefont {Wen}},\
  }\bibfield  {title} {\bibinfo {title} {Symmetry-protected topological orders
  for interacting fermions: Fermionic topological nonlinear
  $\ensuremath{\sigma}$ models and a special group supercohomology theory},\
  }\href {https://doi.org/10.1103/PhysRevB.90.115141} {\bibfield  {journal}
  {\bibinfo  {journal} {Phys. Rev. B}\ }\textbf {\bibinfo {volume} {90}},\
  \bibinfo {pages} {115141} (\bibinfo {year} {2014})}\BibitemShut {NoStop}%
\bibitem [{\citenamefont {Chen}\ \emph {et~al.}(2014)\citenamefont {Chen},
  \citenamefont {Lu},\ and\ \citenamefont {Vishwanath}}]{chen14}%
  \BibitemOpen
  \bibfield  {author} {\bibinfo {author} {\bibfnamefont {X.}~\bibnamefont
  {Chen}}, \bibinfo {author} {\bibfnamefont {Y.-M.}\ \bibnamefont {Lu}},\ and\
  \bibinfo {author} {\bibfnamefont {A.}~\bibnamefont {Vishwanath}},\ }\bibfield
   {title} {\bibinfo {title} {Symmetry-protected topological phases from
  decorated domain walls},\ }\href {https://doi.org/10.1038/ncomms4507}
  {\bibfield  {journal} {\bibinfo  {journal} {Nature Communications}\ }\textbf
  {\bibinfo {volume} {5}},\ \bibinfo {pages} {3507} (\bibinfo {year}
  {2014})}\BibitemShut {NoStop}%
\bibitem [{\citenamefont {Kapustin}(2014)}]{Kapustin:2014tfa}%
  \BibitemOpen
  \bibfield  {author} {\bibinfo {author} {\bibfnamefont {A.}~\bibnamefont
  {Kapustin}},\ }\href@noop {} {\bibinfo {title} {Symmetry protected
  topological phases, anomalies, and cobordisms: Beyond group cohomology}}
  (\bibinfo {year} {2014}),\ \Eprint {https://arxiv.org/abs/1403.1467}
  {arXiv:1403.1467 [cond-mat.str-el]} \BibitemShut {NoStop}%
\bibitem [{\citenamefont {Kapustin}\ \emph {et~al.}(2015)\citenamefont
  {Kapustin}, \citenamefont {Thorngren}, \citenamefont {Turzillo},\ and\
  \citenamefont {Wang}}]{Kapustin:2014dxa}%
  \BibitemOpen
  \bibfield  {author} {\bibinfo {author} {\bibfnamefont {A.}~\bibnamefont
  {Kapustin}}, \bibinfo {author} {\bibfnamefont {R.}~\bibnamefont {Thorngren}},
  \bibinfo {author} {\bibfnamefont {A.}~\bibnamefont {Turzillo}},\ and\
  \bibinfo {author} {\bibfnamefont {Z.}~\bibnamefont {Wang}},\ }\bibfield
  {title} {\bibinfo {title} {{Fermionic Symmetry Protected Topological Phases
  and Cobordisms}},\ }\href {https://doi.org/10.1007/JHEP12(2015)052}
  {\bibfield  {journal} {\bibinfo  {journal} {JHEP}\ }\textbf {\bibinfo
  {volume} {12}},\ \bibinfo {pages} {052}},\ \Eprint
  {https://arxiv.org/abs/1406.7329} {arXiv:1406.7329 [cond-mat.str-el]}
  \BibitemShut {NoStop}%
\bibitem [{\citenamefont {Gaiotto}\ and\ \citenamefont
  {Kapustin}(2016)}]{Gaiotto2016-ba}%
  \BibitemOpen
  \bibfield  {author} {\bibinfo {author} {\bibfnamefont {D.}~\bibnamefont
  {Gaiotto}}\ and\ \bibinfo {author} {\bibfnamefont {A.}~\bibnamefont
  {Kapustin}},\ }\bibfield  {title} {\bibinfo {title} {Spin {TQFTs} and
  fermionic phases of matter},\ }\href@noop {} {\bibfield  {journal} {\bibinfo
  {journal} {Int. J. Mod. Phys. A}\ }\textbf {\bibinfo {volume} {31}},\
  \bibinfo {pages} {1645044} (\bibinfo {year} {2016})}\BibitemShut {NoStop}%
\bibitem [{\citenamefont {Freed}\ and\ \citenamefont
  {Hopkins}(2021)}]{Freed:2016rqq}%
  \BibitemOpen
  \bibfield  {author} {\bibinfo {author} {\bibfnamefont {D.~S.}\ \bibnamefont
  {Freed}}\ and\ \bibinfo {author} {\bibfnamefont {M.~J.}\ \bibnamefont
  {Hopkins}},\ }\bibfield  {title} {\bibinfo {title} {{Reflection positivity
  and invertible topological phases}},\ }\href
  {https://doi.org/10.2140/gt.2021.25.1165} {\bibfield  {journal} {\bibinfo
  {journal} {Geom. Topol.}\ }\textbf {\bibinfo {volume} {25}},\ \bibinfo
  {pages} {1165} (\bibinfo {year} {2021})},\ \Eprint
  {https://arxiv.org/abs/1604.06527} {arXiv:1604.06527 [hep-th]} \BibitemShut
  {NoStop}%
\bibitem [{\citenamefont {Cheng}\ \emph {et~al.}(2018)\citenamefont {Cheng},
  \citenamefont {Bi}, \citenamefont {You},\ and\ \citenamefont
  {Gu}}]{cheng-PRB18}%
  \BibitemOpen
  \bibfield  {author} {\bibinfo {author} {\bibfnamefont {M.}~\bibnamefont
  {Cheng}}, \bibinfo {author} {\bibfnamefont {Z.}~\bibnamefont {Bi}}, \bibinfo
  {author} {\bibfnamefont {Y.-Z.}\ \bibnamefont {You}},\ and\ \bibinfo {author}
  {\bibfnamefont {Z.-C.}\ \bibnamefont {Gu}},\ }\bibfield  {title} {\bibinfo
  {title} {Classification of symmetry-protected phases for interacting fermions
  in two dimensions},\ }\href {https://doi.org/10.1103/PhysRevB.97.205109}
  {\bibfield  {journal} {\bibinfo  {journal} {Phys. Rev. B}\ }\textbf {\bibinfo
  {volume} {97}},\ \bibinfo {pages} {205109} (\bibinfo {year}
  {2018})}\BibitemShut {NoStop}%
\bibitem [{\citenamefont {Wang}\ and\ \citenamefont
  {Gu}(2018)}]{PhysRevX.8.011055}%
  \BibitemOpen
  \bibfield  {author} {\bibinfo {author} {\bibfnamefont {Q.-R.}\ \bibnamefont
  {Wang}}\ and\ \bibinfo {author} {\bibfnamefont {Z.-C.}\ \bibnamefont {Gu}},\
  }\bibfield  {title} {\bibinfo {title} {Towards a complete classification of
  symmetry-protected topological phases for interacting fermions in three
  dimensions and a general group supercohomology theory},\ }\href
  {https://doi.org/10.1103/PhysRevX.8.011055} {\bibfield  {journal} {\bibinfo
  {journal} {Phys. Rev. X}\ }\textbf {\bibinfo {volume} {8}},\ \bibinfo {pages}
  {011055} (\bibinfo {year} {2018})}\BibitemShut {NoStop}%
\bibitem [{\citenamefont {Wang}\ and\ \citenamefont
  {Gu}(2020)}]{PhysRevX.10.031055}%
  \BibitemOpen
  \bibfield  {author} {\bibinfo {author} {\bibfnamefont {Q.-R.}\ \bibnamefont
  {Wang}}\ and\ \bibinfo {author} {\bibfnamefont {Z.-C.}\ \bibnamefont {Gu}},\
  }\bibfield  {title} {\bibinfo {title} {Construction and classification of
  symmetry-protected topological phases in interacting fermion systems},\
  }\href {https://doi.org/10.1103/PhysRevX.10.031055} {\bibfield  {journal}
  {\bibinfo  {journal} {Phys. Rev. X}\ }\textbf {\bibinfo {volume} {10}},\
  \bibinfo {pages} {031055} (\bibinfo {year} {2020})}\BibitemShut {NoStop}%
\bibitem [{\citenamefont {Pai}\ \emph {et~al.}(2019)\citenamefont {Pai},
  \citenamefont {Pretko},\ and\ \citenamefont
  {Nandkishore}}]{PhysRevX.9.021003}%
  \BibitemOpen
  \bibfield  {author} {\bibinfo {author} {\bibfnamefont {S.}~\bibnamefont
  {Pai}}, \bibinfo {author} {\bibfnamefont {M.}~\bibnamefont {Pretko}},\ and\
  \bibinfo {author} {\bibfnamefont {R.~M.}\ \bibnamefont {Nandkishore}},\
  }\bibfield  {title} {\bibinfo {title} {Localization in fractonic random
  circuits},\ }\href {https://doi.org/10.1103/PhysRevX.9.021003} {\bibfield
  {journal} {\bibinfo  {journal} {Phys. Rev. X}\ }\textbf {\bibinfo {volume}
  {9}},\ \bibinfo {pages} {021003} (\bibinfo {year} {2019})}\BibitemShut
  {NoStop}%
\bibitem [{\citenamefont {Feldmeier}\ \emph {et~al.}(2020)\citenamefont
  {Feldmeier}, \citenamefont {Sala}, \citenamefont {De~Tomasi}, \citenamefont
  {Pollmann},\ and\ \citenamefont {Knap}}]{feldmeier20}%
  \BibitemOpen
  \bibfield  {author} {\bibinfo {author} {\bibfnamefont {J.}~\bibnamefont
  {Feldmeier}}, \bibinfo {author} {\bibfnamefont {P.}~\bibnamefont {Sala}},
  \bibinfo {author} {\bibfnamefont {G.}~\bibnamefont {De~Tomasi}}, \bibinfo
  {author} {\bibfnamefont {F.}~\bibnamefont {Pollmann}},\ and\ \bibinfo
  {author} {\bibfnamefont {M.}~\bibnamefont {Knap}},\ }\bibfield  {title}
  {\bibinfo {title} {Anomalous diffusion in dipole- and
  higher-moment-conserving systems},\ }\href
  {https://doi.org/10.1103/PhysRevLett.125.245303} {\bibfield  {journal}
  {\bibinfo  {journal} {Phys. Rev. Lett.}\ }\textbf {\bibinfo {volume} {125}},\
  \bibinfo {pages} {245303} (\bibinfo {year} {2020})}\BibitemShut {NoStop}%
\bibitem [{\citenamefont {Sala}\ \emph {et~al.}(2020)\citenamefont {Sala},
  \citenamefont {Rakovszky}, \citenamefont {Verresen}, \citenamefont {Knap},\
  and\ \citenamefont {Pollmann}}]{sala2020ergodicity}%
  \BibitemOpen
  \bibfield  {author} {\bibinfo {author} {\bibfnamefont {P.}~\bibnamefont
  {Sala}}, \bibinfo {author} {\bibfnamefont {T.}~\bibnamefont {Rakovszky}},
  \bibinfo {author} {\bibfnamefont {R.}~\bibnamefont {Verresen}}, \bibinfo
  {author} {\bibfnamefont {M.}~\bibnamefont {Knap}},\ and\ \bibinfo {author}
  {\bibfnamefont {F.}~\bibnamefont {Pollmann}},\ }\bibfield  {title} {\bibinfo
  {title} {Ergodicity breaking arising from hilbert space fragmentation in
  dipole-conserving hamiltonians},\ }\href
  {https://doi.org/10.1103/PhysRevX.10.011047} {\bibfield  {journal} {\bibinfo
  {journal} {Phys. Rev. X}\ }\textbf {\bibinfo {volume} {10}},\ \bibinfo
  {pages} {011047} (\bibinfo {year} {2020})}\BibitemShut {NoStop}%
\bibitem [{\citenamefont {Gromov}\ \emph {et~al.}(2020)\citenamefont {Gromov},
  \citenamefont {Lucas},\ and\ \citenamefont {Nandkishore}}]{gromov20}%
  \BibitemOpen
  \bibfield  {author} {\bibinfo {author} {\bibfnamefont {A.}~\bibnamefont
  {Gromov}}, \bibinfo {author} {\bibfnamefont {A.}~\bibnamefont {Lucas}},\ and\
  \bibinfo {author} {\bibfnamefont {R.~M.}\ \bibnamefont {Nandkishore}},\
  }\bibfield  {title} {\bibinfo {title} {Fracton hydrodynamics},\ }\href@noop
  {} {\bibfield  {journal} {\bibinfo  {journal} {Physical Review Research}\
  }\textbf {\bibinfo {volume} {2}},\ \bibinfo {pages} {033124} (\bibinfo {year}
  {2020})}\BibitemShut {NoStop}%
\bibitem [{\citenamefont {Iaconis}\ \emph
  {et~al.}(2021{\natexlab{a}})\citenamefont {Iaconis}, \citenamefont {Lucas},\
  and\ \citenamefont {Nandkishore}}]{nandkishore21}%
  \BibitemOpen
  \bibfield  {author} {\bibinfo {author} {\bibfnamefont {J.}~\bibnamefont
  {Iaconis}}, \bibinfo {author} {\bibfnamefont {A.}~\bibnamefont {Lucas}},\
  and\ \bibinfo {author} {\bibfnamefont {R.}~\bibnamefont {Nandkishore}},\
  }\bibfield  {title} {\bibinfo {title} {Multipole conservation laws and
  subdiffusion in any dimension},\ }\href
  {https://doi.org/10.1103/PhysRevE.103.022142} {\bibfield  {journal} {\bibinfo
   {journal} {Phys. Rev. E}\ }\textbf {\bibinfo {volume} {103}},\ \bibinfo
  {pages} {022142} (\bibinfo {year} {2021}{\natexlab{a}})}\BibitemShut
  {NoStop}%
\bibitem [{\citenamefont {Iaconis}\ \emph
  {et~al.}(2021{\natexlab{b}})\citenamefont {Iaconis}, \citenamefont {Lucas},\
  and\ \citenamefont {Nandkishore}}]{Nandkishore-multipole}%
  \BibitemOpen
  \bibfield  {author} {\bibinfo {author} {\bibfnamefont {J.}~\bibnamefont
  {Iaconis}}, \bibinfo {author} {\bibfnamefont {A.}~\bibnamefont {Lucas}},\
  and\ \bibinfo {author} {\bibfnamefont {R.}~\bibnamefont {Nandkishore}},\
  }\bibfield  {title} {\bibinfo {title} {Multipole conservation laws and
  subdiffusion in any dimension},\ }\href
  {https://doi.org/10.1103/PhysRevE.103.022142} {\bibfield  {journal} {\bibinfo
   {journal} {Phys. Rev. E}\ }\textbf {\bibinfo {volume} {103}},\ \bibinfo
  {pages} {022142} (\bibinfo {year} {2021}{\natexlab{b}})}\BibitemShut
  {NoStop}%
\bibitem [{\citenamefont {Sala}\ \emph {et~al.}(2022)\citenamefont {Sala},
  \citenamefont {Lehmann}, \citenamefont {Rakovszky},\ and\ \citenamefont
  {Pollmann}}]{sala2022dynamics}%
  \BibitemOpen
  \bibfield  {author} {\bibinfo {author} {\bibfnamefont {P.}~\bibnamefont
  {Sala}}, \bibinfo {author} {\bibfnamefont {J.}~\bibnamefont {Lehmann}},
  \bibinfo {author} {\bibfnamefont {T.}~\bibnamefont {Rakovszky}},\ and\
  \bibinfo {author} {\bibfnamefont {F.}~\bibnamefont {Pollmann}},\ }\bibfield
  {title} {\bibinfo {title} {Dynamics in systems with modulated symmetries},\
  }\href {https://doi.org/10.1103/PhysRevLett.129.170601} {\bibfield  {journal}
  {\bibinfo  {journal} {Phys. Rev. Lett.}\ }\textbf {\bibinfo {volume} {129}},\
  \bibinfo {pages} {170601} (\bibinfo {year} {2022})}\BibitemShut {NoStop}%
\bibitem [{\citenamefont {Glorioso}\ \emph {et~al.}(2022)\citenamefont
  {Glorioso}, \citenamefont {Guo}, \citenamefont {Rodriguez-Nieva},\ and\
  \citenamefont {Lucas}}]{glorioso22}%
  \BibitemOpen
  \bibfield  {author} {\bibinfo {author} {\bibfnamefont {P.}~\bibnamefont
  {Glorioso}}, \bibinfo {author} {\bibfnamefont {J.}~\bibnamefont {Guo}},
  \bibinfo {author} {\bibfnamefont {J.~F.}\ \bibnamefont {Rodriguez-Nieva}},\
  and\ \bibinfo {author} {\bibfnamefont {A.}~\bibnamefont {Lucas}},\ }\bibfield
   {title} {\bibinfo {title} {Breakdown of hydrodynamics below four dimensions
  in a fracton fluid},\ }\href {https://doi.org/10.1038/s41567-022-01631-x}
  {\bibfield  {journal} {\bibinfo  {journal} {Nature Physics}\ }\textbf
  {\bibinfo {volume} {18}},\ \bibinfo {pages} {912} (\bibinfo {year}
  {2022})}\BibitemShut {NoStop}%
\bibitem [{\citenamefont {Glorioso}\ \emph {et~al.}(2023)\citenamefont
  {Glorioso}, \citenamefont {Huang}, \citenamefont {Guo}, \citenamefont
  {Rodriguez-Nieva},\ and\ \citenamefont {Lucas}}]{glorioso2023goldstone}%
  \BibitemOpen
  \bibfield  {author} {\bibinfo {author} {\bibfnamefont {P.}~\bibnamefont
  {Glorioso}}, \bibinfo {author} {\bibfnamefont {X.}~\bibnamefont {Huang}},
  \bibinfo {author} {\bibfnamefont {J.}~\bibnamefont {Guo}}, \bibinfo {author}
  {\bibfnamefont {J.}~\bibnamefont {Rodriguez-Nieva}},\ and\ \bibinfo {author}
  {\bibfnamefont {A.}~\bibnamefont {Lucas}},\ }\bibfield  {title} {\bibinfo
  {title} {Goldstone bosons and fluctuating hydrodynamics with dipole and
  momentum conservation},\ }\href@noop {} {\bibfield  {journal} {\bibinfo
  {journal} {arXiv preprint arXiv:2301.02680}\ } (\bibinfo {year}
  {2023})}\BibitemShut {NoStop}%
\bibitem [{\citenamefont {Han}\ \emph {et~al.}(2023{\natexlab{a}})\citenamefont
  {Han}, \citenamefont {Lake},\ and\ \citenamefont {Ro}}]{han2023scaling}%
  \BibitemOpen
  \bibfield  {author} {\bibinfo {author} {\bibfnamefont {J.~H.}\ \bibnamefont
  {Han}}, \bibinfo {author} {\bibfnamefont {E.}~\bibnamefont {Lake}},\ and\
  \bibinfo {author} {\bibfnamefont {S.}~\bibnamefont {Ro}},\ }\href@noop {}
  {\bibinfo {title} {Scaling and localization in kinetically constrained
  diffusion}} (\bibinfo {year} {2023}{\natexlab{a}}),\ \Eprint
  {https://arxiv.org/abs/2304.03276} {arXiv:2304.03276 [cond-mat.stat-mech]}
  \BibitemShut {NoStop}%
\bibitem [{\citenamefont {Jain}\ \emph {et~al.}(2023)\citenamefont {Jain},
  \citenamefont {Jensen}, \citenamefont {Liu},\ and\ \citenamefont
  {Mefford}}]{Jain:2023nbf}%
  \BibitemOpen
  \bibfield  {author} {\bibinfo {author} {\bibfnamefont {A.}~\bibnamefont
  {Jain}}, \bibinfo {author} {\bibfnamefont {K.}~\bibnamefont {Jensen}},
  \bibinfo {author} {\bibfnamefont {R.}~\bibnamefont {Liu}},\ and\ \bibinfo
  {author} {\bibfnamefont {E.}~\bibnamefont {Mefford}},\ }\bibfield  {title}
  {\bibinfo {title} {{Dipole superfluid hydrodynamics}},\ }\href@noop {} {\
  (\bibinfo {year} {2023})},\ \Eprint {https://arxiv.org/abs/2304.09852}
  {arXiv:2304.09852 [hep-th]} \BibitemShut {NoStop}%
\bibitem [{\citenamefont {Gliozzi}\ \emph {et~al.}(2023)\citenamefont
  {Gliozzi}, \citenamefont {May-Mann}, \citenamefont {Hughes},\ and\
  \citenamefont {De~Tomasi}}]{Gliozzi:2023zth}%
  \BibitemOpen
  \bibfield  {author} {\bibinfo {author} {\bibfnamefont {J.}~\bibnamefont
  {Gliozzi}}, \bibinfo {author} {\bibfnamefont {J.}~\bibnamefont {May-Mann}},
  \bibinfo {author} {\bibfnamefont {T.~L.}\ \bibnamefont {Hughes}},\ and\
  \bibinfo {author} {\bibfnamefont {G.}~\bibnamefont {De~Tomasi}},\ }\bibfield
  {title} {\bibinfo {title} {{Hierarchical hydrodynamics in long-range
  multipole-conserving systems}},\ }\href@noop {} {\  (\bibinfo {year}
  {2023})},\ \Eprint {https://arxiv.org/abs/2304.12342} {arXiv:2304.12342
  [cond-mat.stat-mech]} \BibitemShut {NoStop}%
\bibitem [{\citenamefont {Chamon}(2005)}]{Chamon2005-fc}%
  \BibitemOpen
  \bibfield  {author} {\bibinfo {author} {\bibfnamefont {C.}~\bibnamefont
  {Chamon}},\ }\bibfield  {title} {\bibinfo {title} {Quantum glassiness in
  strongly correlated clean systems: an example of topological
  overprotection},\ }\href@noop {} {\bibfield  {journal} {\bibinfo  {journal}
  {Phys. Rev. Lett.}\ }\textbf {\bibinfo {volume} {94}},\ \bibinfo {pages}
  {040402} (\bibinfo {year} {2005})}\BibitemShut {NoStop}%
\bibitem [{\citenamefont {Haah}(2011)}]{Haah2011-ny}%
  \BibitemOpen
  \bibfield  {author} {\bibinfo {author} {\bibfnamefont {J.}~\bibnamefont
  {Haah}},\ }\bibfield  {title} {\bibinfo {title} {Local stabilizer codes in
  three dimensions without string logical operators},\ }\href@noop {}
  {\bibfield  {journal} {\bibinfo  {journal} {Phys. Rev. A}\ }\textbf {\bibinfo
  {volume} {83}},\ \bibinfo {pages} {042330} (\bibinfo {year}
  {2011})}\BibitemShut {NoStop}%
\bibitem [{\citenamefont {Vijay}\ \emph {et~al.}(2016)\citenamefont {Vijay},
  \citenamefont {Haah},\ and\ \citenamefont {Fu}}]{PhysRevB.94.235157}%
  \BibitemOpen
  \bibfield  {author} {\bibinfo {author} {\bibfnamefont {S.}~\bibnamefont
  {Vijay}}, \bibinfo {author} {\bibfnamefont {J.}~\bibnamefont {Haah}},\ and\
  \bibinfo {author} {\bibfnamefont {L.}~\bibnamefont {Fu}},\ }\bibfield
  {title} {\bibinfo {title} {Fracton topological order, generalized lattice
  gauge theory, and duality},\ }\href
  {https://doi.org/10.1103/PhysRevB.94.235157} {\bibfield  {journal} {\bibinfo
  {journal} {Phys. Rev. B}\ }\textbf {\bibinfo {volume} {94}},\ \bibinfo
  {pages} {235157} (\bibinfo {year} {2016})}\BibitemShut {NoStop}%
\bibitem [{\citenamefont {Prem}\ \emph {et~al.}(2019)\citenamefont {Prem},
  \citenamefont {Huang}, \citenamefont {Song},\ and\ \citenamefont
  {Hermele}}]{Prem_2019}%
  \BibitemOpen
  \bibfield  {author} {\bibinfo {author} {\bibfnamefont {A.}~\bibnamefont
  {Prem}}, \bibinfo {author} {\bibfnamefont {S.-J.}\ \bibnamefont {Huang}},
  \bibinfo {author} {\bibfnamefont {H.}~\bibnamefont {Song}},\ and\ \bibinfo
  {author} {\bibfnamefont {M.}~\bibnamefont {Hermele}},\ }\bibfield  {title}
  {\bibinfo {title} {Cage-net fracton models},\ }\bibfield  {journal} {\bibinfo
   {journal} {Physical Review X}\ }\textbf {\bibinfo {volume} {9}},\ \href
  {https://doi.org/10.1103/physrevx.9.021010} {10.1103/physrevx.9.021010}
  (\bibinfo {year} {2019})\BibitemShut {NoStop}%
\bibitem [{\citenamefont {Paramekanti}\ \emph {et~al.}(2002)\citenamefont
  {Paramekanti}, \citenamefont {Balents},\ and\ \citenamefont
  {Fisher}}]{Paramekanti_2002}%
  \BibitemOpen
  \bibfield  {author} {\bibinfo {author} {\bibfnamefont {A.}~\bibnamefont
  {Paramekanti}}, \bibinfo {author} {\bibfnamefont {L.}~\bibnamefont
  {Balents}},\ and\ \bibinfo {author} {\bibfnamefont {M.~P.~A.}\ \bibnamefont
  {Fisher}},\ }\bibfield  {title} {\bibinfo {title} {Ring exchange, the exciton
  bose liquid, and bosonization in two dimensions},\ }\bibfield  {journal}
  {\bibinfo  {journal} {Physical Review B}\ }\textbf {\bibinfo {volume} {66}},\
  \href {https://doi.org/10.1103/physrevb.66.054526}
  {10.1103/physrevb.66.054526} (\bibinfo {year} {2002})\BibitemShut {NoStop}%
\bibitem [{\citenamefont {Prem}\ \emph {et~al.}(2018)\citenamefont {Prem},
  \citenamefont {Pretko},\ and\ \citenamefont {Nandkishore}}]{Prem-Pretko}%
  \BibitemOpen
  \bibfield  {author} {\bibinfo {author} {\bibfnamefont {A.}~\bibnamefont
  {Prem}}, \bibinfo {author} {\bibfnamefont {M.}~\bibnamefont {Pretko}},\ and\
  \bibinfo {author} {\bibfnamefont {R.~M.}\ \bibnamefont {Nandkishore}},\
  }\bibfield  {title} {\bibinfo {title} {Emergent phases of fractonic matter},\
  }\href {https://doi.org/10.1103/PhysRevB.97.085116} {\bibfield  {journal}
  {\bibinfo  {journal} {Phys. Rev. B}\ }\textbf {\bibinfo {volume} {97}},\
  \bibinfo {pages} {085116} (\bibinfo {year} {2018})}\BibitemShut {NoStop}%
\bibitem [{\citenamefont {He}\ \emph {et~al.}(2020)\citenamefont {He},
  \citenamefont {You},\ and\ \citenamefont {Prem}}]{He:2019ktp}%
  \BibitemOpen
  \bibfield  {author} {\bibinfo {author} {\bibfnamefont {H.}~\bibnamefont
  {He}}, \bibinfo {author} {\bibfnamefont {Y.}~\bibnamefont {You}},\ and\
  \bibinfo {author} {\bibfnamefont {A.}~\bibnamefont {Prem}},\ }\bibfield
  {title} {\bibinfo {title} {{Lieb-Schultz-Mattis\textendash{}type constraints
  on fractonic matter}},\ }\href {https://doi.org/10.1103/PhysRevB.101.165145}
  {\bibfield  {journal} {\bibinfo  {journal} {Phys. Rev. B}\ }\textbf {\bibinfo
  {volume} {101}},\ \bibinfo {pages} {165145} (\bibinfo {year} {2020})},\
  \Eprint {https://arxiv.org/abs/1912.10520} {arXiv:1912.10520
  [cond-mat.str-el]} \BibitemShut {NoStop}%
\bibitem [{\citenamefont {Yuan}\ \emph {et~al.}(2020)\citenamefont {Yuan},
  \citenamefont {Chen},\ and\ \citenamefont {Ye}}]{Yuan_2020}%
  \BibitemOpen
  \bibfield  {author} {\bibinfo {author} {\bibfnamefont {J.-K.}\ \bibnamefont
  {Yuan}}, \bibinfo {author} {\bibfnamefont {S.~A.}\ \bibnamefont {Chen}},\
  and\ \bibinfo {author} {\bibfnamefont {P.}~\bibnamefont {Ye}},\ }\bibfield
  {title} {\bibinfo {title} {Fractonic superfluids},\ }\bibfield  {journal}
  {\bibinfo  {journal} {Physical Review Research}\ }\textbf {\bibinfo {volume}
  {2}},\ \href {https://doi.org/10.1103/physrevresearch.2.023267}
  {10.1103/physrevresearch.2.023267} (\bibinfo {year} {2020})\BibitemShut
  {NoStop}%
\bibitem [{\citenamefont {Gorantla}\ \emph
  {et~al.}(2022{\natexlab{a}})\citenamefont {Gorantla}, \citenamefont {Lam},
  \citenamefont {Seiberg},\ and\ \citenamefont {Shao}}]{Gorantla:2022eem}%
  \BibitemOpen
  \bibfield  {author} {\bibinfo {author} {\bibfnamefont {P.}~\bibnamefont
  {Gorantla}}, \bibinfo {author} {\bibfnamefont {H.~T.}\ \bibnamefont {Lam}},
  \bibinfo {author} {\bibfnamefont {N.}~\bibnamefont {Seiberg}},\ and\ \bibinfo
  {author} {\bibfnamefont {S.-H.}\ \bibnamefont {Shao}},\ }\bibfield  {title}
  {\bibinfo {title} {{Global dipole symmetry, compact Lifshitz theory, tensor
  gauge theory, and fractons}},\ }\href
  {https://doi.org/10.1103/PhysRevB.106.045112} {\bibfield  {journal} {\bibinfo
   {journal} {Phys. Rev. B}\ }\textbf {\bibinfo {volume} {106}},\ \bibinfo
  {pages} {045112} (\bibinfo {year} {2022}{\natexlab{a}})},\ \Eprint
  {https://arxiv.org/abs/2201.10589} {arXiv:2201.10589 [cond-mat.str-el]}
  \BibitemShut {NoStop}%
\bibitem [{\citenamefont {Lake}\ \emph
  {et~al.}(2022{\natexlab{a}})\citenamefont {Lake}, \citenamefont {Hermele},\
  and\ \citenamefont {Senthil}}]{lake1}%
  \BibitemOpen
  \bibfield  {author} {\bibinfo {author} {\bibfnamefont {E.}~\bibnamefont
  {Lake}}, \bibinfo {author} {\bibfnamefont {M.}~\bibnamefont {Hermele}},\ and\
  \bibinfo {author} {\bibfnamefont {T.}~\bibnamefont {Senthil}},\ }\bibfield
  {title} {\bibinfo {title} {Dipolar bose-hubbard model},\ }\href
  {https://doi.org/10.1103/PhysRevB.106.064511} {\bibfield  {journal} {\bibinfo
   {journal} {Phys. Rev. B}\ }\textbf {\bibinfo {volume} {106}},\ \bibinfo
  {pages} {064511} (\bibinfo {year} {2022}{\natexlab{a}})}\BibitemShut
  {NoStop}%
\bibitem [{\citenamefont {Jensen}\ and\ \citenamefont
  {Raz}(2022)}]{Jensen:2022iww}%
  \BibitemOpen
  \bibfield  {author} {\bibinfo {author} {\bibfnamefont {K.}~\bibnamefont
  {Jensen}}\ and\ \bibinfo {author} {\bibfnamefont {A.}~\bibnamefont {Raz}},\
  }\bibfield  {title} {\bibinfo {title} {{Large $N$ fractons}},\ }\href@noop {}
  {\  (\bibinfo {year} {2022})},\ \Eprint {https://arxiv.org/abs/2205.01132}
  {arXiv:2205.01132 [hep-th]} \BibitemShut {NoStop}%
\bibitem [{\citenamefont {Lake}\ \emph
  {et~al.}(2022{\natexlab{b}})\citenamefont {Lake}, \citenamefont {Lee},
  \citenamefont {Han},\ and\ \citenamefont {Senthil}}]{lake2}%
  \BibitemOpen
  \bibfield  {author} {\bibinfo {author} {\bibfnamefont {E.}~\bibnamefont
  {Lake}}, \bibinfo {author} {\bibfnamefont {H.-Y.}\ \bibnamefont {Lee}},
  \bibinfo {author} {\bibfnamefont {J.~H.}\ \bibnamefont {Han}},\ and\ \bibinfo
  {author} {\bibfnamefont {T.}~\bibnamefont {Senthil}},\ }\bibfield  {title}
  {\bibinfo {title} {Dipole condensates in tilted bose-hubbard chains},\
  }\href@noop {} {\bibfield  {journal} {\bibinfo  {journal} {arXiv:2210.02470}\
  } (\bibinfo {year} {2022}{\natexlab{b}})}\BibitemShut {NoStop}%
\bibitem [{\citenamefont {Zechmann}\ \emph {et~al.}(2022)\citenamefont
  {Zechmann}, \citenamefont {Altman}, \citenamefont {Knap},\ and\ \citenamefont
  {Feldmeier}}]{feldmeier}%
  \BibitemOpen
  \bibfield  {author} {\bibinfo {author} {\bibfnamefont {P.}~\bibnamefont
  {Zechmann}}, \bibinfo {author} {\bibfnamefont {E.}~\bibnamefont {Altman}},
  \bibinfo {author} {\bibfnamefont {M.}~\bibnamefont {Knap}},\ and\ \bibinfo
  {author} {\bibfnamefont {J.}~\bibnamefont {Feldmeier}},\ }\bibfield  {title}
  {\bibinfo {title} {Fractonic luttinger liquids and supersolids in a
  constrained bose-hubbard model},\ }\href@noop {} {\bibfield  {journal}
  {\bibinfo  {journal} {arXiv:2210.11072}\ } (\bibinfo {year}
  {2022})}\BibitemShut {NoStop}%
\bibitem [{\citenamefont {Lake}\ and\ \citenamefont
  {Senthil}(2023)}]{lake2023non}%
  \BibitemOpen
  \bibfield  {author} {\bibinfo {author} {\bibfnamefont {E.}~\bibnamefont
  {Lake}}\ and\ \bibinfo {author} {\bibfnamefont {T.}~\bibnamefont {Senthil}},\
  }\bibfield  {title} {\bibinfo {title} {Non-fermi liquids from kinetic
  constraints in tilted optical lattices},\ }\href@noop {} {\bibfield
  {journal} {\bibinfo  {journal} {arXiv preprint arXiv:2302.08499}\ } (\bibinfo
  {year} {2023})}\BibitemShut {NoStop}%
\bibitem [{\citenamefont {Chen}\ and\ \citenamefont
  {Ye}(2023)}]{chen2023manybody}%
  \BibitemOpen
  \bibfield  {author} {\bibinfo {author} {\bibfnamefont {S.~A.}\ \bibnamefont
  {Chen}}\ and\ \bibinfo {author} {\bibfnamefont {P.}~\bibnamefont {Ye}},\
  }\href@noop {} {\bibinfo {title} {Many-body physics of spontaneously broken
  higher-rank symmetry: from fractonic superfluids to dipolar hubbard model}}
  (\bibinfo {year} {2023}),\ \Eprint {https://arxiv.org/abs/2305.00941}
  {arXiv:2305.00941 [cond-mat.str-el]} \BibitemShut {NoStop}%
\bibitem [{\citenamefont {Nandkishore}\ and\ \citenamefont
  {Hermele}(2019)}]{Nandkishore_2019}%
  \BibitemOpen
  \bibfield  {author} {\bibinfo {author} {\bibfnamefont {R.~M.}\ \bibnamefont
  {Nandkishore}}\ and\ \bibinfo {author} {\bibfnamefont {M.}~\bibnamefont
  {Hermele}},\ }\bibfield  {title} {\bibinfo {title} {Fractons},\ }\href
  {https://doi.org/10.1146/annurev-conmatphys-031218-013604} {\bibfield
  {journal} {\bibinfo  {journal} {Annual Review of Condensed Matter Physics}\
  }\textbf {\bibinfo {volume} {10}},\ \bibinfo {pages} {295} (\bibinfo {year}
  {2019})}\BibitemShut {NoStop}%
\bibitem [{\citenamefont {Pretko}\ \emph {et~al.}(2020)\citenamefont {Pretko},
  \citenamefont {Chen},\ and\ \citenamefont {You}}]{Pretko_2020}%
  \BibitemOpen
  \bibfield  {author} {\bibinfo {author} {\bibfnamefont {M.}~\bibnamefont
  {Pretko}}, \bibinfo {author} {\bibfnamefont {X.}~\bibnamefont {Chen}},\ and\
  \bibinfo {author} {\bibfnamefont {Y.}~\bibnamefont {You}},\ }\bibfield
  {title} {\bibinfo {title} {Fracton phases of matter},\ }\href
  {https://doi.org/10.1142/s0217751x20300033} {\bibfield  {journal} {\bibinfo
  {journal} {International Journal of Modern Physics A}\ }\textbf {\bibinfo
  {volume} {35}},\ \bibinfo {pages} {2030003} (\bibinfo {year}
  {2020})}\BibitemShut {NoStop}%
\bibitem [{\citenamefont {Sachdev}\ \emph {et~al.}(2002)\citenamefont
  {Sachdev}, \citenamefont {Sengupta},\ and\ \citenamefont
  {Girvin}}]{sachdev02}%
  \BibitemOpen
  \bibfield  {author} {\bibinfo {author} {\bibfnamefont {S.}~\bibnamefont
  {Sachdev}}, \bibinfo {author} {\bibfnamefont {K.}~\bibnamefont {Sengupta}},\
  and\ \bibinfo {author} {\bibfnamefont {S.}~\bibnamefont {Girvin}},\
  }\bibfield  {title} {\bibinfo {title} {Mott insulators in strong electric
  fields},\ }\href@noop {} {\bibfield  {journal} {\bibinfo  {journal} {Physical
  Review B}\ }\textbf {\bibinfo {volume} {66}},\ \bibinfo {pages} {075128}
  (\bibinfo {year} {2002})}\BibitemShut {NoStop}%
\bibitem [{\citenamefont {Pielawa}\ \emph {et~al.}(2011)\citenamefont
  {Pielawa}, \citenamefont {Kitagawa}, \citenamefont {Berg},\ and\
  \citenamefont {Sachdev}}]{pielawa11}%
  \BibitemOpen
  \bibfield  {author} {\bibinfo {author} {\bibfnamefont {S.}~\bibnamefont
  {Pielawa}}, \bibinfo {author} {\bibfnamefont {T.}~\bibnamefont {Kitagawa}},
  \bibinfo {author} {\bibfnamefont {E.}~\bibnamefont {Berg}},\ and\ \bibinfo
  {author} {\bibfnamefont {S.}~\bibnamefont {Sachdev}},\ }\bibfield  {title}
  {\bibinfo {title} {Correlated phases of bosons in tilted frustrated
  lattices},\ }\href@noop {} {\bibfield  {journal} {\bibinfo  {journal}
  {Physical Review B}\ }\textbf {\bibinfo {volume} {83}},\ \bibinfo {pages}
  {205135} (\bibinfo {year} {2011})}\BibitemShut {NoStop}%
\bibitem [{\citenamefont {Guardado-Sanchez}\ \emph {et~al.}(2020)\citenamefont
  {Guardado-Sanchez}, \citenamefont {Morningstar}, \citenamefont {Spar},
  \citenamefont {Brown}, \citenamefont {Huse},\ and\ \citenamefont
  {Bakr}}]{bakr20}%
  \BibitemOpen
  \bibfield  {author} {\bibinfo {author} {\bibfnamefont {E.}~\bibnamefont
  {Guardado-Sanchez}}, \bibinfo {author} {\bibfnamefont {A.}~\bibnamefont
  {Morningstar}}, \bibinfo {author} {\bibfnamefont {B.~M.}\ \bibnamefont
  {Spar}}, \bibinfo {author} {\bibfnamefont {P.~T.}\ \bibnamefont {Brown}},
  \bibinfo {author} {\bibfnamefont {D.~A.}\ \bibnamefont {Huse}},\ and\
  \bibinfo {author} {\bibfnamefont {W.~S.}\ \bibnamefont {Bakr}},\ }\bibfield
  {title} {\bibinfo {title} {Subdiffusion and heat transport in a tilted
  two-dimensional fermi-hubbard system},\ }\href@noop {} {\bibfield  {journal}
  {\bibinfo  {journal} {Physical Review X}\ }\textbf {\bibinfo {volume} {10}},\
  \bibinfo {pages} {011042} (\bibinfo {year} {2020})}\BibitemShut {NoStop}%
\bibitem [{\citenamefont {Scherg}\ \emph {et~al.}(2021)\citenamefont {Scherg},
  \citenamefont {Kohlert}, \citenamefont {Sala}, \citenamefont {Pollmann},
  \citenamefont {Madhusudhana}, \citenamefont {Bloch},\ and\ \citenamefont
  {Aidelsburger}}]{aidelsburger21}%
  \BibitemOpen
  \bibfield  {author} {\bibinfo {author} {\bibfnamefont {S.}~\bibnamefont
  {Scherg}}, \bibinfo {author} {\bibfnamefont {T.}~\bibnamefont {Kohlert}},
  \bibinfo {author} {\bibfnamefont {P.}~\bibnamefont {Sala}}, \bibinfo {author}
  {\bibfnamefont {F.}~\bibnamefont {Pollmann}}, \bibinfo {author}
  {\bibfnamefont {B.~H.}\ \bibnamefont {Madhusudhana}}, \bibinfo {author}
  {\bibfnamefont {I.}~\bibnamefont {Bloch}},\ and\ \bibinfo {author}
  {\bibfnamefont {M.}~\bibnamefont {Aidelsburger}},\ }\bibfield  {title}
  {\bibinfo {title} {Observing non-ergodicity due to kinetic constraints in
  tilted fermi-hubbard chains},\ }\href@noop {} {\bibfield  {journal} {\bibinfo
   {journal} {Nature Communications}\ }\textbf {\bibinfo {volume} {12}},\
  \bibinfo {pages} {1} (\bibinfo {year} {2021})}\BibitemShut {NoStop}%
\bibitem [{\citenamefont {Kohlert}\ \emph {et~al.}(2021)\citenamefont
  {Kohlert}, \citenamefont {Scherg}, \citenamefont {Sala}, \citenamefont
  {Pollmann}, \citenamefont {Madhusudhana}, \citenamefont {Bloch},\ and\
  \citenamefont {Aidelsburger}}]{kohlert2021experimental}%
  \BibitemOpen
  \bibfield  {author} {\bibinfo {author} {\bibfnamefont {T.}~\bibnamefont
  {Kohlert}}, \bibinfo {author} {\bibfnamefont {S.}~\bibnamefont {Scherg}},
  \bibinfo {author} {\bibfnamefont {P.}~\bibnamefont {Sala}}, \bibinfo {author}
  {\bibfnamefont {F.}~\bibnamefont {Pollmann}}, \bibinfo {author}
  {\bibfnamefont {B.~H.}\ \bibnamefont {Madhusudhana}}, \bibinfo {author}
  {\bibfnamefont {I.}~\bibnamefont {Bloch}},\ and\ \bibinfo {author}
  {\bibfnamefont {M.}~\bibnamefont {Aidelsburger}},\ }\bibfield  {title}
  {\bibinfo {title} {Experimental realization of fragmented models in tilted
  fermi-hubbard chains},\ }\href@noop {} {\bibfield  {journal} {\bibinfo
  {journal} {arXiv preprint arXiv:2106.15586}\ } (\bibinfo {year}
  {2021})}\BibitemShut {NoStop}%
\bibitem [{\citenamefont {Zahn}\ \emph {et~al.}(2022)\citenamefont {Zahn},
  \citenamefont {Singh}, \citenamefont {Kosch}, \citenamefont {Asteria},
  \citenamefont {Freystatzky}, \citenamefont {Sengstock}, \citenamefont
  {Mathey},\ and\ \citenamefont {Weitenberg}}]{weitenberg22}%
  \BibitemOpen
  \bibfield  {author} {\bibinfo {author} {\bibfnamefont {H.}~\bibnamefont
  {Zahn}}, \bibinfo {author} {\bibfnamefont {V.}~\bibnamefont {Singh}},
  \bibinfo {author} {\bibfnamefont {M.}~\bibnamefont {Kosch}}, \bibinfo
  {author} {\bibfnamefont {L.}~\bibnamefont {Asteria}}, \bibinfo {author}
  {\bibfnamefont {L.}~\bibnamefont {Freystatzky}}, \bibinfo {author}
  {\bibfnamefont {K.}~\bibnamefont {Sengstock}}, \bibinfo {author}
  {\bibfnamefont {L.}~\bibnamefont {Mathey}},\ and\ \bibinfo {author}
  {\bibfnamefont {C.}~\bibnamefont {Weitenberg}},\ }\bibfield  {title}
  {\bibinfo {title} {Formation of spontaneous density-wave patterns in dc
  driven lattices},\ }\href@noop {} {\bibfield  {journal} {\bibinfo  {journal}
  {Physical Review X}\ }\textbf {\bibinfo {volume} {12}},\ \bibinfo {pages}
  {021014} (\bibinfo {year} {2022})}\BibitemShut {NoStop}%
\bibitem [{\citenamefont {Pretko}(2018)}]{pretko18}%
  \BibitemOpen
  \bibfield  {author} {\bibinfo {author} {\bibfnamefont {M.}~\bibnamefont
  {Pretko}},\ }\bibfield  {title} {\bibinfo {title} {The fracton gauge
  principle},\ }\href@noop {} {\bibfield  {journal} {\bibinfo  {journal}
  {Physical Review B}\ }\textbf {\bibinfo {volume} {98}},\ \bibinfo {pages}
  {115134} (\bibinfo {year} {2018})}\BibitemShut {NoStop}%
\bibitem [{\citenamefont {Gromov}(2019)}]{gromov2019towards}%
  \BibitemOpen
  \bibfield  {author} {\bibinfo {author} {\bibfnamefont {A.}~\bibnamefont
  {Gromov}},\ }\bibfield  {title} {\bibinfo {title} {Towards classification of
  fracton phases: the multipole algebra},\ }\href@noop {} {\bibfield  {journal}
  {\bibinfo  {journal} {Physical Review X}\ }\textbf {\bibinfo {volume} {9}},\
  \bibinfo {pages} {031035} (\bibinfo {year} {2019})}\BibitemShut {NoStop}%
\bibitem [{\citenamefont {Seiberg}(2020)}]{Seiberg:2019vrp}%
  \BibitemOpen
  \bibfield  {author} {\bibinfo {author} {\bibfnamefont {N.}~\bibnamefont
  {Seiberg}},\ }\bibfield  {title} {\bibinfo {title} {{Field Theories With a
  Vector Global Symmetry}},\ }\href
  {https://doi.org/10.21468/SciPostPhys.8.4.050} {\bibfield  {journal}
  {\bibinfo  {journal} {SciPost Phys.}\ }\textbf {\bibinfo {volume} {8}},\
  \bibinfo {pages} {050} (\bibinfo {year} {2020})},\ \Eprint
  {https://arxiv.org/abs/1909.10544} {arXiv:1909.10544 [cond-mat.str-el]}
  \BibitemShut {NoStop}%
\bibitem [{\citenamefont {Dubinkin}\ \emph {et~al.}(2021)\citenamefont
  {Dubinkin}, \citenamefont {May-Mann},\ and\ \citenamefont
  {Hughes}}]{Dubinkin_2021}%
  \BibitemOpen
  \bibfield  {author} {\bibinfo {author} {\bibfnamefont {O.}~\bibnamefont
  {Dubinkin}}, \bibinfo {author} {\bibfnamefont {J.}~\bibnamefont {May-Mann}},\
  and\ \bibinfo {author} {\bibfnamefont {T.~L.}\ \bibnamefont {Hughes}},\
  }\bibfield  {title} {\bibinfo {title} {Lieb-schultz-mattis-type theorems and
  other nonperturbative results for strongly correlated systems with conserved
  dipole moments},\ }\bibfield  {journal} {\bibinfo  {journal} {Physical Review
  B}\ }\textbf {\bibinfo {volume} {103}},\ \href
  {https://doi.org/10.1103/physrevb.103.125133} {10.1103/physrevb.103.125133}
  (\bibinfo {year} {2021})\BibitemShut {NoStop}%
\bibitem [{\citenamefont {Stahl}\ \emph {et~al.}(2022)\citenamefont {Stahl},
  \citenamefont {Lake},\ and\ \citenamefont {Nandkishore}}]{Stahl_2022}%
  \BibitemOpen
  \bibfield  {author} {\bibinfo {author} {\bibfnamefont {C.}~\bibnamefont
  {Stahl}}, \bibinfo {author} {\bibfnamefont {E.}~\bibnamefont {Lake}},\ and\
  \bibinfo {author} {\bibfnamefont {R.}~\bibnamefont {Nandkishore}},\
  }\bibfield  {title} {\bibinfo {title} {Spontaneous breaking of multipole
  symmetries},\ }\bibfield  {journal} {\bibinfo  {journal} {Physical Review B}\
  }\textbf {\bibinfo {volume} {105}},\ \href
  {https://doi.org/10.1103/physrevb.105.155107} {10.1103/physrevb.105.155107}
  (\bibinfo {year} {2022})\BibitemShut {NoStop}%
\bibitem [{\citenamefont {Bulmash}\ \emph {et~al.}(2023)\citenamefont
  {Bulmash}, \citenamefont {Hart},\ and\ \citenamefont
  {Nandkishore}}]{bulmash2023multipole}%
  \BibitemOpen
  \bibfield  {author} {\bibinfo {author} {\bibfnamefont {D.}~\bibnamefont
  {Bulmash}}, \bibinfo {author} {\bibfnamefont {O.}~\bibnamefont {Hart}},\ and\
  \bibinfo {author} {\bibfnamefont {R.}~\bibnamefont {Nandkishore}},\
  }\bibfield  {title} {\bibinfo {title} {Multipole groups and fracton phenomena
  on arbitrary crystalline lattices},\ }\href@noop {} {\bibfield  {journal}
  {\bibinfo  {journal} {arXiv preprint arXiv:2301.10782}\ } (\bibinfo {year}
  {2023})}\BibitemShut {NoStop}%
\bibitem [{\citenamefont {Burnell}\ \emph {et~al.}(2023)\citenamefont
  {Burnell}, \citenamefont {Moudgalya},\ and\ \citenamefont
  {Prem}}]{Burnell:2023fsr}%
  \BibitemOpen
  \bibfield  {author} {\bibinfo {author} {\bibfnamefont {F.~J.}\ \bibnamefont
  {Burnell}}, \bibinfo {author} {\bibfnamefont {S.}~\bibnamefont {Moudgalya}},\
  and\ \bibinfo {author} {\bibfnamefont {A.}~\bibnamefont {Prem}},\ }\bibfield
  {title} {\bibinfo {title} {{Filling constraints on translation invariant
  dipole conserving systems}},\ }\href@noop {} {\  (\bibinfo {year} {2023})},\
  \Eprint {https://arxiv.org/abs/2308.16241} {arXiv:2308.16241
  [cond-mat.str-el]} \BibitemShut {NoStop}%
\bibitem [{\citenamefont {Hu}\ and\ \citenamefont
  {Watanabe}(2023)}]{watanabe23}%
  \BibitemOpen
  \bibfield  {author} {\bibinfo {author} {\bibfnamefont {Y.}~\bibnamefont
  {Hu}}\ and\ \bibinfo {author} {\bibfnamefont {H.}~\bibnamefont {Watanabe}},\
  }\bibfield  {title} {\bibinfo {title} {Spontaneous symmetry breaking without
  ground state degeneracy in generalized n-state clock model},\ }\href@noop {}
  {\bibfield  {journal} {\bibinfo  {journal} {arXiv:2302.01207}\ } (\bibinfo
  {year} {2023})}\BibitemShut {NoStop}%
\bibitem [{\citenamefont {Watanabe}\ \emph {et~al.}(2022)\citenamefont
  {Watanabe}, \citenamefont {Cheng},\ and\ \citenamefont
  {Fuji}}]{watanabe2022ground}%
  \BibitemOpen
  \bibfield  {author} {\bibinfo {author} {\bibfnamefont {H.}~\bibnamefont
  {Watanabe}}, \bibinfo {author} {\bibfnamefont {M.}~\bibnamefont {Cheng}},\
  and\ \bibinfo {author} {\bibfnamefont {Y.}~\bibnamefont {Fuji}},\ }\bibfield
  {title} {\bibinfo {title} {Ground state degeneracy on torus in a family of
  $z_n$ toric code},\ }\href@noop {} {\bibfield  {journal} {\bibinfo  {journal}
  {arXiv preprint arXiv:2211.00299}\ } (\bibinfo {year} {2022})}\BibitemShut
  {NoStop}%
\bibitem [{\citenamefont {Delfino}\ \emph {et~al.}(2023)\citenamefont
  {Delfino}, \citenamefont {Chamon},\ and\ \citenamefont
  {You}}]{delfino20232d}%
  \BibitemOpen
  \bibfield  {author} {\bibinfo {author} {\bibfnamefont {G.}~\bibnamefont
  {Delfino}}, \bibinfo {author} {\bibfnamefont {C.}~\bibnamefont {Chamon}},\
  and\ \bibinfo {author} {\bibfnamefont {Y.}~\bibnamefont {You}},\ }\href@noop
  {} {\bibinfo {title} {2d fractons from gauging exponential symmetries}}
  (\bibinfo {year} {2023}),\ \Eprint {https://arxiv.org/abs/2306.17121}
  {arXiv:2306.17121 [cond-mat.str-el]} \BibitemShut {NoStop}%
\bibitem [{\citenamefont {Batista}\ and\ \citenamefont
  {Nussinov}(2005)}]{PhysRevB.72.045137}%
  \BibitemOpen
  \bibfield  {author} {\bibinfo {author} {\bibfnamefont {C.~D.}\ \bibnamefont
  {Batista}}\ and\ \bibinfo {author} {\bibfnamefont {Z.}~\bibnamefont
  {Nussinov}},\ }\bibfield  {title} {\bibinfo {title} {Generalized elitzur's
  theorem and dimensional reductions},\ }\href
  {https://doi.org/10.1103/PhysRevB.72.045137} {\bibfield  {journal} {\bibinfo
  {journal} {Phys. Rev. B}\ }\textbf {\bibinfo {volume} {72}},\ \bibinfo
  {pages} {045137} (\bibinfo {year} {2005})}\BibitemShut {NoStop}%
\bibitem [{\citenamefont {Nussinov}\ and\ \citenamefont
  {Ortiz}(2009)}]{Nussinov_2009}%
  \BibitemOpen
  \bibfield  {author} {\bibinfo {author} {\bibfnamefont {Z.}~\bibnamefont
  {Nussinov}}\ and\ \bibinfo {author} {\bibfnamefont {G.}~\bibnamefont
  {Ortiz}},\ }\bibfield  {title} {\bibinfo {title} {A symmetry principle for
  topological quantum order},\ }\href
  {https://doi.org/10.1016/j.aop.2008.11.002} {\bibfield  {journal} {\bibinfo
  {journal} {Annals of Physics}\ }\textbf {\bibinfo {volume} {324}},\ \bibinfo
  {pages} {977} (\bibinfo {year} {2009})}\BibitemShut {NoStop}%
\bibitem [{\citenamefont {You}\ \emph {et~al.}(2018)\citenamefont {You},
  \citenamefont {Devakul}, \citenamefont {Burnell},\ and\ \citenamefont
  {Sondhi}}]{you2018subsystem}%
  \BibitemOpen
  \bibfield  {author} {\bibinfo {author} {\bibfnamefont {Y.}~\bibnamefont
  {You}}, \bibinfo {author} {\bibfnamefont {T.}~\bibnamefont {Devakul}},
  \bibinfo {author} {\bibfnamefont {F.~J.}\ \bibnamefont {Burnell}},\ and\
  \bibinfo {author} {\bibfnamefont {S.~L.}\ \bibnamefont {Sondhi}},\ }\bibfield
   {title} {\bibinfo {title} {Subsystem symmetry protected topological order},\
  }\href@noop {} {\bibfield  {journal} {\bibinfo  {journal} {Physical Review
  B}\ }\textbf {\bibinfo {volume} {98}},\ \bibinfo {pages} {035112} (\bibinfo
  {year} {2018})}\BibitemShut {NoStop}%
\bibitem [{\citenamefont {Seiberg}\ and\ \citenamefont
  {Shao}(2021{\natexlab{a}})}]{paper1}%
  \BibitemOpen
  \bibfield  {author} {\bibinfo {author} {\bibfnamefont {N.}~\bibnamefont
  {Seiberg}}\ and\ \bibinfo {author} {\bibfnamefont {S.-H.}\ \bibnamefont
  {Shao}},\ }\bibfield  {title} {\bibinfo {title} {{Exotic Symmetries, Duality,
  and Fractons in 2+1-Dimensional Quantum Field Theory}},\ }\href
  {https://doi.org/10.21468/SciPostPhys.10.2.027} {\bibfield  {journal}
  {\bibinfo  {journal} {SciPost Phys.}\ }\textbf {\bibinfo {volume} {10}},\
  \bibinfo {pages} {027} (\bibinfo {year} {2021}{\natexlab{a}})},\ \Eprint
  {https://arxiv.org/abs/2003.10466} {arXiv:2003.10466 [cond-mat.str-el]}
  \BibitemShut {NoStop}%
\bibitem [{\citenamefont {Seiberg}\ and\ \citenamefont {Shao}(2020)}]{paper2}%
  \BibitemOpen
  \bibfield  {author} {\bibinfo {author} {\bibfnamefont {N.}~\bibnamefont
  {Seiberg}}\ and\ \bibinfo {author} {\bibfnamefont {S.-H.}\ \bibnamefont
  {Shao}},\ }\bibfield  {title} {\bibinfo {title} {{Exotic $U(1)$ Symmetries,
  Duality, and Fractons in 3+1-Dimensional Quantum Field Theory}},\ }\href
  {https://doi.org/10.21468/SciPostPhys.9.4.046} {\bibfield  {journal}
  {\bibinfo  {journal} {SciPost Phys.}\ }\textbf {\bibinfo {volume} {9}},\
  \bibinfo {pages} {046} (\bibinfo {year} {2020})},\ \Eprint
  {https://arxiv.org/abs/2004.00015} {arXiv:2004.00015 [cond-mat.str-el]}
  \BibitemShut {NoStop}%
\bibitem [{\citenamefont {Seiberg}\ and\ \citenamefont
  {Shao}(2021{\natexlab{b}})}]{paper3}%
  \BibitemOpen
  \bibfield  {author} {\bibinfo {author} {\bibfnamefont {N.}~\bibnamefont
  {Seiberg}}\ and\ \bibinfo {author} {\bibfnamefont {S.-H.}\ \bibnamefont
  {Shao}},\ }\bibfield  {title} {\bibinfo {title} {{Exotic $\mathbb{Z}_N$
  Symmetries, Duality, and Fractons in 3+1-Dimensional Quantum Field Theory}},\
  }\href {https://doi.org/10.21468/SciPostPhys.10.1.003} {\bibfield  {journal}
  {\bibinfo  {journal} {SciPost Phys.}\ }\textbf {\bibinfo {volume} {10}},\
  \bibinfo {pages} {003} (\bibinfo {year} {2021}{\natexlab{b}})},\ \Eprint
  {https://arxiv.org/abs/2004.06115} {arXiv:2004.06115 [cond-mat.str-el]}
  \BibitemShut {NoStop}%
\bibitem [{\citenamefont {Gorantla}\ \emph {et~al.}(2020)\citenamefont
  {Gorantla}, \citenamefont {Lam}, \citenamefont {Seiberg},\ and\ \citenamefont
  {Shao}}]{Gorantla:2020xap}%
  \BibitemOpen
  \bibfield  {author} {\bibinfo {author} {\bibfnamefont {P.}~\bibnamefont
  {Gorantla}}, \bibinfo {author} {\bibfnamefont {H.~T.}\ \bibnamefont {Lam}},
  \bibinfo {author} {\bibfnamefont {N.}~\bibnamefont {Seiberg}},\ and\ \bibinfo
  {author} {\bibfnamefont {S.-H.}\ \bibnamefont {Shao}},\ }\bibfield  {title}
  {\bibinfo {title} {{More Exotic Field Theories in 3+1 Dimensions}},\ }\href
  {https://doi.org/10.21468/SciPostPhys.9.5.073} {\bibfield  {journal}
  {\bibinfo  {journal} {SciPost Phys.}\ }\textbf {\bibinfo {volume} {9}},\
  \bibinfo {pages} {073} (\bibinfo {year} {2020})},\ \Eprint
  {https://arxiv.org/abs/2007.04904} {arXiv:2007.04904 [cond-mat.str-el]}
  \BibitemShut {NoStop}%
\bibitem [{\citenamefont {Distler}\ \emph {et~al.}(2022)\citenamefont
  {Distler}, \citenamefont {Karch},\ and\ \citenamefont
  {Raz}}]{Distler:2021qzc}%
  \BibitemOpen
  \bibfield  {author} {\bibinfo {author} {\bibfnamefont {J.}~\bibnamefont
  {Distler}}, \bibinfo {author} {\bibfnamefont {A.}~\bibnamefont {Karch}},\
  and\ \bibinfo {author} {\bibfnamefont {A.}~\bibnamefont {Raz}},\ }\bibfield
  {title} {\bibinfo {title} {{Spontaneously broken subsystem symmetries}},\
  }\href {https://doi.org/10.1007/JHEP03(2022)016} {\bibfield  {journal}
  {\bibinfo  {journal} {JHEP}\ }\textbf {\bibinfo {volume} {03}},\ \bibinfo
  {pages} {016}},\ \Eprint {https://arxiv.org/abs/2110.12611} {arXiv:2110.12611
  [hep-th]} \BibitemShut {NoStop}%
\bibitem [{\citenamefont {Burnell}\ \emph {et~al.}(2022)\citenamefont
  {Burnell}, \citenamefont {Devakul}, \citenamefont {Gorantla}, \citenamefont
  {Lam},\ and\ \citenamefont {Shao}}]{Burnell:2021reh}%
  \BibitemOpen
  \bibfield  {author} {\bibinfo {author} {\bibfnamefont {F.~J.}\ \bibnamefont
  {Burnell}}, \bibinfo {author} {\bibfnamefont {T.}~\bibnamefont {Devakul}},
  \bibinfo {author} {\bibfnamefont {P.}~\bibnamefont {Gorantla}}, \bibinfo
  {author} {\bibfnamefont {H.~T.}\ \bibnamefont {Lam}},\ and\ \bibinfo {author}
  {\bibfnamefont {S.-H.}\ \bibnamefont {Shao}},\ }\bibfield  {title} {\bibinfo
  {title} {{Anomaly inflow for subsystem symmetries}},\ }\href
  {https://doi.org/10.1103/PhysRevB.106.085113} {\bibfield  {journal} {\bibinfo
   {journal} {Phys. Rev. B}\ }\textbf {\bibinfo {volume} {106}},\ \bibinfo
  {pages} {085113} (\bibinfo {year} {2022})},\ \Eprint
  {https://arxiv.org/abs/2110.09529} {arXiv:2110.09529 [cond-mat.str-el]}
  \BibitemShut {NoStop}%
\bibitem [{\citenamefont {Rayhaun}\ and\ \citenamefont
  {Williamson}(2021)}]{Rayhaun:2021ocs}%
  \BibitemOpen
  \bibfield  {author} {\bibinfo {author} {\bibfnamefont {B.~C.}\ \bibnamefont
  {Rayhaun}}\ and\ \bibinfo {author} {\bibfnamefont {D.~J.}\ \bibnamefont
  {Williamson}},\ }\bibfield  {title} {\bibinfo {title} {{Higher-Form Subsystem
  Symmetry Breaking: Subdimensional Criticality and Fracton Phase
  Transitions}},\ }\href@noop {} {\  (\bibinfo {year} {2021})},\ \Eprint
  {https://arxiv.org/abs/2112.12735} {arXiv:2112.12735 [cond-mat.str-el]}
  \BibitemShut {NoStop}%
\bibitem [{\citenamefont {Newman}\ and\ \citenamefont
  {Moore}(1999)}]{PhysRevE.60.5068}%
  \BibitemOpen
  \bibfield  {author} {\bibinfo {author} {\bibfnamefont {M.~E.~J.}\
  \bibnamefont {Newman}}\ and\ \bibinfo {author} {\bibfnamefont
  {C.}~\bibnamefont {Moore}},\ }\bibfield  {title} {\bibinfo {title} {Glassy
  dynamics and aging in an exactly solvable spin model},\ }\href
  {https://doi.org/10.1103/PhysRevE.60.5068} {\bibfield  {journal} {\bibinfo
  {journal} {Phys. Rev. E}\ }\textbf {\bibinfo {volume} {60}},\ \bibinfo
  {pages} {5068} (\bibinfo {year} {1999})}\BibitemShut {NoStop}%
\bibitem [{\citenamefont {Castelnovo}\ and\ \citenamefont
  {Chamon}(2012)}]{Castelnovo_2012}%
  \BibitemOpen
  \bibfield  {author} {\bibinfo {author} {\bibfnamefont {C.}~\bibnamefont
  {Castelnovo}}\ and\ \bibinfo {author} {\bibfnamefont {C.}~\bibnamefont
  {Chamon}},\ }\bibfield  {title} {\bibinfo {title} {Topological quantum
  glassiness},\ }\href {https://doi.org/10.1080/14786435.2011.609152}
  {\bibfield  {journal} {\bibinfo  {journal} {Philosophical Magazine}\ }\textbf
  {\bibinfo {volume} {92}},\ \bibinfo {pages} {304} (\bibinfo {year}
  {2012})}\BibitemShut {NoStop}%
\bibitem [{\citenamefont {Yoshida}(2013)}]{PhysRevB.88.125122}%
  \BibitemOpen
  \bibfield  {author} {\bibinfo {author} {\bibfnamefont {B.}~\bibnamefont
  {Yoshida}},\ }\bibfield  {title} {\bibinfo {title} {Exotic topological order
  in fractal spin liquids},\ }\href
  {https://doi.org/10.1103/PhysRevB.88.125122} {\bibfield  {journal} {\bibinfo
  {journal} {Phys. Rev. B}\ }\textbf {\bibinfo {volume} {88}},\ \bibinfo
  {pages} {125122} (\bibinfo {year} {2013})}\BibitemShut {NoStop}%
\bibitem [{\citenamefont
  {Williamson}(2016{\natexlab{a}})}]{PhysRevB.94.155128}%
  \BibitemOpen
  \bibfield  {author} {\bibinfo {author} {\bibfnamefont {D.~J.}\ \bibnamefont
  {Williamson}},\ }\bibfield  {title} {\bibinfo {title} {Fractal symmetries:
  Ungauging the cubic code},\ }\href
  {https://doi.org/10.1103/PhysRevB.94.155128} {\bibfield  {journal} {\bibinfo
  {journal} {Phys. Rev. B}\ }\textbf {\bibinfo {volume} {94}},\ \bibinfo
  {pages} {155128} (\bibinfo {year} {2016}{\natexlab{a}})}\BibitemShut
  {NoStop}%
\bibitem [{\citenamefont {Devakul}\ \emph {et~al.}(2019)\citenamefont
  {Devakul}, \citenamefont {You}, \citenamefont {Burnell},\ and\ \citenamefont
  {Sondhi}}]{Devakul_2019}%
  \BibitemOpen
  \bibfield  {author} {\bibinfo {author} {\bibfnamefont {T.}~\bibnamefont
  {Devakul}}, \bibinfo {author} {\bibfnamefont {Y.}~\bibnamefont {You}},
  \bibinfo {author} {\bibfnamefont {F.~J.}\ \bibnamefont {Burnell}},\ and\
  \bibinfo {author} {\bibfnamefont {S.}~\bibnamefont {Sondhi}},\ }\bibfield
  {title} {\bibinfo {title} {Fractal symmetric phases of matter},\ }\bibfield
  {journal} {\bibinfo  {journal} {{SciPost} Physics}\ }\textbf {\bibinfo
  {volume} {6}},\ \href {https://doi.org/10.21468/scipostphys.6.1.007}
  {10.21468/scipostphys.6.1.007} (\bibinfo {year} {2019})\BibitemShut {NoStop}%
\bibitem [{\citenamefont {Sauerwein}\ \emph {et~al.}(2019)\citenamefont
  {Sauerwein}, \citenamefont {Molnar}, \citenamefont {Cirac},\ and\
  \citenamefont {Kraus}}]{Lcyclesym}%
  \BibitemOpen
  \bibfield  {author} {\bibinfo {author} {\bibfnamefont {D.}~\bibnamefont
  {Sauerwein}}, \bibinfo {author} {\bibfnamefont {A.}~\bibnamefont {Molnar}},
  \bibinfo {author} {\bibfnamefont {J.~I.}\ \bibnamefont {Cirac}},\ and\
  \bibinfo {author} {\bibfnamefont {B.}~\bibnamefont {Kraus}},\ }\bibfield
  {title} {\bibinfo {title} {Matrix product states: Entanglement, symmetries,
  and state transformations},\ }\href
  {https://doi.org/10.1103/PhysRevLett.123.170504} {\bibfield  {journal}
  {\bibinfo  {journal} {Phys. Rev. Lett.}\ }\textbf {\bibinfo {volume} {123}},\
  \bibinfo {pages} {170504} (\bibinfo {year} {2019})}\BibitemShut {NoStop}%
\bibitem [{\citenamefont {Stephen}\ \emph
  {et~al.}(2019{\natexlab{a}})\citenamefont {Stephen}, \citenamefont {Nautrup},
  \citenamefont {Bermejo-Vega}, \citenamefont {Eisert},\ and\ \citenamefont
  {Raussendorf}}]{Stephen_2019}%
  \BibitemOpen
  \bibfield  {author} {\bibinfo {author} {\bibfnamefont {D.~T.}\ \bibnamefont
  {Stephen}}, \bibinfo {author} {\bibfnamefont {H.~P.}\ \bibnamefont
  {Nautrup}}, \bibinfo {author} {\bibfnamefont {J.}~\bibnamefont
  {Bermejo-Vega}}, \bibinfo {author} {\bibfnamefont {J.}~\bibnamefont
  {Eisert}},\ and\ \bibinfo {author} {\bibfnamefont {R.}~\bibnamefont
  {Raussendorf}},\ }\bibfield  {title} {\bibinfo {title} {Subsystem symmetries,
  quantum cellular automata, and computational phases of quantum matter},\
  }\href {https://doi.org/10.22331/q-2019-05-20-142} {\bibfield  {journal}
  {\bibinfo  {journal} {Quantum}\ }\textbf {\bibinfo {volume} {3}},\ \bibinfo
  {pages} {142} (\bibinfo {year} {2019}{\natexlab{a}})}\BibitemShut {NoStop}%
\bibitem [{\citenamefont {Stephen}\ \emph {et~al.}(2022)\citenamefont
  {Stephen}, \citenamefont {Ho}, \citenamefont {Wei}, \citenamefont
  {Raussendorf},\ and\ \citenamefont {Verresen}}]{Stephen:2022zyt}%
  \BibitemOpen
  \bibfield  {author} {\bibinfo {author} {\bibfnamefont {D.~T.}\ \bibnamefont
  {Stephen}}, \bibinfo {author} {\bibfnamefont {W.~W.}\ \bibnamefont {Ho}},
  \bibinfo {author} {\bibfnamefont {T.-C.}\ \bibnamefont {Wei}}, \bibinfo
  {author} {\bibfnamefont {R.}~\bibnamefont {Raussendorf}},\ and\ \bibinfo
  {author} {\bibfnamefont {R.}~\bibnamefont {Verresen}},\ }\bibfield  {title}
  {\bibinfo {title} {{Universal measurement-based quantum computation in a
  one-dimensional architecture enabled by dual-unitary circuits}},\ }\href@noop
  {} {\  (\bibinfo {year} {2022})},\ \Eprint {https://arxiv.org/abs/2209.06191}
  {arXiv:2209.06191 [quant-ph]} \BibitemShut {NoStop}%
\bibitem [{\citenamefont {Williamson}(2016{\natexlab{b}})}]{Dominic-gauging}%
  \BibitemOpen
  \bibfield  {author} {\bibinfo {author} {\bibfnamefont {D.~J.}\ \bibnamefont
  {Williamson}},\ }\bibfield  {title} {\bibinfo {title} {Fractal symmetries:
  Ungauging the cubic code},\ }\href
  {https://doi.org/10.1103/PhysRevB.94.155128} {\bibfield  {journal} {\bibinfo
  {journal} {Phys. Rev. B}\ }\textbf {\bibinfo {volume} {94}},\ \bibinfo
  {pages} {155128} (\bibinfo {year} {2016}{\natexlab{b}})}\BibitemShut
  {NoStop}%
\bibitem [{\citenamefont {Shirley}\ \emph {et~al.}(2019)\citenamefont
  {Shirley}, \citenamefont {Slagle},\ and\ \citenamefont
  {Chen}}]{Shirley_2019}%
  \BibitemOpen
  \bibfield  {author} {\bibinfo {author} {\bibfnamefont {W.}~\bibnamefont
  {Shirley}}, \bibinfo {author} {\bibfnamefont {K.}~\bibnamefont {Slagle}},\
  and\ \bibinfo {author} {\bibfnamefont {X.}~\bibnamefont {Chen}},\ }\bibfield
  {title} {\bibinfo {title} {Foliated fracton order from gauging subsystem
  symmetries},\ }\bibfield  {journal} {\bibinfo  {journal} {{SciPost} Physics}\
  }\textbf {\bibinfo {volume} {6}},\ \href
  {https://doi.org/10.21468/scipostphys.6.4.041} {10.21468/scipostphys.6.4.041}
  (\bibinfo {year} {2019})\BibitemShut {NoStop}%
\bibitem [{\citenamefont {Dijkgraaf}\ and\ \citenamefont
  {Witten}(1990)}]{Dijkgraaf:1989pz}%
  \BibitemOpen
  \bibfield  {author} {\bibinfo {author} {\bibfnamefont {R.}~\bibnamefont
  {Dijkgraaf}}\ and\ \bibinfo {author} {\bibfnamefont {E.}~\bibnamefont
  {Witten}},\ }\bibfield  {title} {\bibinfo {title} {{Topological Gauge
  Theories and Group Cohomology}},\ }\href {https://doi.org/10.1007/BF02096988}
  {\bibfield  {journal} {\bibinfo  {journal} {Commun. Math. Phys.}\ }\textbf
  {\bibinfo {volume} {129}},\ \bibinfo {pages} {393} (\bibinfo {year}
  {1990})}\BibitemShut {NoStop}%
\bibitem [{\citenamefont {You}\ \emph {et~al.}(2020{\natexlab{a}})\citenamefont
  {You}, \citenamefont {Devakul}, \citenamefont {Burnell},\ and\ \citenamefont
  {Sondhi}}]{You:2018zhj}%
  \BibitemOpen
  \bibfield  {author} {\bibinfo {author} {\bibfnamefont {Y.}~\bibnamefont
  {You}}, \bibinfo {author} {\bibfnamefont {T.}~\bibnamefont {Devakul}},
  \bibinfo {author} {\bibfnamefont {F.~J.}\ \bibnamefont {Burnell}},\ and\
  \bibinfo {author} {\bibfnamefont {S.~L.}\ \bibnamefont {Sondhi}},\ }\bibfield
   {title} {\bibinfo {title} {{Symmetric Fracton Matter: Twisted and
  Enriched}},\ }\href {https://doi.org/10.1016/j.aop.2020.168140} {\bibfield
  {journal} {\bibinfo  {journal} {Annals Phys.}\ }\textbf {\bibinfo {volume}
  {416}},\ \bibinfo {pages} {168140} (\bibinfo {year} {2020}{\natexlab{a}})},\
  \Eprint {https://arxiv.org/abs/1805.09800} {arXiv:1805.09800
  [cond-mat.str-el]} \BibitemShut {NoStop}%
\bibitem [{\citenamefont {Shirley}\ \emph {et~al.}(2020)\citenamefont
  {Shirley}, \citenamefont {Slagle},\ and\ \citenamefont
  {Chen}}]{Shirley_2020_twisted}%
  \BibitemOpen
  \bibfield  {author} {\bibinfo {author} {\bibfnamefont {W.}~\bibnamefont
  {Shirley}}, \bibinfo {author} {\bibfnamefont {K.}~\bibnamefont {Slagle}},\
  and\ \bibinfo {author} {\bibfnamefont {X.}~\bibnamefont {Chen}},\ }\bibfield
  {title} {\bibinfo {title} {Twisted foliated fracton phases},\ }\bibfield
  {journal} {\bibinfo  {journal} {Physical Review B}\ }\textbf {\bibinfo
  {volume} {102}},\ \href {https://doi.org/10.1103/physrevb.102.115103}
  {10.1103/physrevb.102.115103} (\bibinfo {year} {2020})\BibitemShut {NoStop}%
\bibitem [{\citenamefont {Devakul}\ \emph {et~al.}(2018)\citenamefont
  {Devakul}, \citenamefont {Williamson},\ and\ \citenamefont
  {You}}]{devakul2018classification}%
  \BibitemOpen
  \bibfield  {author} {\bibinfo {author} {\bibfnamefont {T.}~\bibnamefont
  {Devakul}}, \bibinfo {author} {\bibfnamefont {D.~J.}\ \bibnamefont
  {Williamson}},\ and\ \bibinfo {author} {\bibfnamefont {Y.}~\bibnamefont
  {You}},\ }\bibfield  {title} {\bibinfo {title} {Classification of subsystem
  symmetry-protected topological phases},\ }\href@noop {} {\bibfield  {journal}
  {\bibinfo  {journal} {Physical Review B}\ }\textbf {\bibinfo {volume} {98}},\
  \bibinfo {pages} {235121} (\bibinfo {year} {2018})}\BibitemShut {NoStop}%
\bibitem [{\citenamefont {Stephen}\ \emph
  {et~al.}(2019{\natexlab{b}})\citenamefont {Stephen}, \citenamefont {Dreyer},
  \citenamefont {Iqbal},\ and\ \citenamefont {Schuch}}]{Stephen_2019_SSPT}%
  \BibitemOpen
  \bibfield  {author} {\bibinfo {author} {\bibfnamefont {D.~T.}\ \bibnamefont
  {Stephen}}, \bibinfo {author} {\bibfnamefont {H.}~\bibnamefont {Dreyer}},
  \bibinfo {author} {\bibfnamefont {M.}~\bibnamefont {Iqbal}},\ and\ \bibinfo
  {author} {\bibfnamefont {N.}~\bibnamefont {Schuch}},\ }\bibfield  {title}
  {\bibinfo {title} {Detecting subsystem symmetry protected topological order
  via entanglement entropy},\ }\bibfield  {journal} {\bibinfo  {journal}
  {Physical Review B}\ }\textbf {\bibinfo {volume} {100}},\ \href
  {https://doi.org/10.1103/physrevb.100.115112} {10.1103/physrevb.100.115112}
  (\bibinfo {year} {2019}{\natexlab{b}})\BibitemShut {NoStop}%
\bibitem [{\citenamefont {Devakul}\ \emph {et~al.}(2020)\citenamefont
  {Devakul}, \citenamefont {Shirley},\ and\ \citenamefont
  {Wang}}]{Devakul_2020}%
  \BibitemOpen
  \bibfield  {author} {\bibinfo {author} {\bibfnamefont {T.}~\bibnamefont
  {Devakul}}, \bibinfo {author} {\bibfnamefont {W.}~\bibnamefont {Shirley}},\
  and\ \bibinfo {author} {\bibfnamefont {J.}~\bibnamefont {Wang}},\ }\bibfield
  {title} {\bibinfo {title} {Strong planar subsystem symmetry-protected
  topological phases and their dual fracton orders},\ }\bibfield  {journal}
  {\bibinfo  {journal} {Physical Review Research}\ }\textbf {\bibinfo {volume}
  {2}},\ \href {https://doi.org/10.1103/physrevresearch.2.012059}
  {10.1103/physrevresearch.2.012059} (\bibinfo {year} {2020})\BibitemShut
  {NoStop}%
\bibitem [{\citenamefont {Hsin}\ and\ \citenamefont
  {Slagle}(2021)}]{Hsin:2021mjn}%
  \BibitemOpen
  \bibfield  {author} {\bibinfo {author} {\bibfnamefont {P.-S.}\ \bibnamefont
  {Hsin}}\ and\ \bibinfo {author} {\bibfnamefont {K.}~\bibnamefont {Slagle}},\
  }\bibfield  {title} {\bibinfo {title} {{Comments on foliated gauge theories
  and dualities in 3+1d}},\ }\href
  {https://doi.org/10.21468/SciPostPhys.11.2.032} {\bibfield  {journal}
  {\bibinfo  {journal} {SciPost Phys.}\ }\textbf {\bibinfo {volume} {11}},\
  \bibinfo {pages} {032} (\bibinfo {year} {2021})},\ \Eprint
  {https://arxiv.org/abs/2105.09363} {arXiv:2105.09363 [cond-mat.str-el]}
  \BibitemShut {NoStop}%
\bibitem [{\citenamefont {Devakul}(2019)}]{devakul2019classifying}%
  \BibitemOpen
  \bibfield  {author} {\bibinfo {author} {\bibfnamefont {T.}~\bibnamefont
  {Devakul}},\ }\bibfield  {title} {\bibinfo {title} {Classifying local fractal
  subsystem symmetry-protected topological phases},\ }\href@noop {} {\bibfield
  {journal} {\bibinfo  {journal} {Physical Review B}\ }\textbf {\bibinfo
  {volume} {99}},\ \bibinfo {pages} {235131} (\bibinfo {year}
  {2019})}\BibitemShut {NoStop}%
\bibitem [{\citenamefont {Han}\ \emph {et~al.}(2023{\natexlab{b}})\citenamefont
  {Han}, \citenamefont {Lake}, \citenamefont {Lam}, \citenamefont {Verresen},\
  and\ \citenamefont {You}}]{Han:2023fas}%
  \BibitemOpen
  \bibfield  {author} {\bibinfo {author} {\bibfnamefont {J.~H.}\ \bibnamefont
  {Han}}, \bibinfo {author} {\bibfnamefont {E.}~\bibnamefont {Lake}}, \bibinfo
  {author} {\bibfnamefont {H.~T.}\ \bibnamefont {Lam}}, \bibinfo {author}
  {\bibfnamefont {R.}~\bibnamefont {Verresen}},\ and\ \bibinfo {author}
  {\bibfnamefont {Y.}~\bibnamefont {You}},\ }\bibfield  {title} {\bibinfo
  {title} {{Topological quantum chains protected by dipolar and other modulated
  symmetries}},\ }\href@noop {} {\  (\bibinfo {year} {2023}{\natexlab{b}})},\
  \Eprint {https://arxiv.org/abs/2309.10036} {arXiv:2309.10036
  [cond-mat.str-el]} \BibitemShut {NoStop}%
\bibitem [{\citenamefont {Sulejmanpasic}\ and\ \citenamefont
  {Gattringer}(2019)}]{Sulejmanpasic:2019ytl}%
  \BibitemOpen
  \bibfield  {author} {\bibinfo {author} {\bibfnamefont {T.}~\bibnamefont
  {Sulejmanpasic}}\ and\ \bibinfo {author} {\bibfnamefont {C.}~\bibnamefont
  {Gattringer}},\ }\bibfield  {title} {\bibinfo {title} {{Abelian gauge
  theories on the lattice: $\theta$-Terms and compact gauge theory with(out)
  monopoles}},\ }\href {https://doi.org/10.1016/j.nuclphysb.2019.114616}
  {\bibfield  {journal} {\bibinfo  {journal} {Nucl. Phys. B}\ }\textbf
  {\bibinfo {volume} {943}},\ \bibinfo {pages} {114616} (\bibinfo {year}
  {2019})},\ \Eprint {https://arxiv.org/abs/1901.02637} {arXiv:1901.02637
  [hep-lat]} \BibitemShut {NoStop}%
\bibitem [{\citenamefont {Gorantla}\ \emph
  {et~al.}(2021{\natexlab{a}})\citenamefont {Gorantla}, \citenamefont {Lam},
  \citenamefont {Seiberg},\ and\ \citenamefont {Shao}}]{Gorantla:2021svj}%
  \BibitemOpen
  \bibfield  {author} {\bibinfo {author} {\bibfnamefont {P.}~\bibnamefont
  {Gorantla}}, \bibinfo {author} {\bibfnamefont {H.~T.}\ \bibnamefont {Lam}},
  \bibinfo {author} {\bibfnamefont {N.}~\bibnamefont {Seiberg}},\ and\ \bibinfo
  {author} {\bibfnamefont {S.-H.}\ \bibnamefont {Shao}},\ }\bibfield  {title}
  {\bibinfo {title} {{A modified Villain formulation of fractons and other
  exotic theories}},\ }\href {https://doi.org/10.1063/5.0060808} {\bibfield
  {journal} {\bibinfo  {journal} {J. Math. Phys.}\ }\textbf {\bibinfo {volume}
  {62}},\ \bibinfo {pages} {102301} (\bibinfo {year} {2021}{\natexlab{a}})},\
  \Eprint {https://arxiv.org/abs/2103.01257} {arXiv:2103.01257
  [cond-mat.str-el]} \BibitemShut {NoStop}%
\bibitem [{\citenamefont {Fazza}\ and\ \citenamefont
  {Sulejmanpasic}(2023)}]{Fazza:2022fss}%
  \BibitemOpen
  \bibfield  {author} {\bibinfo {author} {\bibfnamefont {L.}~\bibnamefont
  {Fazza}}\ and\ \bibinfo {author} {\bibfnamefont {T.}~\bibnamefont
  {Sulejmanpasic}},\ }\bibfield  {title} {\bibinfo {title} {{Lattice quantum
  Villain Hamiltonians: compact scalars, U(1) gauge theories, fracton models
  and quantum Ising model dualities}},\ }\href
  {https://doi.org/10.1007/JHEP05(2023)017} {\bibfield  {journal} {\bibinfo
  {journal} {JHEP}\ }\textbf {\bibinfo {volume} {05}},\ \bibinfo {pages}
  {017}},\ \Eprint {https://arxiv.org/abs/2211.13047} {arXiv:2211.13047
  [hep-th]} \BibitemShut {NoStop}%
\bibitem [{\citenamefont {Fannes}\ \emph {et~al.}(1992)\citenamefont {Fannes},
  \citenamefont {Nachtergaele},\ and\ \citenamefont {Werner}}]{Fannes1992}%
  \BibitemOpen
  \bibfield  {author} {\bibinfo {author} {\bibfnamefont {M.}~\bibnamefont
  {Fannes}}, \bibinfo {author} {\bibfnamefont {B.}~\bibnamefont
  {Nachtergaele}},\ and\ \bibinfo {author} {\bibfnamefont {R.~F.}\ \bibnamefont
  {Werner}},\ }\bibfield  {title} {\bibinfo {title} {Finitely correlated states
  on quantum spin chains},\ }\href {https://doi.org/10.1007/BF02099178}
  {\bibfield  {journal} {\bibinfo  {journal} {Communications in Mathematical
  Physics}\ }\textbf {\bibinfo {volume} {144}},\ \bibinfo {pages} {443}
  (\bibinfo {year} {1992})}\BibitemShut {NoStop}%
\bibitem [{\citenamefont {Bridgeman}\ and\ \citenamefont
  {Chubb}(2017)}]{bridgeman17}%
  \BibitemOpen
  \bibfield  {author} {\bibinfo {author} {\bibfnamefont {J.~C.}\ \bibnamefont
  {Bridgeman}}\ and\ \bibinfo {author} {\bibfnamefont {C.~T.}\ \bibnamefont
  {Chubb}},\ }\bibfield  {title} {\bibinfo {title} {Hand-waving and
  interpretive dance: an introductory course on tensor networks},\ }\href
  {https://doi.org/10.1088/1751-8121/aa6dc3} {\bibfield  {journal} {\bibinfo
  {journal} {Journal of Physics A: Mathematical and Theoretical}\ }\textbf
  {\bibinfo {volume} {50}},\ \bibinfo {pages} {223001} (\bibinfo {year}
  {2017})}\BibitemShut {NoStop}%
\bibitem [{\citenamefont {Cirac}\ \emph {et~al.}(2021)\citenamefont {Cirac},
  \citenamefont {P\'erez-Garc\'{\i}a}, \citenamefont {Schuch},\ and\
  \citenamefont {Verstraete}}]{Cirac21}%
  \BibitemOpen
  \bibfield  {author} {\bibinfo {author} {\bibfnamefont {J.~I.}\ \bibnamefont
  {Cirac}}, \bibinfo {author} {\bibfnamefont {D.}~\bibnamefont
  {P\'erez-Garc\'{\i}a}}, \bibinfo {author} {\bibfnamefont {N.}~\bibnamefont
  {Schuch}},\ and\ \bibinfo {author} {\bibfnamefont {F.}~\bibnamefont
  {Verstraete}},\ }\bibfield  {title} {\bibinfo {title} {Matrix product states
  and projected entangled pair states: Concepts, symmetries, theorems},\ }\href
  {https://doi.org/10.1103/RevModPhys.93.045003} {\bibfield  {journal}
  {\bibinfo  {journal} {Rev. Mod. Phys.}\ }\textbf {\bibinfo {volume} {93}},\
  \bibinfo {pages} {045003} (\bibinfo {year} {2021})}\BibitemShut {NoStop}%
\bibitem [{\citenamefont {P{\'{e} }rez-Garc{\'{\i}}a}\ \emph
  {et~al.}(2008)\citenamefont {P{\'{e} }rez-Garc{\'{\i}}a}, \citenamefont
  {Wolf}, \citenamefont {Sanz}, \citenamefont {Verstraete},\ and\ \citenamefont
  {Cirac}}]{P_rez_Garc_a_2008}%
  \BibitemOpen
  \bibfield  {author} {\bibinfo {author} {\bibfnamefont {D.}~\bibnamefont
  {P{\'{e} }rez-Garc{\'{\i}}a}}, \bibinfo {author} {\bibfnamefont {M.~M.}\
  \bibnamefont {Wolf}}, \bibinfo {author} {\bibfnamefont {M.}~\bibnamefont
  {Sanz}}, \bibinfo {author} {\bibfnamefont {F.}~\bibnamefont {Verstraete}},\
  and\ \bibinfo {author} {\bibfnamefont {J.~I.}\ \bibnamefont {Cirac}},\
  }\bibfield  {title} {\bibinfo {title} {String order and symmetries in quantum
  spin lattices},\ }\bibfield  {journal} {\bibinfo  {journal} {Physical Review
  Letters}\ }\textbf {\bibinfo {volume} {100}},\ \href
  {https://doi.org/10.1103/physrevlett.100.167202}
  {10.1103/physrevlett.100.167202} (\bibinfo {year} {2008})\BibitemShut
  {NoStop}%
\bibitem [{\citenamefont {Xu}(2006{\natexlab{a}})}]{xu2006novel}%
  \BibitemOpen
  \bibfield  {author} {\bibinfo {author} {\bibfnamefont {C.}~\bibnamefont
  {Xu}},\ }\href@noop {} {\bibinfo {title} {Novel algebraic boson liquid phase
  with soft graviton excitations}} (\bibinfo {year} {2006}{\natexlab{a}}),\
  \Eprint {https://arxiv.org/abs/cond-mat/0602443} {arXiv:cond-mat/0602443
  [cond-mat.str-el]} \BibitemShut {NoStop}%
\bibitem [{\citenamefont {Gu}\ and\ \citenamefont {Wen}(2006)}]{Gu:2006vw}%
  \BibitemOpen
  \bibfield  {author} {\bibinfo {author} {\bibfnamefont {Z.-C.}\ \bibnamefont
  {Gu}}\ and\ \bibinfo {author} {\bibfnamefont {X.-G.}\ \bibnamefont {Wen}},\
  }\bibfield  {title} {\bibinfo {title} {{A Lattice bosonic model as a quantum
  theory of gravity}},\ }\href@noop {} {\  (\bibinfo {year} {2006})},\ \Eprint
  {https://arxiv.org/abs/gr-qc/0606100} {arXiv:gr-qc/0606100} \BibitemShut
  {NoStop}%
\bibitem [{\citenamefont {Xu}(2006{\natexlab{b}})}]{Xu_2006}%
  \BibitemOpen
  \bibfield  {author} {\bibinfo {author} {\bibfnamefont {C.}~\bibnamefont
  {Xu}},\ }\bibfield  {title} {\bibinfo {title} {Gapless bosonic excitation
  without symmetry breaking: An algebraic spin liquid with soft gravitons},\
  }\bibfield  {journal} {\bibinfo  {journal} {Physical Review B}\ }\textbf
  {\bibinfo {volume} {74}},\ \href {https://doi.org/10.1103/physrevb.74.224433}
  {10.1103/physrevb.74.224433} (\bibinfo {year}
  {2006}{\natexlab{b}})\BibitemShut {NoStop}%
\bibitem [{\citenamefont {Xu}\ and\ \citenamefont {Wu}(2008)}]{Xu_2008}%
  \BibitemOpen
  \bibfield  {author} {\bibinfo {author} {\bibfnamefont {C.}~\bibnamefont
  {Xu}}\ and\ \bibinfo {author} {\bibfnamefont {C.}~\bibnamefont {Wu}},\
  }\bibfield  {title} {\bibinfo {title} {Resonating plaquette phases in {SU}(4)
  heisenberg antiferromagnet},\ }\bibfield  {journal} {\bibinfo  {journal}
  {Physical Review B}\ }\textbf {\bibinfo {volume} {77}},\ \href
  {https://doi.org/10.1103/physrevb.77.134449} {10.1103/physrevb.77.134449}
  (\bibinfo {year} {2008})\BibitemShut {NoStop}%
\bibitem [{\citenamefont {Gu}\ and\ \citenamefont {Wen}(2012)}]{Gu:2009jh}%
  \BibitemOpen
  \bibfield  {author} {\bibinfo {author} {\bibfnamefont {Z.-C.}\ \bibnamefont
  {Gu}}\ and\ \bibinfo {author} {\bibfnamefont {X.-G.}\ \bibnamefont {Wen}},\
  }\bibfield  {title} {\bibinfo {title} {{Emergence of helicity +- 2 modes
  (gravitons) from qbit models}},\ }\href
  {https://doi.org/10.1016/j.nuclphysb.2012.05.010} {\bibfield  {journal}
  {\bibinfo  {journal} {Nucl. Phys. B}\ }\textbf {\bibinfo {volume} {863}},\
  \bibinfo {pages} {90} (\bibinfo {year} {2012})},\ \Eprint
  {https://arxiv.org/abs/0907.1203} {arXiv:0907.1203 [gr-qc]} \BibitemShut
  {NoStop}%
\bibitem [{\citenamefont {Rasmussen}\ \emph {et~al.}(2016)\citenamefont
  {Rasmussen}, \citenamefont {You},\ and\ \citenamefont
  {Xu}}]{rasmussen2016stable}%
  \BibitemOpen
  \bibfield  {author} {\bibinfo {author} {\bibfnamefont {A.}~\bibnamefont
  {Rasmussen}}, \bibinfo {author} {\bibfnamefont {Y.-Z.}\ \bibnamefont {You}},\
  and\ \bibinfo {author} {\bibfnamefont {C.}~\bibnamefont {Xu}},\ }\href@noop
  {} {\bibinfo {title} {Stable gapless bose liquid phases without any
  symmetry}} (\bibinfo {year} {2016}),\ \Eprint
  {https://arxiv.org/abs/1601.08235} {arXiv:1601.08235 [cond-mat.str-el]}
  \BibitemShut {NoStop}%
\bibitem [{\citenamefont {Pretko}(2017{\natexlab{a}})}]{pretko17}%
  \BibitemOpen
  \bibfield  {author} {\bibinfo {author} {\bibfnamefont {M.}~\bibnamefont
  {Pretko}},\ }\bibfield  {title} {\bibinfo {title} {Subdimensional particle
  structure of higher rank u (1) spin liquids},\ }\href@noop {} {\bibfield
  {journal} {\bibinfo  {journal} {Physical Review B}\ }\textbf {\bibinfo
  {volume} {95}},\ \bibinfo {pages} {115139} (\bibinfo {year}
  {2017}{\natexlab{a}})}\BibitemShut {NoStop}%
\bibitem [{\citenamefont
  {Pretko}(2017{\natexlab{b}})}]{Pretko_2017_generalizeEM}%
  \BibitemOpen
  \bibfield  {author} {\bibinfo {author} {\bibfnamefont {M.}~\bibnamefont
  {Pretko}},\ }\bibfield  {title} {\bibinfo {title} {Generalized
  electromagnetism of subdimensional particles: A spin liquid story},\
  }\bibfield  {journal} {\bibinfo  {journal} {Physical Review B}\ }\textbf
  {\bibinfo {volume} {96}},\ \href {https://doi.org/10.1103/physrevb.96.035119}
  {10.1103/physrevb.96.035119} (\bibinfo {year}
  {2017}{\natexlab{b}})\BibitemShut {NoStop}%
\bibitem [{\citenamefont {Slagle}\ and\ \citenamefont
  {Kim}(2017)}]{Slagle:2017wrc}%
  \BibitemOpen
  \bibfield  {author} {\bibinfo {author} {\bibfnamefont {K.}~\bibnamefont
  {Slagle}}\ and\ \bibinfo {author} {\bibfnamefont {Y.~B.}\ \bibnamefont
  {Kim}},\ }\bibfield  {title} {\bibinfo {title} {{Quantum Field Theory of
  X-Cube Fracton Topological Order and Robust Degeneracy from Geometry}},\
  }\href {https://doi.org/10.1103/PhysRevB.96.195139} {\bibfield  {journal}
  {\bibinfo  {journal} {Phys. Rev. B}\ }\textbf {\bibinfo {volume} {96}},\
  \bibinfo {pages} {195139} (\bibinfo {year} {2017})},\ \Eprint
  {https://arxiv.org/abs/1708.04619} {arXiv:1708.04619 [cond-mat.str-el]}
  \BibitemShut {NoStop}%
\bibitem [{\citenamefont {Pretko}\ and\ \citenamefont
  {Radzihovsky}(2017)}]{Pretko2017-ej}%
  \BibitemOpen
  \bibfield  {author} {\bibinfo {author} {\bibfnamefont {M.}~\bibnamefont
  {Pretko}}\ and\ \bibinfo {author} {\bibfnamefont {L.}~\bibnamefont
  {Radzihovsky}},\ }\bibfield  {title} {\bibinfo {title} {{Fracton-Elasticity}
  duality},\ }\href@noop {} {\  (\bibinfo {year} {2017})},\ \Eprint
  {https://arxiv.org/abs/1711.11044} {arXiv:1711.11044 [cond-mat.str-el]}
  \BibitemShut {NoStop}%
\bibitem [{\citenamefont {Ma}\ \emph {et~al.}(2018)\citenamefont {Ma},
  \citenamefont {Hermele},\ and\ \citenamefont {Chen}}]{Ma:2018nhd}%
  \BibitemOpen
  \bibfield  {author} {\bibinfo {author} {\bibfnamefont {H.}~\bibnamefont
  {Ma}}, \bibinfo {author} {\bibfnamefont {M.}~\bibnamefont {Hermele}},\ and\
  \bibinfo {author} {\bibfnamefont {X.}~\bibnamefont {Chen}},\ }\bibfield
  {title} {\bibinfo {title} {{Fracton topological order from the Higgs and
  partial-confinement mechanisms of rank-two gauge theory}},\ }\href
  {https://doi.org/10.1103/PhysRevB.98.035111} {\bibfield  {journal} {\bibinfo
  {journal} {Phys. Rev. B}\ }\textbf {\bibinfo {volume} {98}},\ \bibinfo
  {pages} {035111} (\bibinfo {year} {2018})},\ \Eprint
  {https://arxiv.org/abs/1802.10108} {arXiv:1802.10108 [cond-mat.str-el]}
  \BibitemShut {NoStop}%
\bibitem [{\citenamefont {Bulmash}\ and\ \citenamefont
  {Barkeshli}(2018{\natexlab{a}})}]{Bulmash:2018lid}%
  \BibitemOpen
  \bibfield  {author} {\bibinfo {author} {\bibfnamefont {D.}~\bibnamefont
  {Bulmash}}\ and\ \bibinfo {author} {\bibfnamefont {M.}~\bibnamefont
  {Barkeshli}},\ }\bibfield  {title} {\bibinfo {title} {{The Higgs Mechanism in
  Higher-Rank Symmetric $U(1)$ Gauge Theories}},\ }\href
  {https://doi.org/10.1103/PhysRevB.97.235112} {\bibfield  {journal} {\bibinfo
  {journal} {Phys. Rev. B}\ }\textbf {\bibinfo {volume} {97}},\ \bibinfo
  {pages} {235112} (\bibinfo {year} {2018}{\natexlab{a}})},\ \Eprint
  {https://arxiv.org/abs/1802.10099} {arXiv:1802.10099 [cond-mat.str-el]}
  \BibitemShut {NoStop}%
\bibitem [{\citenamefont {Bulmash}\ and\ \citenamefont
  {Barkeshli}(2018{\natexlab{b}})}]{Bulmash:2018knk}%
  \BibitemOpen
  \bibfield  {author} {\bibinfo {author} {\bibfnamefont {D.}~\bibnamefont
  {Bulmash}}\ and\ \bibinfo {author} {\bibfnamefont {M.}~\bibnamefont
  {Barkeshli}},\ }\bibfield  {title} {\bibinfo {title} {{Generalized $U(1)$
  Gauge Field Theories and Fractal Dynamics}},\ }\href@noop {} {\  (\bibinfo
  {year} {2018}{\natexlab{b}})},\ \Eprint {https://arxiv.org/abs/1806.01855}
  {arXiv:1806.01855 [cond-mat.str-el]} \BibitemShut {NoStop}%
\bibitem [{\citenamefont {Slagle}\ \emph {et~al.}(2019)\citenamefont {Slagle},
  \citenamefont {Prem},\ and\ \citenamefont {Pretko}}]{Slagle:2018kqf}%
  \BibitemOpen
  \bibfield  {author} {\bibinfo {author} {\bibfnamefont {K.}~\bibnamefont
  {Slagle}}, \bibinfo {author} {\bibfnamefont {A.}~\bibnamefont {Prem}},\ and\
  \bibinfo {author} {\bibfnamefont {M.}~\bibnamefont {Pretko}},\ }\bibfield
  {title} {\bibinfo {title} {{Symmetric Tensor Gauge Theories on Curved
  Spaces}},\ }\href {https://doi.org/10.1016/j.aop.2019.167910} {\bibfield
  {journal} {\bibinfo  {journal} {Annals Phys.}\ }\textbf {\bibinfo {volume}
  {410}},\ \bibinfo {pages} {167910} (\bibinfo {year} {2019})},\ \Eprint
  {https://arxiv.org/abs/1807.00827} {arXiv:1807.00827 [cond-mat.str-el]}
  \BibitemShut {NoStop}%
\bibitem [{\citenamefont {You}\ \emph {et~al.}(2021)\citenamefont {You},
  \citenamefont {Burnell},\ and\ \citenamefont {Hughes}}]{You:2019bvu}%
  \BibitemOpen
  \bibfield  {author} {\bibinfo {author} {\bibfnamefont {Y.}~\bibnamefont
  {You}}, \bibinfo {author} {\bibfnamefont {F.~J.}\ \bibnamefont {Burnell}},\
  and\ \bibinfo {author} {\bibfnamefont {T.~L.}\ \bibnamefont {Hughes}},\
  }\bibfield  {title} {\bibinfo {title} {{Multipolar Topological Field
  Theories: Bridging Higher Order Topological Insulators and Fractons}},\
  }\href {https://doi.org/10.1103/PhysRevB.103.245128} {\bibfield  {journal}
  {\bibinfo  {journal} {Phys. Rev. B}\ }\textbf {\bibinfo {volume} {103}},\
  \bibinfo {pages} {245128} (\bibinfo {year} {2021})},\ \Eprint
  {https://arxiv.org/abs/1909.05868} {arXiv:1909.05868 [cond-mat.str-el]}
  \BibitemShut {NoStop}%
\bibitem [{\citenamefont {You}\ \emph {et~al.}(2020{\natexlab{b}})\citenamefont
  {You}, \citenamefont {Devakul}, \citenamefont {Sondhi},\ and\ \citenamefont
  {Burnell}}]{fractonCS}%
  \BibitemOpen
  \bibfield  {author} {\bibinfo {author} {\bibfnamefont {Y.}~\bibnamefont
  {You}}, \bibinfo {author} {\bibfnamefont {T.}~\bibnamefont {Devakul}},
  \bibinfo {author} {\bibfnamefont {S.~L.}\ \bibnamefont {Sondhi}},\ and\
  \bibinfo {author} {\bibfnamefont {F.~J.}\ \bibnamefont {Burnell}},\
  }\bibfield  {title} {\bibinfo {title} {Fractonic chern-simons and {BF}
  theories},\ }\bibfield  {journal} {\bibinfo  {journal} {Physical Review
  Research}\ }\textbf {\bibinfo {volume} {2}},\ \href
  {https://doi.org/10.1103/physrevresearch.2.023249}
  {10.1103/physrevresearch.2.023249} (\bibinfo {year}
  {2020}{\natexlab{b}})\BibitemShut {NoStop}%
\bibitem [{\citenamefont {Wang}\ and\ \citenamefont {Xu}(2021)}]{Wang:2019aiq}%
  \BibitemOpen
  \bibfield  {author} {\bibinfo {author} {\bibfnamefont {J.}~\bibnamefont
  {Wang}}\ and\ \bibinfo {author} {\bibfnamefont {K.}~\bibnamefont {Xu}},\
  }\bibfield  {title} {\bibinfo {title} {{Higher-Rank Tensor Field Theory of
  Non-Abelian Fracton and Embeddon}},\ }\href
  {https://doi.org/10.1016/j.aop.2020.168370} {\bibfield  {journal} {\bibinfo
  {journal} {Annals Phys.}\ }\textbf {\bibinfo {volume} {424}},\ \bibinfo
  {pages} {168370} (\bibinfo {year} {2021})},\ \Eprint
  {https://arxiv.org/abs/1909.13879} {arXiv:1909.13879 [hep-th]} \BibitemShut
  {NoStop}%
\bibitem [{\citenamefont {Wang}\ \emph {et~al.}(2021)\citenamefont {Wang},
  \citenamefont {Xu},\ and\ \citenamefont {Yau}}]{Wang:2019cbj}%
  \BibitemOpen
  \bibfield  {author} {\bibinfo {author} {\bibfnamefont {J.}~\bibnamefont
  {Wang}}, \bibinfo {author} {\bibfnamefont {K.}~\bibnamefont {Xu}},\ and\
  \bibinfo {author} {\bibfnamefont {S.-T.}\ \bibnamefont {Yau}},\ }\bibfield
  {title} {\bibinfo {title} {{Higher-rank tensor non-Abelian field theory:
  Higher-moment or subdimensional polynomial global symmetry, algebraic
  variety, Noether's theorem, and gauging}},\ }\href
  {https://doi.org/10.1103/PhysRevResearch.3.013185} {\bibfield  {journal}
  {\bibinfo  {journal} {Phys. Rev. Res.}\ }\textbf {\bibinfo {volume} {3}},\
  \bibinfo {pages} {013185} (\bibinfo {year} {2021})},\ \Eprint
  {https://arxiv.org/abs/1911.01804} {arXiv:1911.01804 [hep-th]} \BibitemShut
  {NoStop}%
\bibitem [{\citenamefont {Nguyen}\ \emph {et~al.}(2020)\citenamefont {Nguyen},
  \citenamefont {Gromov},\ and\ \citenamefont {Moroz}}]{Nguyen:2020yve}%
  \BibitemOpen
  \bibfield  {author} {\bibinfo {author} {\bibfnamefont {D.~X.}\ \bibnamefont
  {Nguyen}}, \bibinfo {author} {\bibfnamefont {A.}~\bibnamefont {Gromov}},\
  and\ \bibinfo {author} {\bibfnamefont {S.}~\bibnamefont {Moroz}},\ }\bibfield
   {title} {\bibinfo {title} {{Fracton-elasticity duality of two-dimensional
  superfluid vortex crystals: defect interactions and quantum melting}},\
  }\href {https://doi.org/10.21468/SciPostPhys.9.5.076} {\bibfield  {journal}
  {\bibinfo  {journal} {SciPost Phys.}\ }\textbf {\bibinfo {volume} {9}},\
  \bibinfo {pages} {076} (\bibinfo {year} {2020})},\ \Eprint
  {https://arxiv.org/abs/2005.12317} {arXiv:2005.12317 [cond-mat.quant-gas]}
  \BibitemShut {NoStop}%
\bibitem [{\citenamefont {Dubinkin}\ \emph {et~al.}(2020)\citenamefont
  {Dubinkin}, \citenamefont {Rasmussen},\ and\ \citenamefont
  {Hughes}}]{Dubinkin:2020kxo}%
  \BibitemOpen
  \bibfield  {author} {\bibinfo {author} {\bibfnamefont {O.}~\bibnamefont
  {Dubinkin}}, \bibinfo {author} {\bibfnamefont {A.}~\bibnamefont
  {Rasmussen}},\ and\ \bibinfo {author} {\bibfnamefont {T.~L.}\ \bibnamefont
  {Hughes}},\ }\bibfield  {title} {\bibinfo {title} {{Higher-form Gauge
  Symmetries in Multipole Topological Phases}},\ }\href
  {https://doi.org/10.1016/j.aop.2020.168297} {\bibfield  {journal} {\bibinfo
  {journal} {Annals Phys.}\ }\textbf {\bibinfo {volume} {422}},\ \bibinfo
  {pages} {168297} (\bibinfo {year} {2020})},\ \Eprint
  {https://arxiv.org/abs/2007.05539} {arXiv:2007.05539 [cond-mat.str-el]}
  \BibitemShut {NoStop}%
\bibitem [{\citenamefont {Qi}\ \emph {et~al.}(2021)\citenamefont {Qi},
  \citenamefont {Radzihovsky},\ and\ \citenamefont {Hermele}}]{Qi_2021}%
  \BibitemOpen
  \bibfield  {author} {\bibinfo {author} {\bibfnamefont {M.}~\bibnamefont
  {Qi}}, \bibinfo {author} {\bibfnamefont {L.}~\bibnamefont {Radzihovsky}},\
  and\ \bibinfo {author} {\bibfnamefont {M.}~\bibnamefont {Hermele}},\
  }\bibfield  {title} {\bibinfo {title} {Fracton phases via exotic higher-form
  symmetry-breaking},\ }\href {https://doi.org/10.1016/j.aop.2020.168360}
  {\bibfield  {journal} {\bibinfo  {journal} {Annals of Physics}\ }\textbf
  {\bibinfo {volume} {424}},\ \bibinfo {pages} {168360} (\bibinfo {year}
  {2021})}\BibitemShut {NoStop}%
\bibitem [{\citenamefont {Gorantla}\ \emph
  {et~al.}(2021{\natexlab{b}})\citenamefont {Gorantla}, \citenamefont {Lam},
  \citenamefont {Seiberg},\ and\ \citenamefont {Shao}}]{Gorantla:2020jpy}%
  \BibitemOpen
  \bibfield  {author} {\bibinfo {author} {\bibfnamefont {P.}~\bibnamefont
  {Gorantla}}, \bibinfo {author} {\bibfnamefont {H.~T.}\ \bibnamefont {Lam}},
  \bibinfo {author} {\bibfnamefont {N.}~\bibnamefont {Seiberg}},\ and\ \bibinfo
  {author} {\bibfnamefont {S.-H.}\ \bibnamefont {Shao}},\ }\bibfield  {title}
  {\bibinfo {title} {{fcc lattice, checkerboards, fractons, and quantum field
  theory}},\ }\href {https://doi.org/10.1103/PhysRevB.103.205116} {\bibfield
  {journal} {\bibinfo  {journal} {Phys. Rev. B}\ }\textbf {\bibinfo {volume}
  {103}},\ \bibinfo {pages} {205116} (\bibinfo {year} {2021}{\natexlab{b}})},\
  \Eprint {https://arxiv.org/abs/2010.16414} {arXiv:2010.16414
  [cond-mat.str-el]} \BibitemShut {NoStop}%
\bibitem [{\citenamefont {Rudelius}\ \emph {et~al.}(2021)\citenamefont
  {Rudelius}, \citenamefont {Seiberg},\ and\ \citenamefont
  {Shao}}]{Rudelius:2020kta}%
  \BibitemOpen
  \bibfield  {author} {\bibinfo {author} {\bibfnamefont {T.}~\bibnamefont
  {Rudelius}}, \bibinfo {author} {\bibfnamefont {N.}~\bibnamefont {Seiberg}},\
  and\ \bibinfo {author} {\bibfnamefont {S.-H.}\ \bibnamefont {Shao}},\
  }\bibfield  {title} {\bibinfo {title} {{Fractons with Twisted Boundary
  Conditions and Their Symmetries}},\ }\href
  {https://doi.org/10.1103/PhysRevB.103.195113} {\bibfield  {journal} {\bibinfo
   {journal} {Phys. Rev. B}\ }\textbf {\bibinfo {volume} {103}},\ \bibinfo
  {pages} {195113} (\bibinfo {year} {2021})},\ \Eprint
  {https://arxiv.org/abs/2012.11592} {arXiv:2012.11592 [cond-mat.str-el]}
  \BibitemShut {NoStop}%
\bibitem [{\citenamefont {Du}\ \emph {et~al.}(2022)\citenamefont {Du},
  \citenamefont {Mehta}, \citenamefont {Nguyen},\ and\ \citenamefont
  {Son}}]{Du:2021pbc}%
  \BibitemOpen
  \bibfield  {author} {\bibinfo {author} {\bibfnamefont {Y.-H.}\ \bibnamefont
  {Du}}, \bibinfo {author} {\bibfnamefont {U.}~\bibnamefont {Mehta}}, \bibinfo
  {author} {\bibfnamefont {D.~X.}\ \bibnamefont {Nguyen}},\ and\ \bibinfo
  {author} {\bibfnamefont {D.~T.}\ \bibnamefont {Son}},\ }\bibfield  {title}
  {\bibinfo {title} {{Volume-preserving diffeomorphism as nonabelian
  higher-rank gauge symmetry}},\ }\href
  {https://doi.org/10.21468/SciPostPhys.12.2.050} {\bibfield  {journal}
  {\bibinfo  {journal} {SciPost Phys.}\ }\textbf {\bibinfo {volume} {12}},\
  \bibinfo {pages} {050} (\bibinfo {year} {2022})},\ \Eprint
  {https://arxiv.org/abs/2103.09826} {arXiv:2103.09826 [cond-mat.str-el]}
  \BibitemShut {NoStop}%
\bibitem [{\citenamefont {Oh}\ \emph {et~al.}(2022{\natexlab{a}})\citenamefont
  {Oh}, \citenamefont {Kim}, \citenamefont {Moon},\ and\ \citenamefont
  {Han}}]{Oh_2022_rank2toric}%
  \BibitemOpen
  \bibfield  {author} {\bibinfo {author} {\bibfnamefont {Y.-T.}\ \bibnamefont
  {Oh}}, \bibinfo {author} {\bibfnamefont {J.}~\bibnamefont {Kim}}, \bibinfo
  {author} {\bibfnamefont {E.-G.}\ \bibnamefont {Moon}},\ and\ \bibinfo
  {author} {\bibfnamefont {J.~H.}\ \bibnamefont {Han}},\ }\bibfield  {title}
  {\bibinfo {title} {Rank-2 toric code in two dimensions},\ }\bibfield
  {journal} {\bibinfo  {journal} {Physical Review B}\ }\textbf {\bibinfo
  {volume} {105}},\ \href {https://doi.org/10.1103/physrevb.105.045128}
  {10.1103/physrevb.105.045128} (\bibinfo {year}
  {2022}{\natexlab{a}})\BibitemShut {NoStop}%
\bibitem [{\citenamefont {Oh}\ \emph {et~al.}(2022{\natexlab{b}})\citenamefont
  {Oh}, \citenamefont {Kim},\ and\ \citenamefont
  {Han}}]{Oh_2022_dipolarfieldtheory}%
  \BibitemOpen
  \bibfield  {author} {\bibinfo {author} {\bibfnamefont {Y.-T.}\ \bibnamefont
  {Oh}}, \bibinfo {author} {\bibfnamefont {J.}~\bibnamefont {Kim}},\ and\
  \bibinfo {author} {\bibfnamefont {J.~H.}\ \bibnamefont {Han}},\ }\bibfield
  {title} {\bibinfo {title} {Effective field theory of dipolar braiding
  statistics in two dimensions},\ }\bibfield  {journal} {\bibinfo  {journal}
  {Physical Review B}\ }\textbf {\bibinfo {volume} {106}},\ \href
  {https://doi.org/10.1103/physrevb.106.155150} {10.1103/physrevb.106.155150}
  (\bibinfo {year} {2022}{\natexlab{b}})\BibitemShut {NoStop}%
\bibitem [{\citenamefont {Pace}\ and\ \citenamefont
  {Wen}(2022)}]{Pace:2022wgl}%
  \BibitemOpen
  \bibfield  {author} {\bibinfo {author} {\bibfnamefont {S.~D.}\ \bibnamefont
  {Pace}}\ and\ \bibinfo {author} {\bibfnamefont {X.-G.}\ \bibnamefont {Wen}},\
  }\bibfield  {title} {\bibinfo {title} {{Position-dependent excitations and
  UV/IR mixing in the ZN rank-2 toric code and its low-energy effective field
  theory}},\ }\href {https://doi.org/10.1103/PhysRevB.106.045145} {\bibfield
  {journal} {\bibinfo  {journal} {Phys. Rev. B}\ }\textbf {\bibinfo {volume}
  {106}},\ \bibinfo {pages} {045145} (\bibinfo {year} {2022})},\ \Eprint
  {https://arxiv.org/abs/2204.07111} {arXiv:2204.07111 [cond-mat.str-el]}
  \BibitemShut {NoStop}%
\bibitem [{\citenamefont {Ebisu}(2023)}]{Ebisu:2023ayu}%
  \BibitemOpen
  \bibfield  {author} {\bibinfo {author} {\bibfnamefont {H.}~\bibnamefont
  {Ebisu}},\ }\bibfield  {title} {\bibinfo {title} {{Symmetric higher rank
  topological phases on generic graphs}},\ }\href
  {https://doi.org/10.1103/PhysRevB.107.125154} {\bibfield  {journal} {\bibinfo
   {journal} {Phys. Rev. B}\ }\textbf {\bibinfo {volume} {107}},\ \bibinfo
  {pages} {125154} (\bibinfo {year} {2023})},\ \Eprint
  {https://arxiv.org/abs/2302.03747} {arXiv:2302.03747 [cond-mat.str-el]}
  \BibitemShut {NoStop}%
\bibitem [{\citenamefont {Gorantla}\ \emph
  {et~al.}(2022{\natexlab{b}})\citenamefont {Gorantla}, \citenamefont {Lam},\
  and\ \citenamefont {Shao}}]{Gorantla:2022mrp}%
  \BibitemOpen
  \bibfield  {author} {\bibinfo {author} {\bibfnamefont {P.}~\bibnamefont
  {Gorantla}}, \bibinfo {author} {\bibfnamefont {H.~T.}\ \bibnamefont {Lam}},\
  and\ \bibinfo {author} {\bibfnamefont {S.-H.}\ \bibnamefont {Shao}},\
  }\bibfield  {title} {\bibinfo {title} {{Fractons on graphs and complexity}},\
  }\href {https://doi.org/10.1103/PhysRevB.106.195139} {\bibfield  {journal}
  {\bibinfo  {journal} {Phys. Rev. B}\ }\textbf {\bibinfo {volume} {106}},\
  \bibinfo {pages} {195139} (\bibinfo {year} {2022}{\natexlab{b}})},\ \Eprint
  {https://arxiv.org/abs/2207.08585} {arXiv:2207.08585 [cond-mat.str-el]}
  \BibitemShut {NoStop}%
\bibitem [{\citenamefont {Gorantla}\ \emph
  {et~al.}(2023{\natexlab{a}})\citenamefont {Gorantla}, \citenamefont {Lam},
  \citenamefont {Seiberg},\ and\ \citenamefont {Shao}}]{Gorantla:2022ssr}%
  \BibitemOpen
  \bibfield  {author} {\bibinfo {author} {\bibfnamefont {P.}~\bibnamefont
  {Gorantla}}, \bibinfo {author} {\bibfnamefont {H.~T.}\ \bibnamefont {Lam}},
  \bibinfo {author} {\bibfnamefont {N.}~\bibnamefont {Seiberg}},\ and\ \bibinfo
  {author} {\bibfnamefont {S.-H.}\ \bibnamefont {Shao}},\ }\bibfield  {title}
  {\bibinfo {title} {{(2+1)-dimensional compact Lifshitz theory, tensor gauge
  theory, and fractons}},\ }\href {https://doi.org/10.1103/PhysRevB.108.075106}
  {\bibfield  {journal} {\bibinfo  {journal} {Phys. Rev. B}\ }\textbf {\bibinfo
  {volume} {108}},\ \bibinfo {pages} {075106} (\bibinfo {year}
  {2023}{\natexlab{a}})},\ \Eprint {https://arxiv.org/abs/2209.10030}
  {arXiv:2209.10030 [cond-mat.str-el]} \BibitemShut {NoStop}%
\bibitem [{\citenamefont {Radzihovsky}(2022)}]{Radzihovsky_lifshitz_2022}%
  \BibitemOpen
  \bibfield  {author} {\bibinfo {author} {\bibfnamefont {L.}~\bibnamefont
  {Radzihovsky}},\ }\bibfield  {title} {\bibinfo {title} {Lifshitz gauge
  duality},\ }\bibfield  {journal} {\bibinfo  {journal} {Physical Review B}\
  }\textbf {\bibinfo {volume} {106}},\ \href
  {https://doi.org/10.1103/physrevb.106.224510} {10.1103/physrevb.106.224510}
  (\bibinfo {year} {2022})\BibitemShut {NoStop}%
\bibitem [{\citenamefont {Gorantla}\ \emph
  {et~al.}(2023{\natexlab{b}})\citenamefont {Gorantla}, \citenamefont {Lam},
  \citenamefont {Seiberg},\ and\ \citenamefont {Shao}}]{Gorantla:2022pii}%
  \BibitemOpen
  \bibfield  {author} {\bibinfo {author} {\bibfnamefont {P.}~\bibnamefont
  {Gorantla}}, \bibinfo {author} {\bibfnamefont {H.~T.}\ \bibnamefont {Lam}},
  \bibinfo {author} {\bibfnamefont {N.}~\bibnamefont {Seiberg}},\ and\ \bibinfo
  {author} {\bibfnamefont {S.-H.}\ \bibnamefont {Shao}},\ }\bibfield  {title}
  {\bibinfo {title} {{Gapped lineon and fracton models on graphs}},\ }\href
  {https://doi.org/10.1103/PhysRevB.107.125121} {\bibfield  {journal} {\bibinfo
   {journal} {Phys. Rev. B}\ }\textbf {\bibinfo {volume} {107}},\ \bibinfo
  {pages} {125121} (\bibinfo {year} {2023}{\natexlab{b}})},\ \Eprint
  {https://arxiv.org/abs/2210.03727} {arXiv:2210.03727 [cond-mat.str-el]}
  \BibitemShut {NoStop}%
\bibitem [{\citenamefont {Ohmori}\ and\ \citenamefont
  {Shimamura}(2023)}]{Ohmori:2022rzz}%
  \BibitemOpen
  \bibfield  {author} {\bibinfo {author} {\bibfnamefont {K.}~\bibnamefont
  {Ohmori}}\ and\ \bibinfo {author} {\bibfnamefont {S.}~\bibnamefont
  {Shimamura}},\ }\bibfield  {title} {\bibinfo {title} {{Foliated-exotic
  duality in fractonic BF theories}},\ }\href
  {https://doi.org/10.21468/SciPostPhys.14.6.164} {\bibfield  {journal}
  {\bibinfo  {journal} {SciPost Phys.}\ }\textbf {\bibinfo {volume} {14}},\
  \bibinfo {pages} {164} (\bibinfo {year} {2023})},\ \Eprint
  {https://arxiv.org/abs/2210.11001} {arXiv:2210.11001 [hep-th]} \BibitemShut
  {NoStop}%
\bibitem [{\citenamefont {Oh}\ \emph {et~al.}(2023)\citenamefont {Oh},
  \citenamefont {Pace}, \citenamefont {Han}, \citenamefont {You},\ and\
  \citenamefont {Lee}}]{Oh:2023bnk}%
  \BibitemOpen
  \bibfield  {author} {\bibinfo {author} {\bibfnamefont {Y.-T.}\ \bibnamefont
  {Oh}}, \bibinfo {author} {\bibfnamefont {S.~D.}\ \bibnamefont {Pace}},
  \bibinfo {author} {\bibfnamefont {J.~H.}\ \bibnamefont {Han}}, \bibinfo
  {author} {\bibfnamefont {Y.}~\bibnamefont {You}},\ and\ \bibinfo {author}
  {\bibfnamefont {H.-Y.}\ \bibnamefont {Lee}},\ }\bibfield  {title} {\bibinfo
  {title} {{Aspects of ZN rank-2 gauge theory in (2+1) dimensions: Construction
  schemes, holonomies, and sublattice one-form symmetries}},\ }\href
  {https://doi.org/10.1103/PhysRevB.107.155151} {\bibfield  {journal} {\bibinfo
   {journal} {Phys. Rev. B}\ }\textbf {\bibinfo {volume} {107}},\ \bibinfo
  {pages} {155151} (\bibinfo {year} {2023})},\ \Eprint
  {https://arxiv.org/abs/2301.04706} {arXiv:2301.04706 [cond-mat.str-el]}
  \BibitemShut {NoStop}%
\bibitem [{\citenamefont {Du}\ \emph {et~al.}(2023)\citenamefont {Du},
  \citenamefont {Lam},\ and\ \citenamefont {Radzihovsky}}]{Du:2023zdq}%
  \BibitemOpen
  \bibfield  {author} {\bibinfo {author} {\bibfnamefont {Y.-H.}\ \bibnamefont
  {Du}}, \bibinfo {author} {\bibfnamefont {H.~T.}\ \bibnamefont {Lam}},\ and\
  \bibinfo {author} {\bibfnamefont {L.}~\bibnamefont {Radzihovsky}},\
  }\bibfield  {title} {\bibinfo {title} {{Quantum vortex lattice via Lifshitz
  duality}},\ }\href@noop {} {\  (\bibinfo {year} {2023})},\ \Eprint
  {https://arxiv.org/abs/2310.13794} {arXiv:2310.13794 [cond-mat.str-el]}
  \BibitemShut {NoStop}%
\bibitem [{\citenamefont {Nguyen}\ and\ \citenamefont
  {Moroz}(2023)}]{Nguyen:2023quf}%
  \BibitemOpen
  \bibfield  {author} {\bibinfo {author} {\bibfnamefont {D.~X.}\ \bibnamefont
  {Nguyen}}\ and\ \bibinfo {author} {\bibfnamefont {S.}~\bibnamefont {Moroz}},\
  }\bibfield  {title} {\bibinfo {title} {{On quantum melting of superfluid
  vortex crystals: from Lifshitz scalar to dual gravity}},\ }\href@noop {} {\
  (\bibinfo {year} {2023})},\ \Eprint {https://arxiv.org/abs/2310.13741}
  {arXiv:2310.13741 [cond-mat.quant-gas]} \BibitemShut {NoStop}%
\bibitem [{\citenamefont {Else}\ and\ \citenamefont
  {Nayak}(2014{\natexlab{b}})}]{Else:2014vma}%
  \BibitemOpen
  \bibfield  {author} {\bibinfo {author} {\bibfnamefont {D.~V.}\ \bibnamefont
  {Else}}\ and\ \bibinfo {author} {\bibfnamefont {C.}~\bibnamefont {Nayak}},\
  }\bibfield  {title} {\bibinfo {title} {{Classifying symmetry-protected
  topological phases through the anomalous action of the symmetry on the
  edge}},\ }\href {https://doi.org/10.1103/PhysRevB.90.235137} {\bibfield
  {journal} {\bibinfo  {journal} {Phys. Rev. B}\ }\textbf {\bibinfo {volume}
  {90}},\ \bibinfo {pages} {235137} (\bibinfo {year} {2014}{\natexlab{b}})},\
  \Eprint {https://arxiv.org/abs/1409.5436} {arXiv:1409.5436 [cond-mat.str-el]}
  \BibitemShut {NoStop}%
\bibitem [{\citenamefont {Kapustin}\ and\ \citenamefont
  {Seiberg}(2014)}]{Kapustin:2014gua}%
  \BibitemOpen
  \bibfield  {author} {\bibinfo {author} {\bibfnamefont {A.}~\bibnamefont
  {Kapustin}}\ and\ \bibinfo {author} {\bibfnamefont {N.}~\bibnamefont
  {Seiberg}},\ }\bibfield  {title} {\bibinfo {title} {{Coupling a QFT to a TQFT
  and Duality}},\ }\href {https://doi.org/10.1007/JHEP04(2014)001} {\bibfield
  {journal} {\bibinfo  {journal} {JHEP}\ }\textbf {\bibinfo {volume} {04}},\
  \bibinfo {pages} {001}},\ \Eprint {https://arxiv.org/abs/1401.0740}
  {arXiv:1401.0740 [hep-th]} \BibitemShut {NoStop}%
\bibitem [{\citenamefont {Villain}(1975)}]{Villain:1974ir}%
  \BibitemOpen
  \bibfield  {author} {\bibinfo {author} {\bibfnamefont {J.}~\bibnamefont
  {Villain}},\ }\bibfield  {title} {\bibinfo {title} {{Theory of
  one-dimensional and two-dimensional magnets with an easy magnetization plane.
  2. The Planar, classical, two-dimensional magnet}},\ }\href
  {https://doi.org/10.1051/jphys:01975003606058100} {\bibfield  {journal}
  {\bibinfo  {journal} {J. Phys.(France)}\ }\textbf {\bibinfo {volume} {36}},\
  \bibinfo {pages} {581} (\bibinfo {year} {1975})}\BibitemShut {NoStop}%
\end{thebibliography}%

\end{document}